\documentclass[fleqn,usenatbib]{mnras}

\usepackage[T1]{fontenc}
\RequirePackage{rotating}
\usepackage{graphicx}	
\usepackage{amsmath}	
\usepackage{subcaption}
\usepackage{xcolor}
\usepackage{placeins}
\usepackage{pdflscape}
\setcitestyle{citesep={,}}
\usepackage{hyperref}
\DeclareRobustCommand{\VAN}[3]{#2}
\let\VANthebibliography\thebibliography
\def\thebibliography{\DeclareRobustCommand{\VAN}[3]{##3}\VANthebibliography}

\usepackage{graphicx}	
\usepackage{subcaption}
\usepackage{hyperref}
\usepackage{lscape}

\usepackage{xcolor}
\usepackage{placeins}
\setcitestyle{citesep={,}}

\usepackage{amssymb}
\usepackage{newtxtext,newtxmath}

\newcommand{\todo}{\textcolor{black}}
\newcommand{\degree}{$^{\circ}$}

\newcommand{\hhh}{$^{\mathrm{h}}$}
\newcommand{\mmm}{$^{\mathrm{m}}$}
\newcommand{\sss}{$^{\mathrm{s}}$}
\newcommand{\ddd}{$^{\mathrm{\circ}}$}
\newcommand{\dmm}{$^{\prime}$}

\title[MIGHTEE Continuum DR1]{{MIGHTEE: The Continuum Survey Data Release 1}}

\author[Hale, Heywood, Jarvis, et al.]{C. L. Hale,$^{1,2}$\thanks{E-mail: Catherine.Hale@physics.ox.ac.uk}
I. Heywood,$^{1,3,4}$\thanks{E-mail: Ian.Heywood@physics.ox.ac.uk}
M. J. Jarvis,$^{1,5}$\thanks{E-mail: Matt.Jarvis@physics.ox.ac.uk}
I. H. Whittam,$^{1}$
P. N. Best,$^{2}$
Fangxia An,$^{6, 7}$
R. A. A. Bowler,$^{8}$ 
\and
I. Harrison,$^{9}$
A. Matthews,$^{10}$
D. J. B. Smith,$^{11}$
A. R. Taylor,$^{12, 13, 5}$
M. Vaccari$^{12, 7, 14}$
\\
\noindent 
$^{1}$ Astrophysics, Department of Physics, University of Oxford, Keble Road, Oxford, OX1 3RH, UK 
\\
$^{2}$ Institute for Astronomy, Royal Observatory Edinburgh, Blackford Hill, Edinburgh, EH9 3HJ, UK
\\
$^{3}$ Centre for Radio Astronomy Techniques and Technologies, Department of Physics and Electronics, Rhodes University, PO Box 94, Makhanda, 6140, South Africa. 
\\
$^{4}$ South African Radio Astronomy Observatory, 2 Fir Street, Black River Park, Observatory, Cape Town, 7925, South Africa.
\\
$^{5}$ Department of Physics and Astronomy, University of the Western Cape, Robert Sobukwe Road, 7535 Bellville, Cape Town, South Africa. \\
$^{6}$ Purple Mountain Observatory, Chinese Academy of Sciences, 10 Yuanhua Road, Qixia District, Nanjing 210023, People's Republic of China \\
$^{7}$Inter-University Institute for Data Intensive Astronomy, and Department of Physics and Astronomy, University of the Western Cape, Robert Sobukwe Road, \\ \ \ 7535 Bellville, Cape Town, South Africa \\
$^{8}$ Jodrell Bank Centre for Astrophysics, Department of Physics and Astronomy, School of Natural Sciences, The University of Manchester, Manchester, M13 9PL, UK \\
$^{9}$ School of Physics and Astronomy, Cardiff University, The Parade, Cardiff, Wales CF24 3AA, UK \\
$^{10}$ Carnegie Observatories, 813 Santa Barbara Street, Pasadena, CA 91101, USA \\
$^{11}$ Centre for Astrophysics Research, University of Hertfordshire, College Lane, Hatfield AL10 9AB, UK \\
$^{12}$ Inter-University Institute for Data Intensive Astronomy, Department of Astronomy, University of Cape Town, 7701 Rondebosch, Cape Town, South Africa \\
$^{13}$ Department of Astronomy, University of Cape Town, Rondebosch, Cape Town, 7701, South Africa \\
$^{14}$ INAF - Istituto di Radioastronomia, via Gobetti 101, 40129 Bologna, Italy \\
}

\date{Accepted XXX. Received YYY; in original form ZZZ}

\pubyear{2024}

\begin{document}
\label{firstpage}
\pagerange{\pageref{firstpage}--\pageref{lastpage}}
\maketitle

\begin{abstract}
The MeerKAT International {GHz} Tiered Extragalactic Exploration Survey (MIGHTEE) is one of the large survey projects using the MeerKAT telescope, covering four fields that have a wealth of ancillary data {available}. We present Data Release 1 of the MIGHTEE continuum survey, releasing total intensity images and catalogues over $\sim$20 deg$^2$, across three fields {at $\sim$1.2-1.3 GHz}. This includes 4.2 deg$^2$ over the {Cosmic Evolution Survey (COSMOS)} field, 14.4 deg$^2$ over the XMM Large-Scale Structure (XMM-LSS) field and deeper imaging over {1.5} deg$^2$ of the Extended Chandra Deep Field South (CDFS). We release images at {both a lower} resolution (7--9~arcsec) and higher resolution ($\sim 5$\,arcsec). {These images have central rms sensitivities of $\sim$1.3$-$2.7 $\muup$Jy beam$^{-1}$ ($\sim$1.2$-$3.6 $\muup$Jy beam$^{-1}$) in the lower (higher) resolution images respectively.}
We also release catalogues comprised of $\sim$144~000 ($\sim$114~000) sources using the {lower (higher) resolution} images. We compare the astrometry and flux-density calibration with the Early Science data in the COSMOS and XMM-LSS fields and previous radio observations in the CDFS field, finding {broad agreement}. {Furthermore, we} extend the source counts at the {$\sim$10 $\muup$Jy} level to these larger {areas} ($\sim 20$~deg$^2$) {and,} using the areal coverage of MIGHTEE we measure the sample variance {for differing areas of sky}. {We find a typical sample variance of 10-20 per cent for 0.3 and 0.5 sq. deg. sub-regions at $S_{1.4} \leq 200$ $\muup$Jy, which increases at brighter flux densities, given the lower source density and expected higher galaxy bias for these sources.}
\end{abstract}



\section{Introduction}
Radio surveys provide a dust-free view of the active Universe. At relatively high radio flux density ({e.g. at 1.4 GHz}, $S_{1.4} \gtrsim$ 1 mJy), radio surveys can cover huge swathes of sky quickly \citep[e.g.][]{NVSS,FIRST,SUMSS,RACS, VLASS}, providing a census of the rare, bright radio source population over the vast majority of cosmic time. {Surveys at these relatively high flux densities are dominated} by sources where the radio emission arises from synchrotron radiation from jets and lobes as a result of accretion on to the central supermassive black holes in galaxies {(known as Active Galactic Nuclei, or AGN) and observations of these sources} provides a plethora of information on the physical conditions of these high-energy phenomena. The fact that we are able to observe these AGN across the vast majority of cosmic time \citep[e.g.][]{debreuck2000,Jarvis2009,Saxena2018}, means we are also able to trace the evolution of radio-loud AGN \citep[e.g.][]{Willott2001,Jarvis2001,ClewleyJarvis2004,Rigby2011}. {This allows insights} into their relation to their host galaxy properties \citep[e.g.][]{McLure1999,Dunlop2003,Gurkan2014,Saxena2019} and their larger-scale environment \citep[e.g.][]{RawlingsJarvis2004,McNamaraNulsen2007,Fabian2012,Magliocchetti2022} {to be studied}. {Such surveys have revealed how radio AGN populations are related to their accretion rate} \citep[e.g.][]{BestHeckman2012,HeckmanBest2014,Mingo2014,Whittam2018} through both positive and negative feedback processes \citep[e.g.][]{Bower2006,Croton2006,Hardcastle2007,Fernandes2015,Kalfountzou2017}.

In addition to the synchrotron radiation that is emitted due to the accretion activity on to supermassive black holes, radio surveys are also able to detect the integrated synchrotron emission from the cosmic rays accelerated due to the young massive stars becoming supernovae \cite[e.g.][]{Condon1992}. Given that the radio emission has a long wavelength in relation to the size of dust grains, the radio waves are able to penetrate any dusty regions in the host galaxies, or along the line of sight, and thus potentially provide a dust-free measurement of the star-formation rate in galaxies. As such, {radio surveys} provide an important tracer of both star-formation and AGN activity across cosmic time. {The wide and shallow surveys, highlighted above, are only able to detect these star forming galaxies {(SFGs)} at relatively low redshift \citep[e.g.][]{Yun2001,Bell2003,MauchSadler2007,Jarvis2010}. In order to {probe star}-forming galaxies (and low-luminosity AGN) across a wide redshift range, radio surveys {much deeper than the mJy regime (at 1.4 GHz)} are required. {This is demonstrated by recent cross-matched and classified radio data from both the LOFAR Deep Fields \citep{Kondapally2022, Best2023} and MIGHTEE Early Science \citep{Whittam2022, Whittam2023}, which show that SFGs have a typical median redshift of $z$$\sim$$0.1$ for $S_{1.4 \textrm{GHz}}\geq$1 mJy, increasing to $z$$\sim$0.8$-$0.9 across the full, deeper, samples.}}
 
Indeed, {we have} seen significant progress in our understanding of how radio emission may trace star formation and low-luminosity AGN in distant galaxies. {This is from both wider surveys which probe the sub-mJy regime, e.g. the Faint Images of the Radio Sky Survey {\citep[FIRST; ][]{White1997} and the LOw Frequency Array (LOFAR) Two-metre Sky Survey \citep[LoTSS;][]{Shimwell2022} and deep surveys over smaller areas \citep[e.g.][]{Bondi2003,Smolcic2017,Heywood2020,Sabater2021}.}} {SFGs become a more dominant population for deeper surveys, and their star-formation} rate is often investigated through the study of the far-infrared--radio correlation \citep[FIRC e.g.][]{Garrett2002,Appleton2004,Ivison2010,Delhaize2017}. Several recent studies have investigated the mass-dependence of the {FIRC or the relationship of radio luminosity-to-star formation rate. These studies} demonstrate that the observed evolution in the {such relationships} with redshift could be largely explained by {introducing a stellar-mass dependence} \citep[e.g.][]{Gurkan2018,Smith2021,Delvecchio2021}, {with other work considering its luminosity dependence \citep[see][]{Matthews2021b}}. Due to the fact that the ancillary data that {are} required to measure the stellar masses of galaxies detected at both far-infrared and radio wavelengths is flux-limited, those galaxies that are observed at higher redshifts necessarily also have higher average stellar masses than the galaxies observed in the low-redshift Universe. Alternatively, the dependence could also be related to the morphology of the galaxy \citep{Molnar2018}, which is in turn also correlated with the mass, with ellipticals generally being more massive than spirals in flux-limited samples. {Thus, the combination of deep radio data and deep multi-wavelength data is essential for such studies, allowing the possibility of redshifts, host galaxies and galaxy properties to be identified for the radio sources. This is the main reason why the new generation of deep radio surveys, such as the focus of this paper namely the MeerKAT International {GHz} Tiered Extragalactic Exploration \citep[MIGHTEE; ][]{Jarvis2016} survey, are designed to overlap with those extragalactic deep fields with the best multi-wavelength data.}

{These recent deeper radio surveys} have also been used to measure the radio emission due to AGN activity to much lower {luminosities}, demonstrating that the space-density evolution is luminosity dependent \citep[e.g.][]{Rigby2015,Yuan2017,Ceraj2018,Slaus2020,Kondapally2022} and that the correspondence between radio-luminosity and accretion rate \citep[e.g.][]{Mingo2014, Whittam2018}, host galaxy properties \citep[e.g.][]{Radcliffe2021,Delvecchio2022,Ji2022,Best2023} and the underlying dark matter haloes \citep[e.g.][]{Lindsay2014,Hale2018,Alonso2021} is complicated. Furthermore, these surveys are deep enough to begin to probe the radio emission from large samples of those AGN classified as radio quiet \citep[e.g.][]{White2015,White2017,Panessa2019,Ceraj2020,Mcfarlane2021}. 
 
Thus, we are entering a realm of detailed understanding of evolution of star formation and AGN activity in galaxies from their radio emission. However, {the inter-dependencies between star formation and AGN activity also highlights} the need to combine deep radio observations with deep multi-wavelength data covering a wide enough area to fully sample the radio luminosity function for star-forming galaxies and all types of AGN, above and below the knee in respective luminosity functions. {These inter-dependencies also need to be studied in the context of} whilst enabling the role of the environment to be investigated, {whilst mitigating the effects of cosmic variance}.

{The MIGHTEE Survey} \citep{Jarvis2016} is a radio survey {using the MeerKAT telescope \citep[][]{Jonas2009, Jonas2016}} aiming to provide this combination of depth and area, over some of the most well-studied extragalactic deep fields accessible from the southern hemisphere, in order to accelerate {our} understanding of galaxy and AGN evolution. The Early Science Data Release from this survey covered an area of 1.6\,deg$^2$ in the COSMOS field and 3.5\,deg$^2$ in the XMM-LSS field \citep{Heywood2022}, and provided the basis for investigations across the full range of science outlined above, including the discovery of two giant radio galaxies \citep{Delhaize2021}, investigations of the FIRC to high redshift \citep{Delvecchio2021,An2021}, probing the accretion rates to much lower radio luminosities than previously possible \citep{Whittam2022} and measuring the contribution of star-forming galaxies and AGN to the radio sky background temperature \citep{Hale2023}.

In this paper we describe the processing of the MIGHTEE data over an extended area in each of the COSMOS and XMM-LSS fields {(which are the final MIGHTEE L-band images of these fields)}, along with a very deep single pointing exposure centred on the Chandra-Deep-Field South (CDFS, denoted in this work as CDFS-DEEP). {This is known as MIGHTEE Data Release 1 (DR1). The deep observations of the CDFS field are} taken as part of the Looking At the Distant Universe with the MeerKAT Array \citep[LADUMA; ][]{Blyth2016} survey, whose main science goals are to determine the role of neutral atomic {H}ydrogen in galaxy evolution. We describe the observations and data processing, including a description of the effective frequency maps over all fields, in Section~\ref{sec:radiodata}. We {outline} how the radio source and component catalogues are constructed in Section~\ref{sec:radiocatalogues}. 
 In Section~\ref{sec:validation} we provide an overview of the quality of the data in terms of astrometric uncertainties and the flux-density scale and in Section~\ref{sec:sourcecounts} we present the source counts across the three fields, which comprise the deepest radio continuum data ever released over several degree-scale areas of sky{. We also} present results on the level of sample variance between the three fields as a function of flux-density and survey area. We summarise the results of these analyses and draw conclusions in Section \ref{sec:conclusions}.

\section{Observations and data processing}
\label{sec:radiodata}

This work makes use of a total of 86 individual MeerKAT tracks {at L-band (856 to 1711 MHz)} totalling 709.2 h, with 22, 45, and 19 pointings respectively in COSMOS, XMM-LSS and CDFS-DEEP. The total on-target time in each field is 139.6~h (COSMOS), 297.9~h (XMM-LSS), and 126.7~h (CDFS-DEEP). For the COSMOS and XMM-LSS fields, the images released in this Data Release 1 are the final L-band observations for these fields. A larger area of the CDFS field will be released in a future L-band data release, for which the expected coverage is shown in Appendix C1 of \cite{Heywood2022} {and will consist of 291~h of on source time. Alongside this there will additionally be L-band observations of the ELAIS S1 field, totalling 72~h on source.} Further details of the observations {and pointings} for each field are provided in the Appendix in Tables \ref{tab:cosmos_obs}, \ref{tab:cdfs_obs} and {\ref{tab:xmm_obs}}. The data processing took place on a per-track basis up to the mosaicking point, and followed a similar procedure to that described in detail by \citet{Heywood2022}, with scripts available online in the {\sc oxkat} repository \citep{heywood2020a}. {As for the Early Science data of \cite{Heywood2022}, primary calibrators were observed at least twice within a block for 5-10 minutes, with secondary calibrators visited regularly throughout the target observations but for a shorter duration ($\sim$2-3 minutes).} The underlying software packages were containerised using {\sc singularity} \citep{kurtzer2017}, and the majority of the data processing took place on the \emph{ilifu} cluster\footnote{\url{https://www.ilifu.ac.za/}} in Cape Town. A brief summary of the procedure is given as follows. 

\subsection{Flagging, calibration and imaging}
The data were retrieved in Measurement Set format from the South African Radio Astronomy Observatory (SARAO) archive\footnote{\url{https://archive.sarao.ac.za/}}, and {were} immediately averaged down to 1024 frequency channels {(836~kHz in width)} in the process\footnote{{The total number of channels in each observations are given in Tables \ref{tab:cosmos_obs}-\ref{tab:xmm_obs}.}}. Following initial flagging of the data, bandpass, delay and time-dependent gain solutions were derived using the primary and secondary calibrators using {\sc casa} \citep{casa2022}. The solutions were applied to the target data, which {were} flagged using the {\sc tricolour} software \citep{hugo2022}, which implements the {\sc sumthreshold} algorithm \citep{offringa2010} optimised for MeerKAT and accelerated using {\sc dask-ms} \citep{perkins2022}.

The flagged and reference-calibrated target data were then imaged using {\sc wsclean} \citep{offringa2014}, with deconvolution proceeding blindly, {terminating with a relatively shallow threshold where the peak residual drops below 20~$\muup$Jy~beam$^{-1}$}. The same algorithm that is implemented in the {\sc breizorro} \citep{ramaila2023} tool was then used to construct a mask from this initial image for a subsequent round of constrained deconvolution. {The second round is deeper, forming a highly-complete sky model, with the deconvolution within the masked region terminating at a peak residual of 1~$\muup$Jy~beam$^{-1}$.} The model visibilities corresponding to the spectral clean component model derived from this second round of deconvolution were used as the basis for a single round of phase and delay self-calibration using the {\sc cubical} package \citep{kenyon2018}, and the self-calibrated data were re-imaged.

Visual inspection occurs at this stage, and if it is deemed necessary to subtract a strong problematic source from the field then the data were re-imaged at higher spectral resolution with a refined deconvolution mask, and a direction-dependent peeling\footnote{The `peeling' process has a firm original definition \citep{noordam2004} however it is commonly (mis)used to describe any process that removes a problematic source from the visibility database via a method that is more advanced than a straightforward model subtraction. We are also guilty of misusing the term here, as the subtraction of problematic sources from the field of view in our processing is more accurately described as a differential gains approach \citep{smirnov2011} with only a single additional solvable gain term in the Measurement Equation.} step was performed using {\sc cubical}. In total, peeling was required for 14 of the 45 pointings in XMM-LSS, only one of the 22 pointings in COSMOS, and all of the 19 pointings in CDFS-DEEP. 

Whether a source has been peeled or not, the final stage of the processing for each MeerKAT track involved imaging of the data using {\sc ddfacet} \citep{tasse2018,tasse2023a}. Residual direction-dependent errors (DDEs) were solved for by manually identifying 10--20 sources across the field of view. {These sources are generally ones that are bright enough to provide high signal to noise ratio gain solutions, and exhibit residual PSF-like structures due to the the DDEs. At L-band the DDEs are dominated by the antenna primary beam pattern coupled with stochastic pointing errors that differ from antenna to antenna, thus this subset of sources is generally uniformly distributed around the flank of the main lobe of the primary beam, and into the sidelobes.} A tesselation algorithm partitions the sky according to their positions, and direction-dependent gain terms are solved for on a per-tessel basis, {with solution intervals of 5 minutes in time and 107~MHz in frequency,} using the {\sc killms} package \citep{smirnov2015,tasse2023b}. Two subsequent rounds of imaging with {\sc ddfacet} applying these directional solutions resulted in the final images, with each run producing an image at two different resolutions via two different values (0.0 and {$-$1.2}) of the \citet{briggs1995} robust parameter. 

\subsection{Mosaicking}

For each target field (including CDFS-DEEP), the images produced from each individual MeerKAT observation were combined in the image domain. First, the constituent pointings were all homogenised to a common resolution, namely a circular Gaussian beam with a full-width half-maximum dictated by the largest value of the fitted beam major axis from the set of images of a given field. The clean component model image was directly convolved with this Gaussian, whereas the residual image was convolved with a homogenisation kernel computed using the {\sc pypher} package \citep{boucaud2016} prior to being summed with the convolved model. Following this, each convolved image was primary beam corrected, using a model of the MeerKAT Stokes I beam evaluated at the nominal band-centre frequency using the {\sc katbeam}\footnote{\url{https://github.com/ska-sa/katbeam}} library. The Stokes I beam model was also azimuthally-averaged in order to average out the asymmetries in the MeerKAT main lobe. The resulting homogenised, primary beam corrected images were then linearly mosaicked using the {\sc montage}\footnote{\url{http://montage.ipac.caltech.edu/}} toolkit, using the square of the primary beam (equivalent to variance weighting) as a weighting function.

\begin{figure*}
     \begin{minipage}[b]{0.5\textwidth}
 \includegraphics[width=\textwidth]{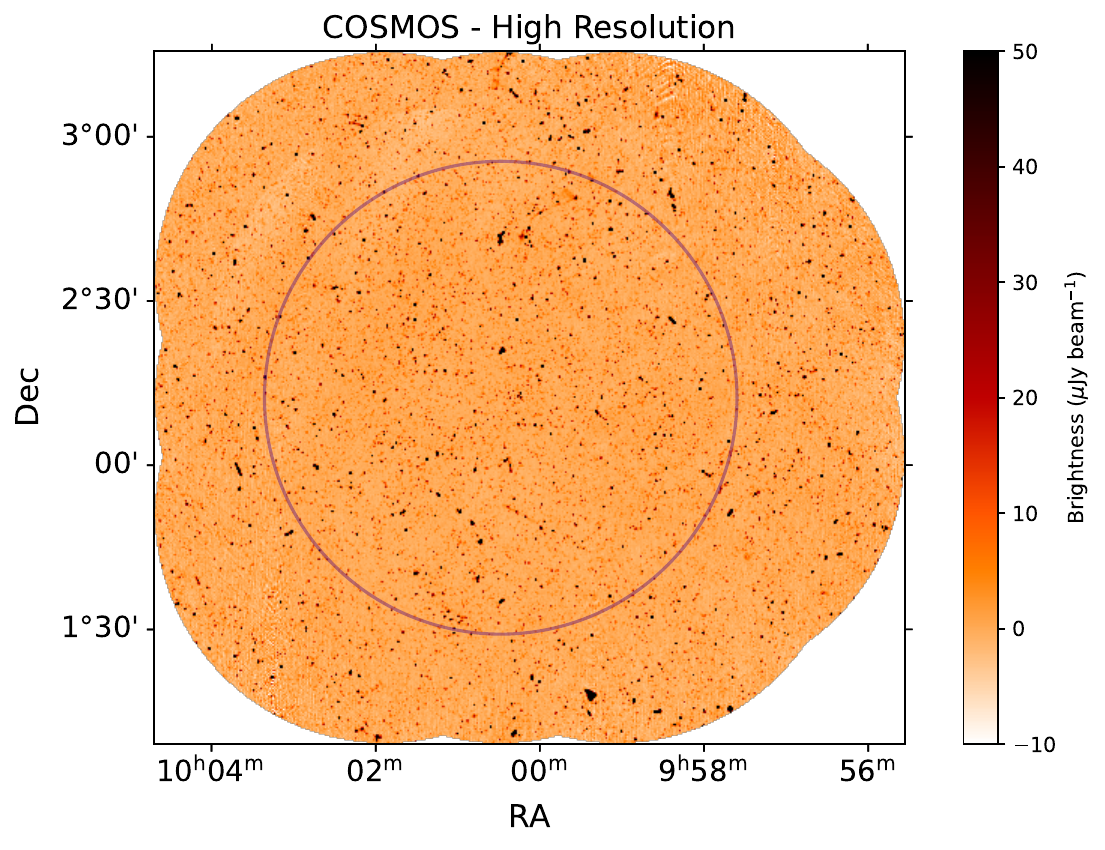}
    \end{minipage}%
    \begin{minipage}[b]{0.5\textwidth}
\includegraphics[width=\textwidth]{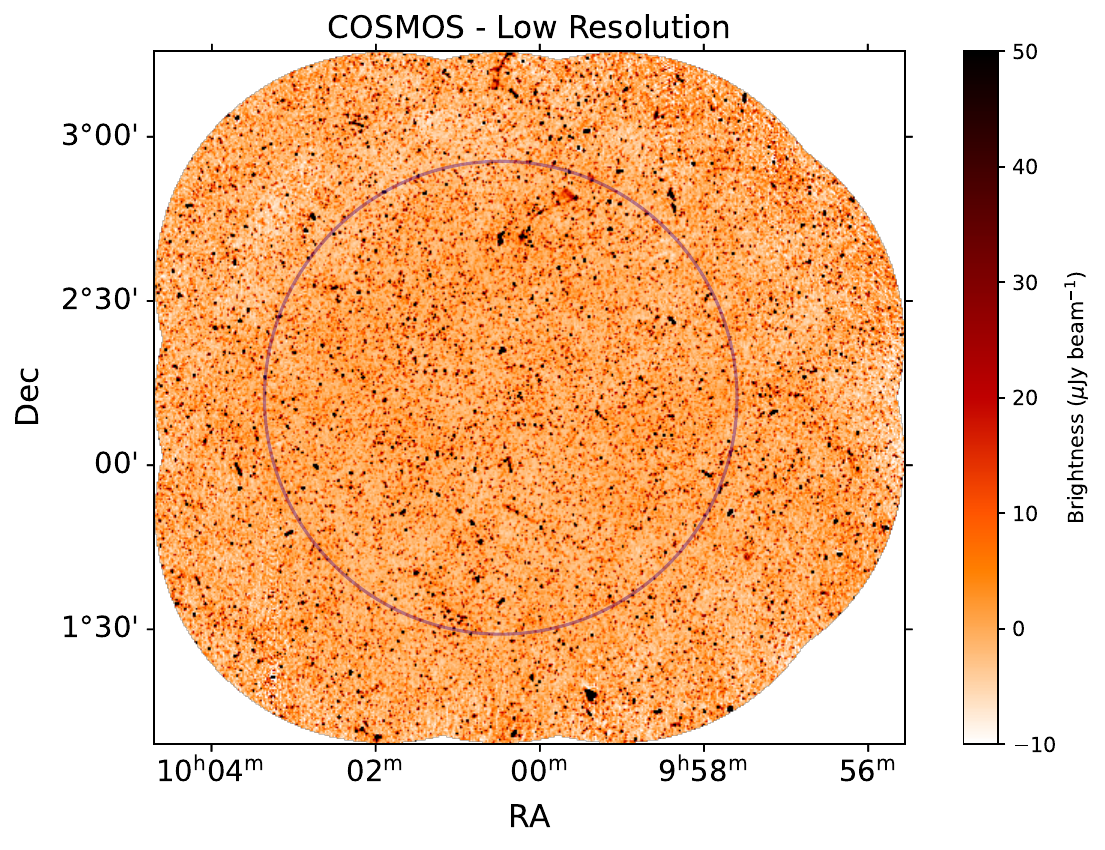}
    \end{minipage}%

    \begin{minipage}[b]{0.5\textwidth}
 \includegraphics[width=\textwidth]{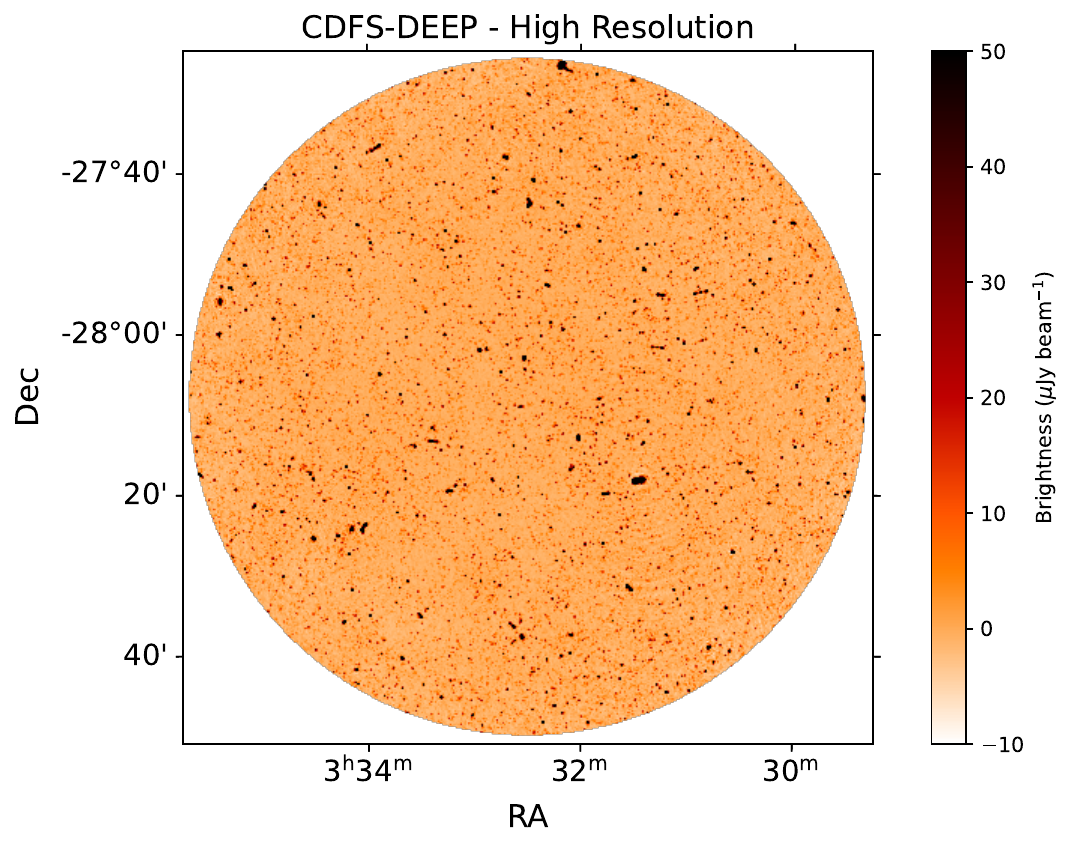}
    \end{minipage}%
    \begin{minipage}[b]{0.5\textwidth}
\includegraphics[width=\textwidth]{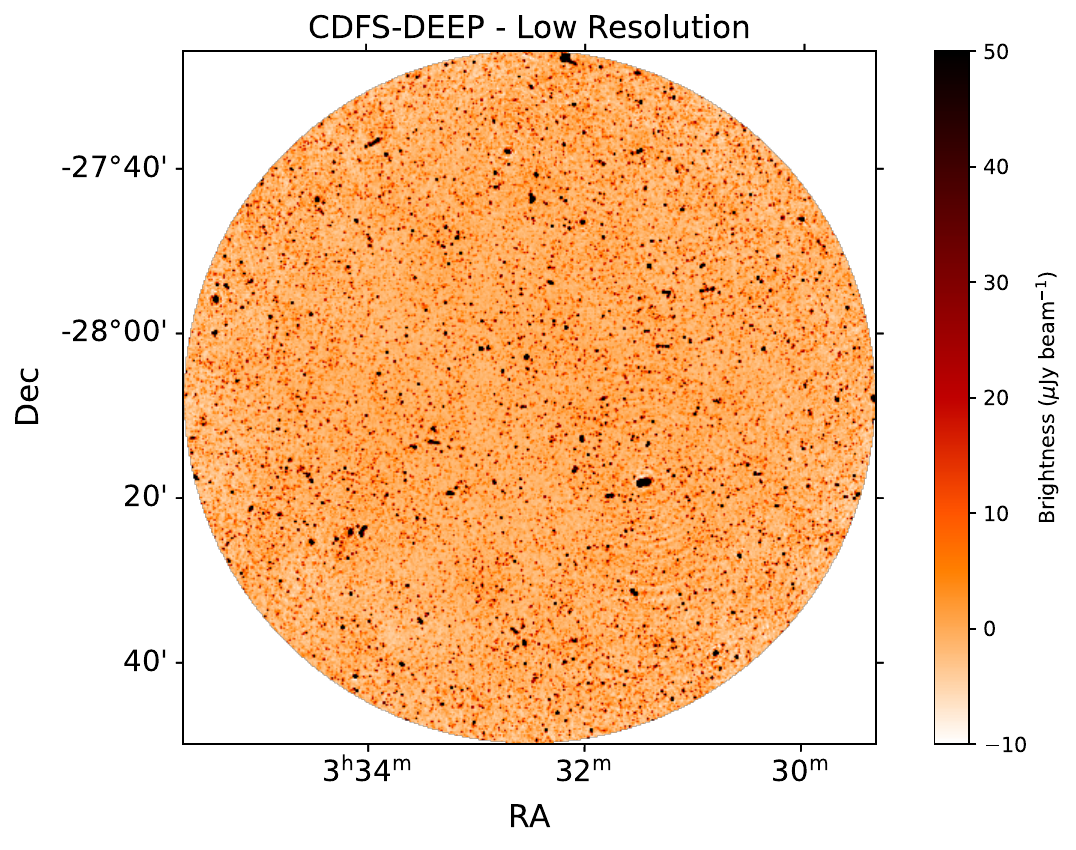}
    \end{minipage}%
    
         \begin{minipage}[b]{0.5\textwidth}
 \includegraphics[width=\textwidth]{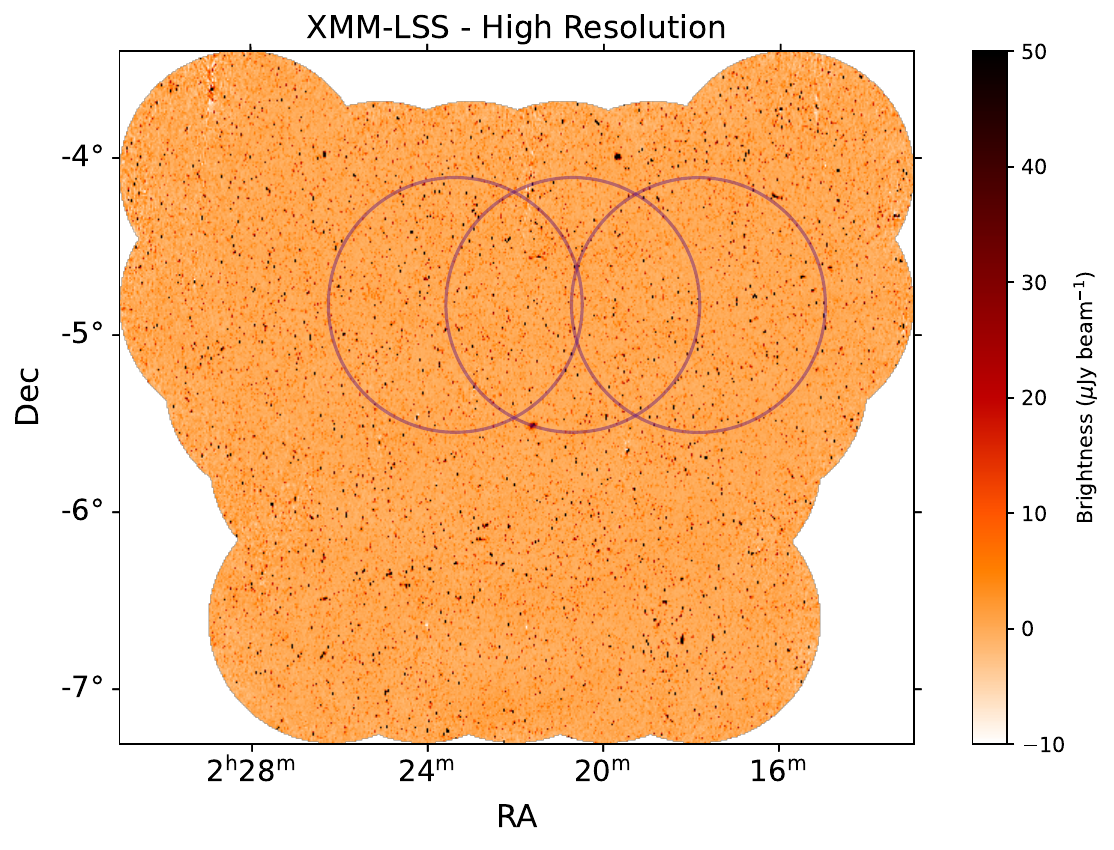}
    \end{minipage}%
    \begin{minipage}[b]{0.5\textwidth}
\includegraphics[width=\textwidth]{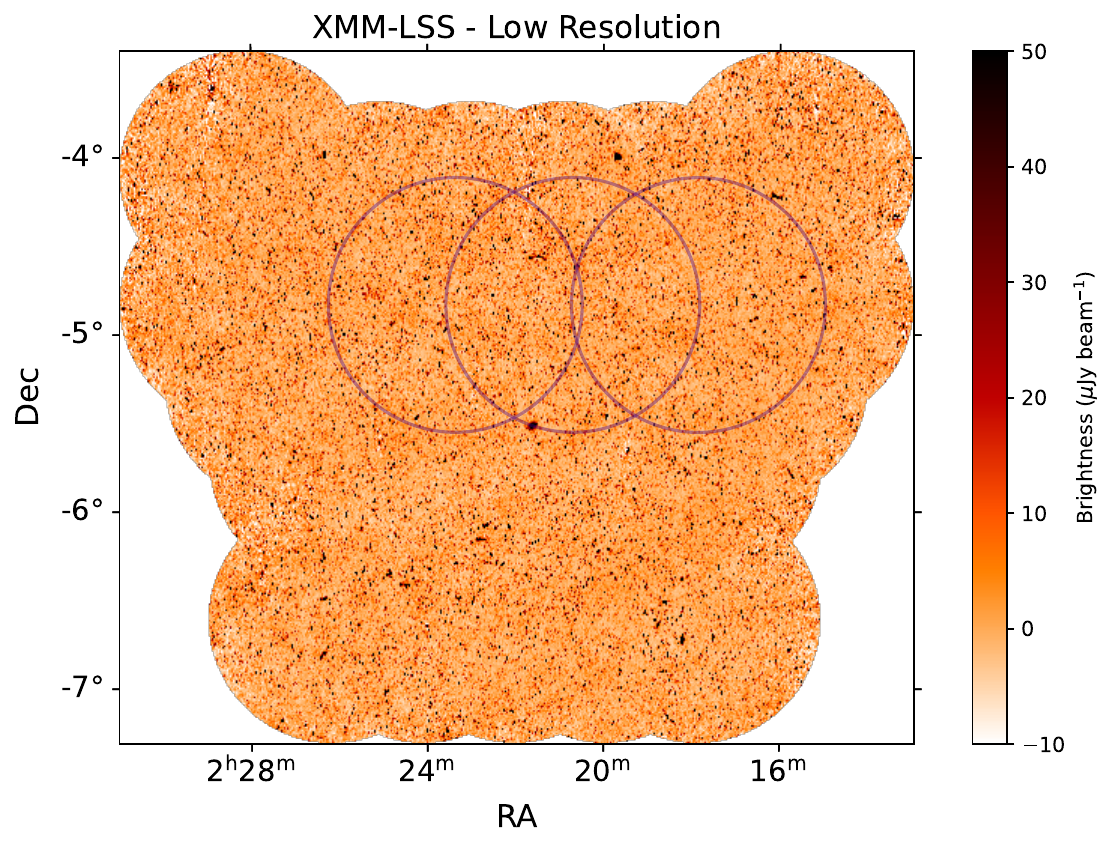}
    \end{minipage}%
     \caption{{The brightness distributions} for each of the MIGHTEE Data Release 1 images which are released with this work. This is shown for COSMOS (top row), the deep pointing in CDFS (middle row) and XMM-LSS (bottom row) for the high resolution images (left, 5.2\arcsec, 5.5\arcsec \ and 5.0\arcsec \ beam full-width half max, FWHM, each field respectively) and the low resolution images (right, 8.9\arcsec, 7.3\arcsec \ and 8.9\arcsec \ {beam FWHM} respectively). For the COSMOS and XMM-LSS {fields} the extent of the pointings of the Early Science data of \protect \cite{Heywood2022} are indicated by {purple} circles.  }
     \label{fig:images}
 \end{figure*}

\begin{figure*}
     \begin{minipage}[b]{0.74\textwidth}
 \includegraphics[width=\textwidth]{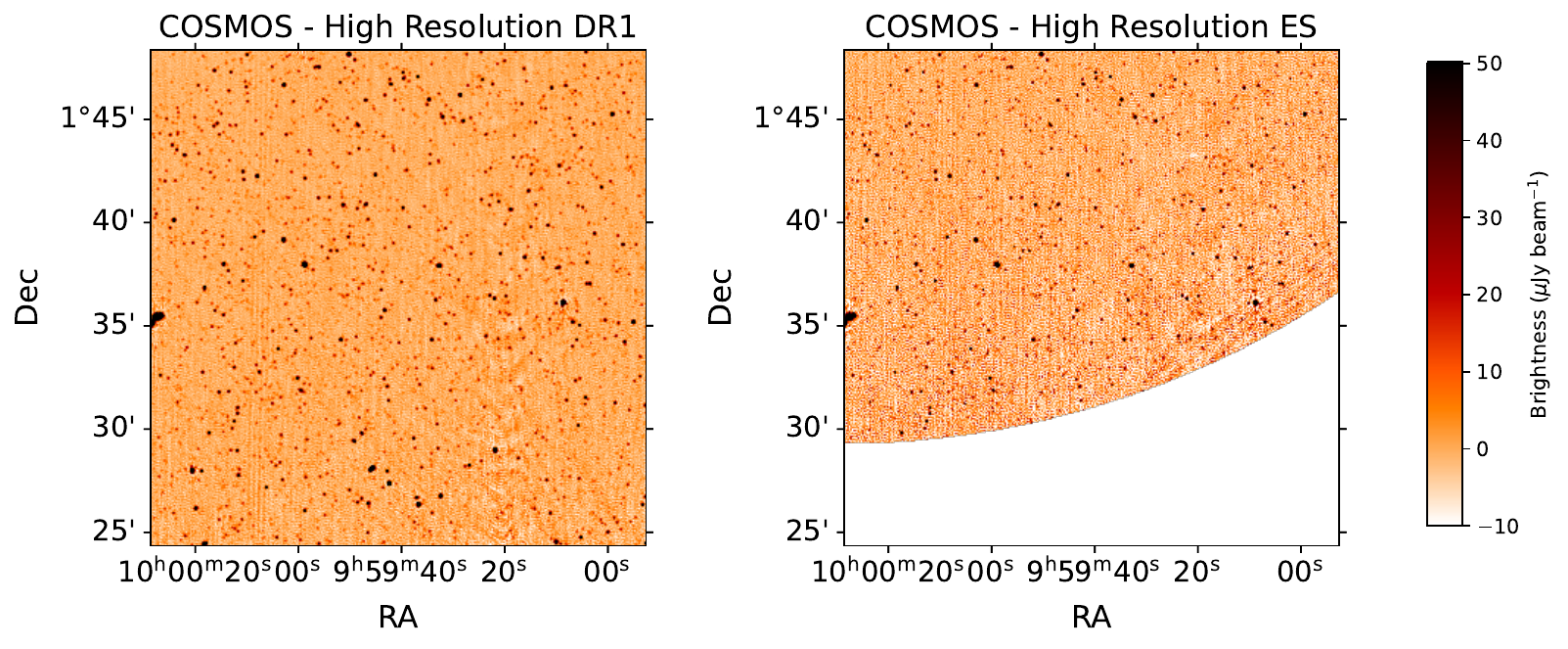}
    \end{minipage}%
    
    \begin{minipage}[b]{0.74\textwidth}
\includegraphics[width=\textwidth]{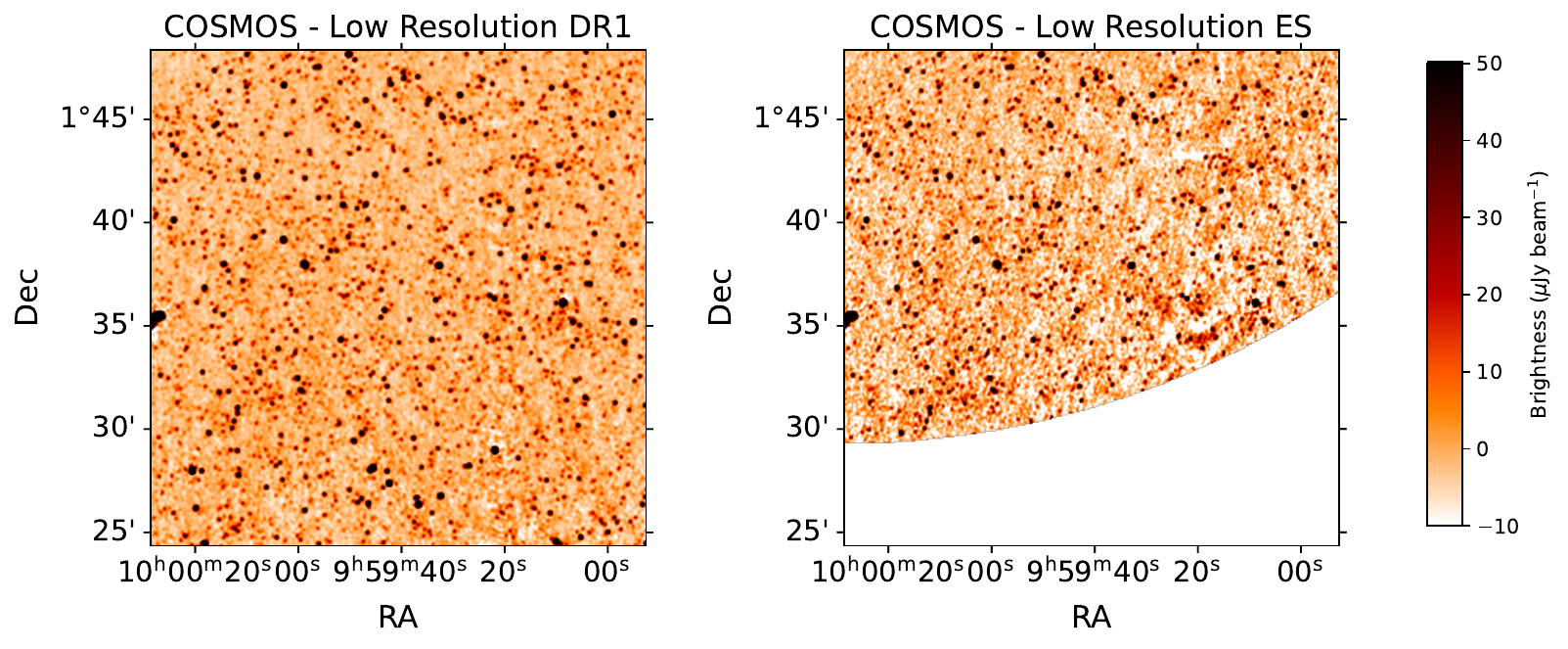}
    \end{minipage}%
    
         \begin{minipage}[b]{0.74\textwidth}
 \includegraphics[width=\textwidth]{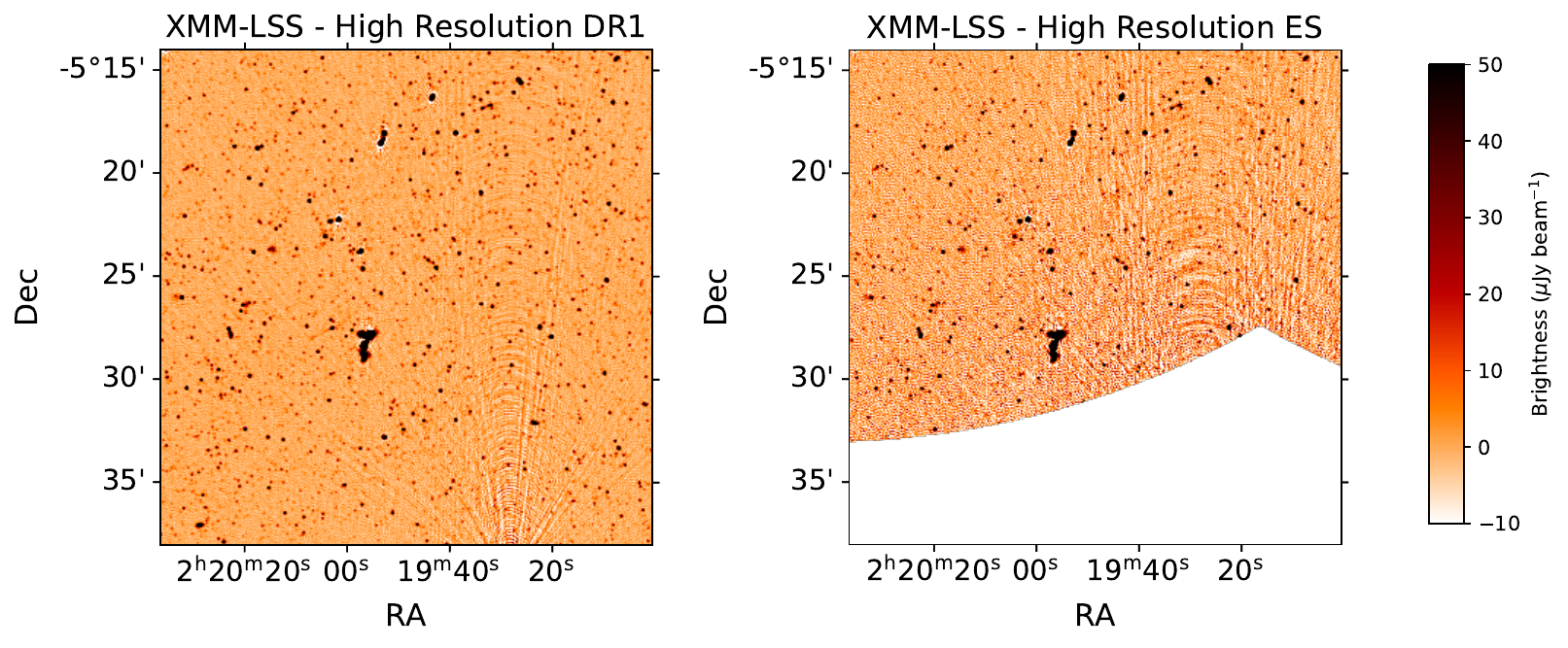}
    \end{minipage}%

    \begin{minipage}[b]{0.74\textwidth}
\includegraphics[width=\textwidth]{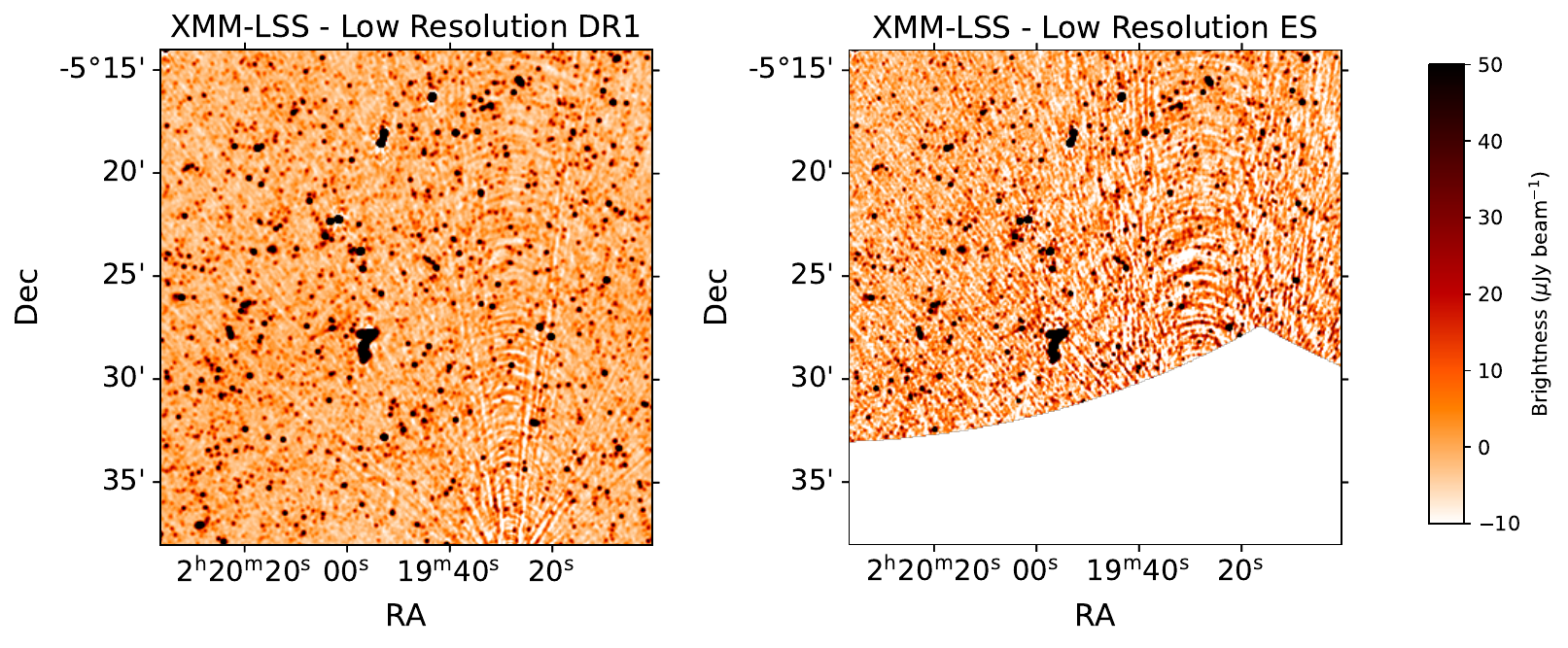}
    \end{minipage}%

     \caption{Comparisons of the brightness images for a zoomed in region of the Data Release 1 images (left) released with this work compared to the same region within the Early Science data (right) of \protect \cite{Heywood2022} for the COSMOS high resolution images (1st row) and low resolution images (2nd row) and the high resolution images of the XMM-LSS field (3rd row) and the XMM-LSS low resolution images (4th row). The regions are chosen to cover {$\sim$0.4\degree $\times$ 0.4\degree} over a region that is towards the outer regions of a pointing in the Early Science data to the south of the field which has not been mosaicked with neighbouring pointings, as it has been in DR1. }
     \label{fig:images_es_comp}
 \end{figure*}

A comparison of the final images for each of the three fields at both Briggs' weightings are presented in Figure~\ref{fig:images}. 
We also show a comparison of selected regions within this data release compared to the MIGHTEE Early Science data in Figure~\ref{fig:images_es_comp}. {These regions were chosen to include areas near the edge of the \cite{Heywood2022} Early Science image, as these areas are mosaicked with additional data in this data release and therefore can show an improvement in sensitivity.} For COSMOS and XMM-LSS the DR1 images are the final L-band continuum images from the MIGHTEE survey. For further details of these additional pointings see Appendix C of \cite{Heywood2022}.

\subsection{Effective Frequency Maps}
\label{sec:efffreq}

Following \cite{Heywood2022}, effective frequency maps are generated for each field to provide information of the typical frequency at a source location when effects such as the primary beam and flagging are accounted for. Such maps were generated using the same methodology as for the Early Science data by using linear mosaics of the frequency values in eight sub-bands, with a weighting function that is the primary beam response evaluated at that sub-band frequency, and weighted further by the inverse of the square of the sub-band image noise \citep[please refer to Section 3.2 of][for a more detailed explanation]{Heywood2022}.

\begin{figure*}
     \begin{minipage}[b]{0.33\textwidth}
 \includegraphics[width=\textwidth]{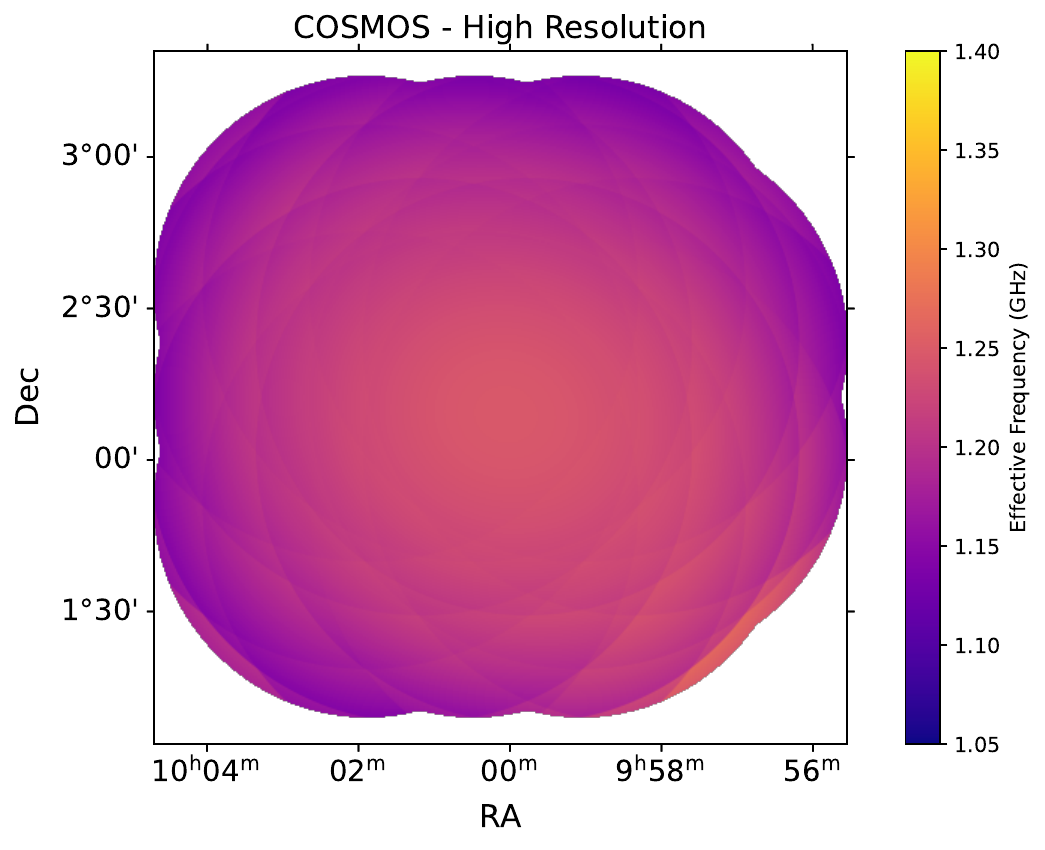}
    \end{minipage}%
    \begin{minipage}[b]{0.33\textwidth}
\includegraphics[width=\textwidth]{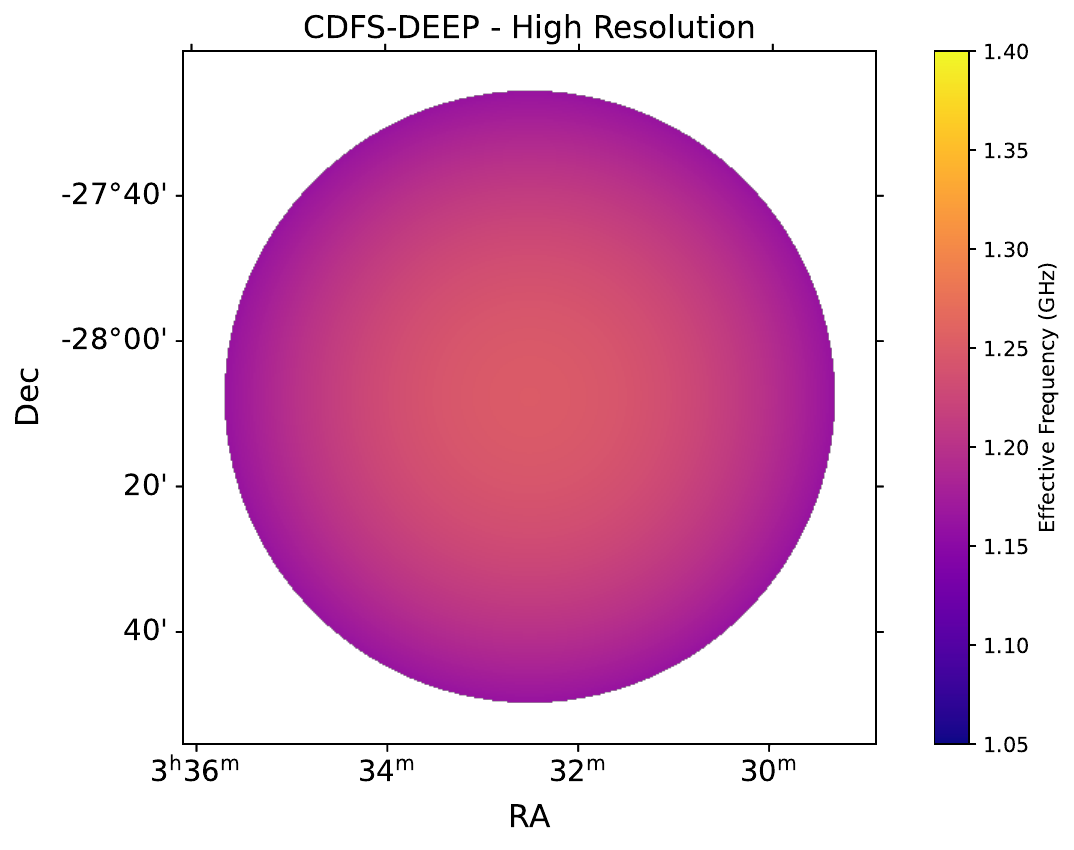}
    \end{minipage}%
    \begin{minipage}[b]{0.33\textwidth}
\includegraphics[width=\textwidth]{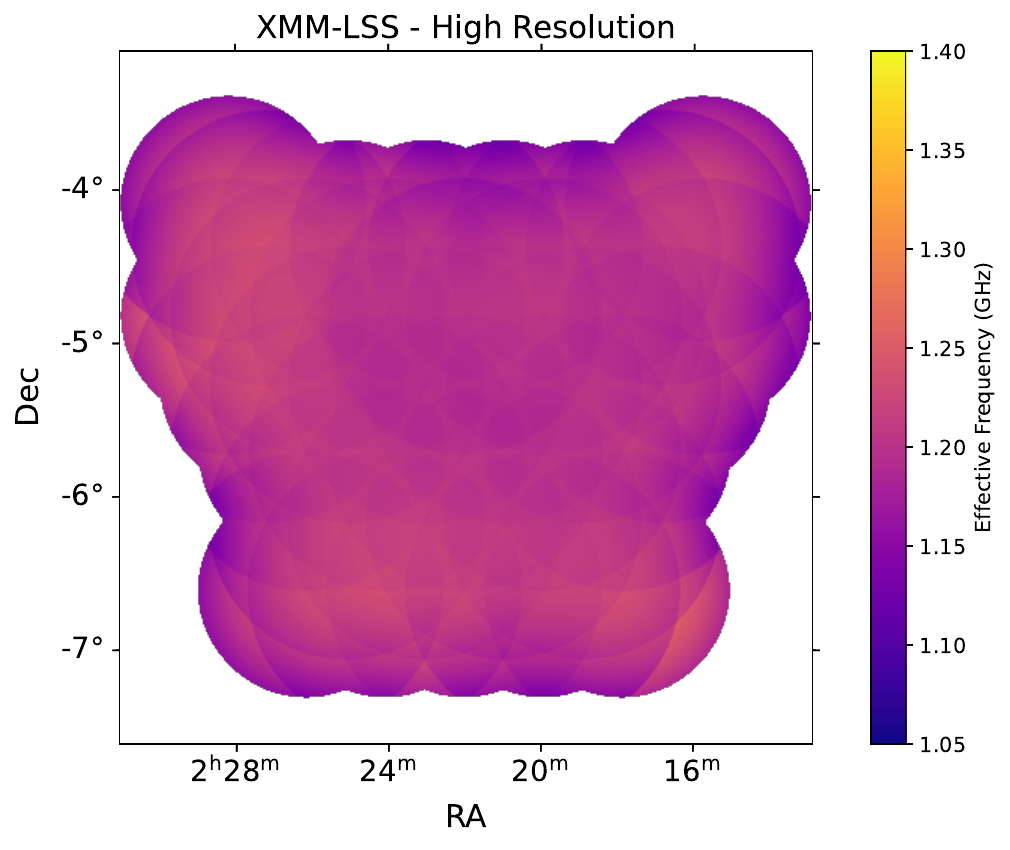}
    \end{minipage}%

    \begin{minipage}[b]{0.33\textwidth}
 \includegraphics[width=\textwidth]{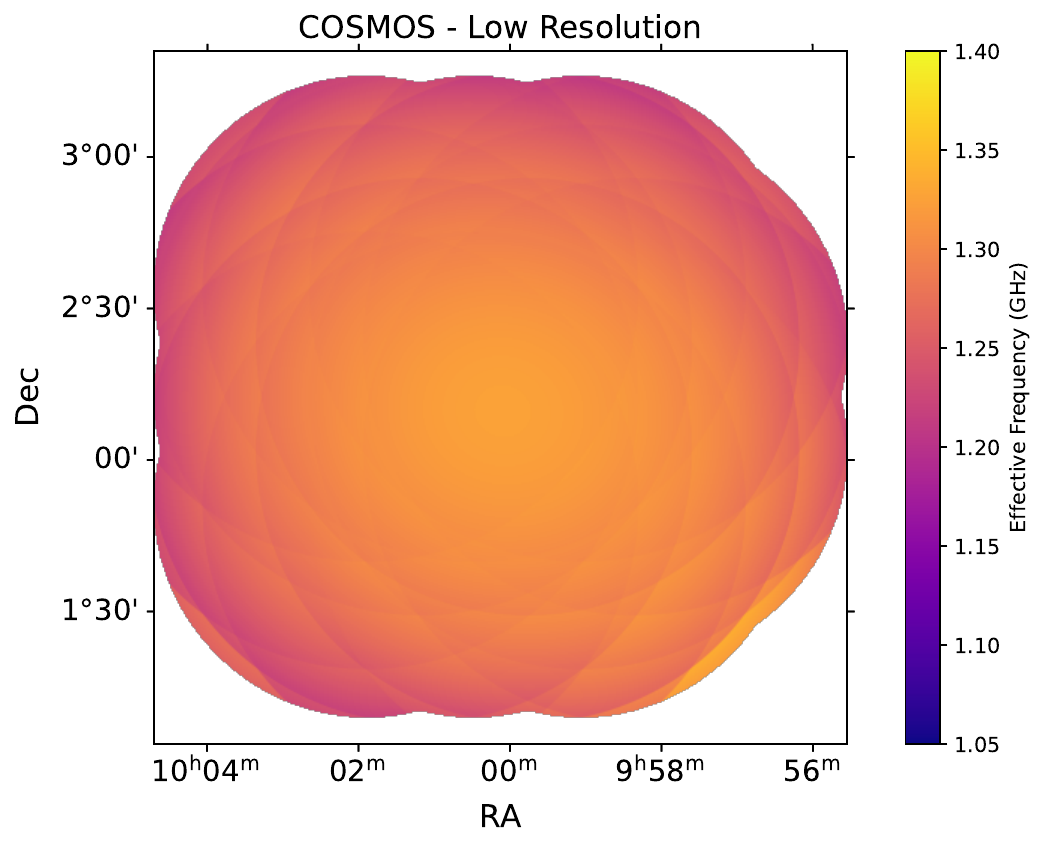}
    \end{minipage}%
    \begin{minipage}[b]{0.33\textwidth}
\includegraphics[width=\textwidth]{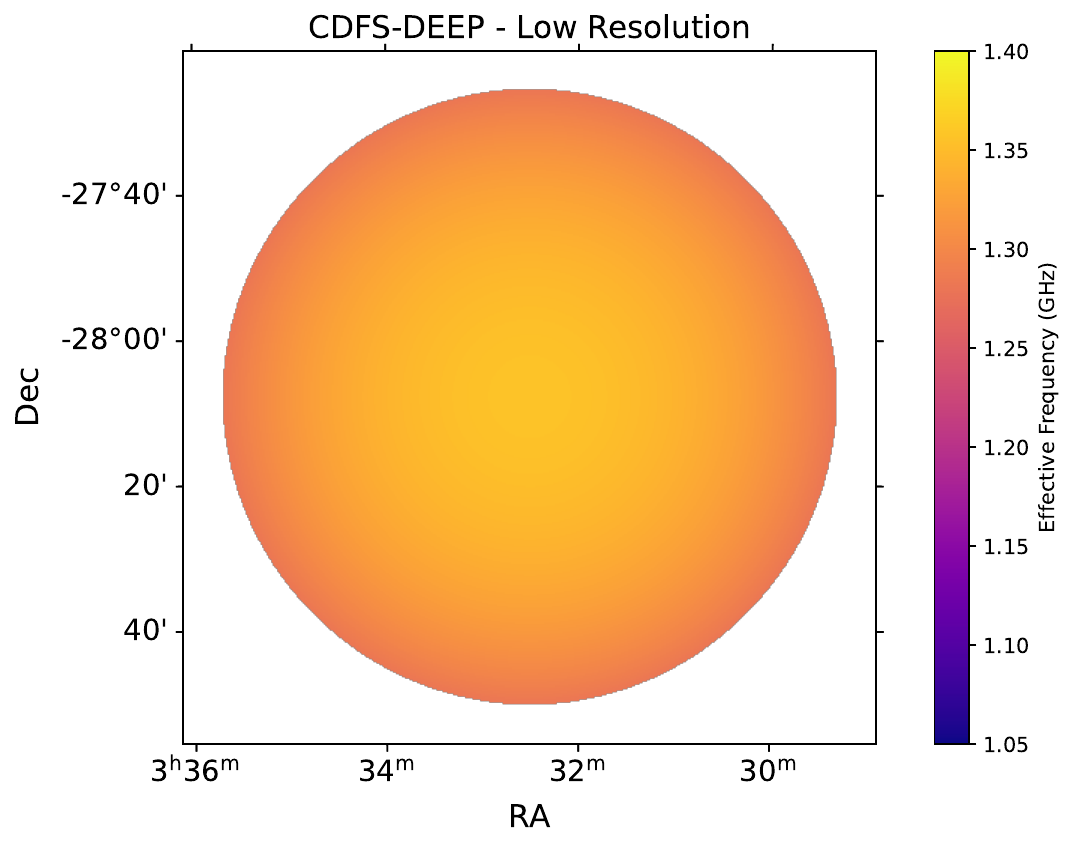}
    \end{minipage}%
    \begin{minipage}[b]{0.33\textwidth}
\includegraphics[width=\textwidth]{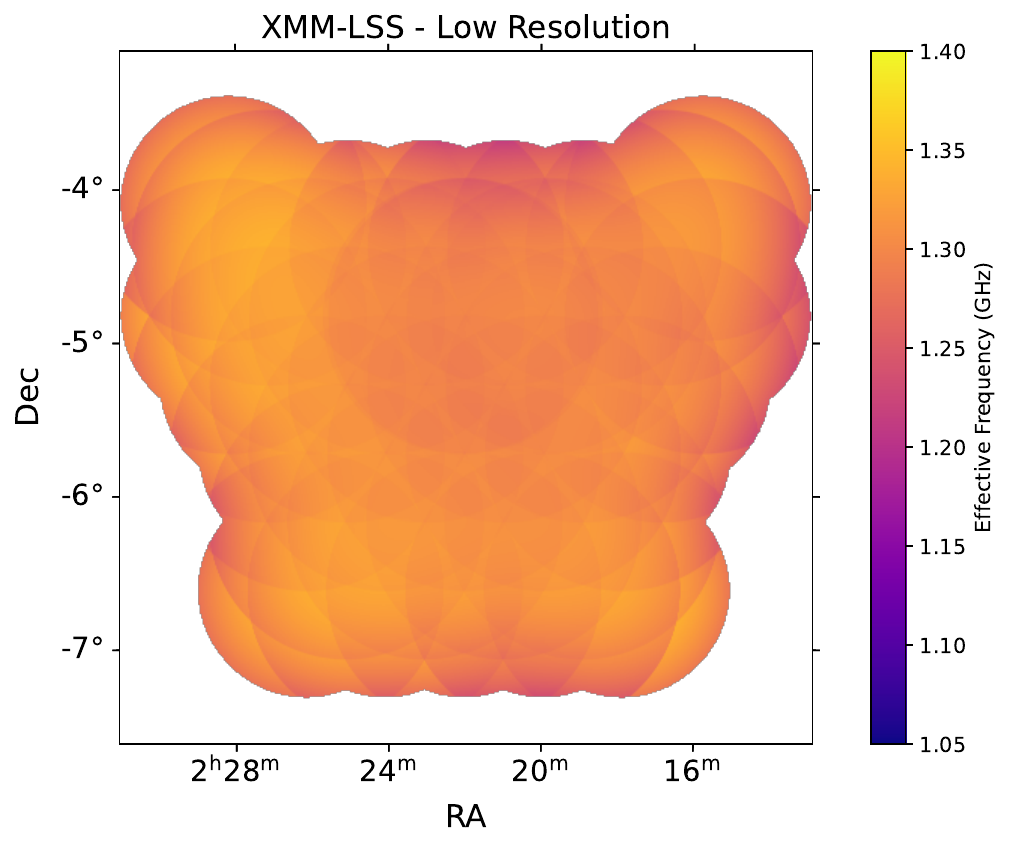}
    \end{minipage}%

         \begin{minipage}[b]{0.33\textwidth}
 \includegraphics[width=\textwidth]{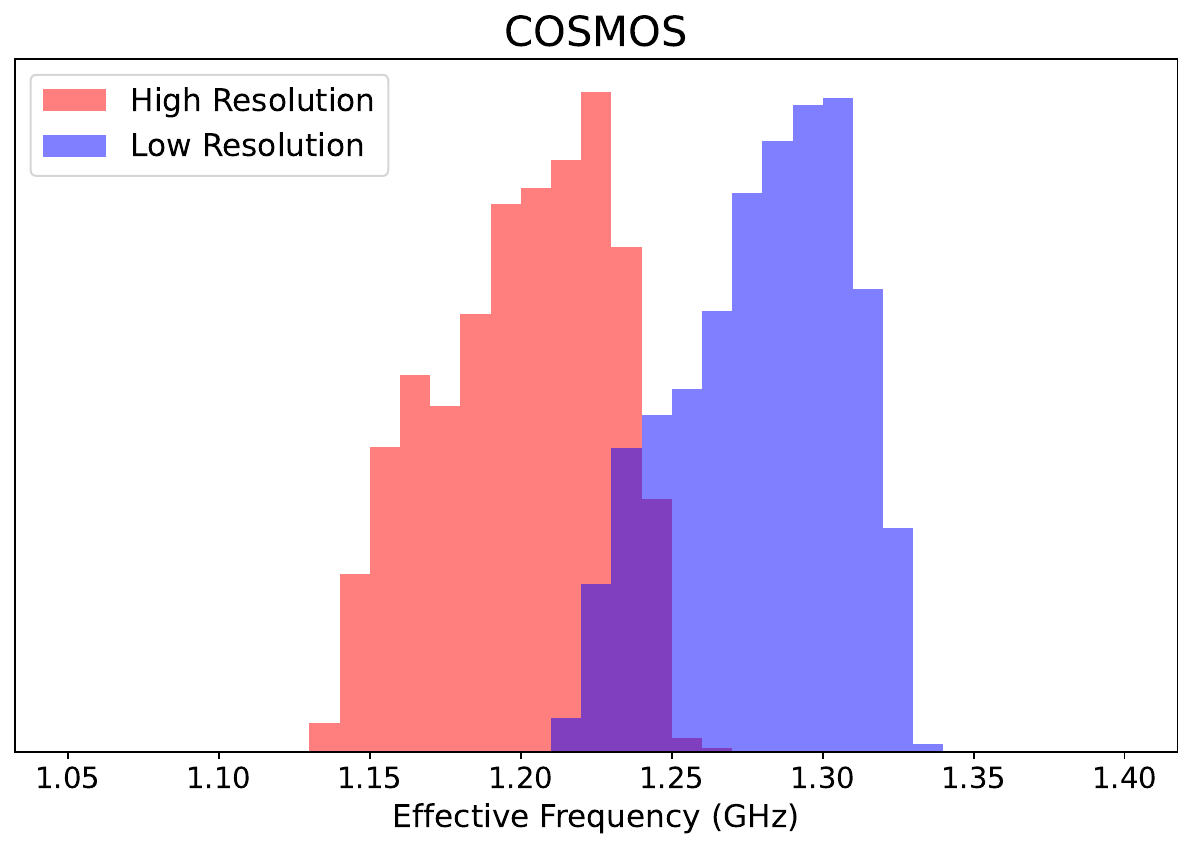}
    \end{minipage}%
    \begin{minipage}[b]{0.33\textwidth}
\includegraphics[width=\textwidth]{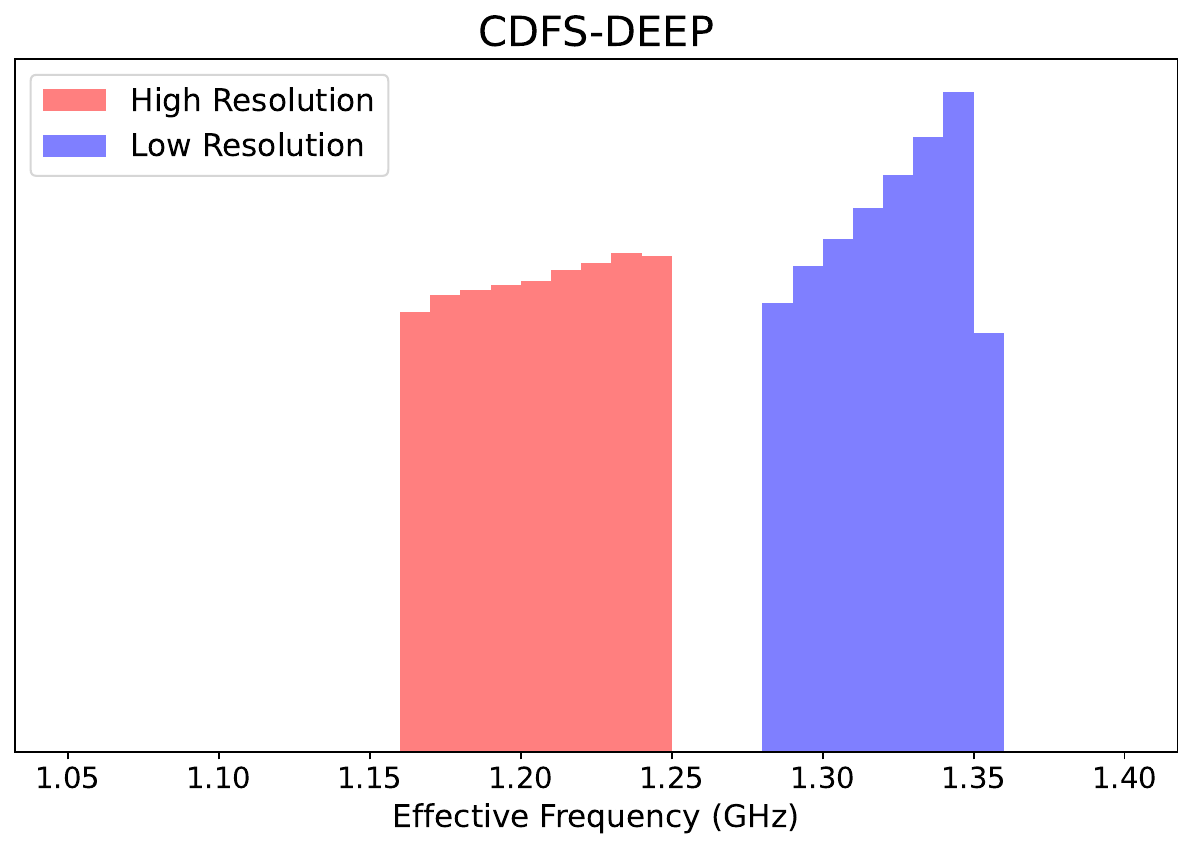}
    \end{minipage}%
    \begin{minipage}[b]{0.33\textwidth}
\includegraphics[width=\textwidth]{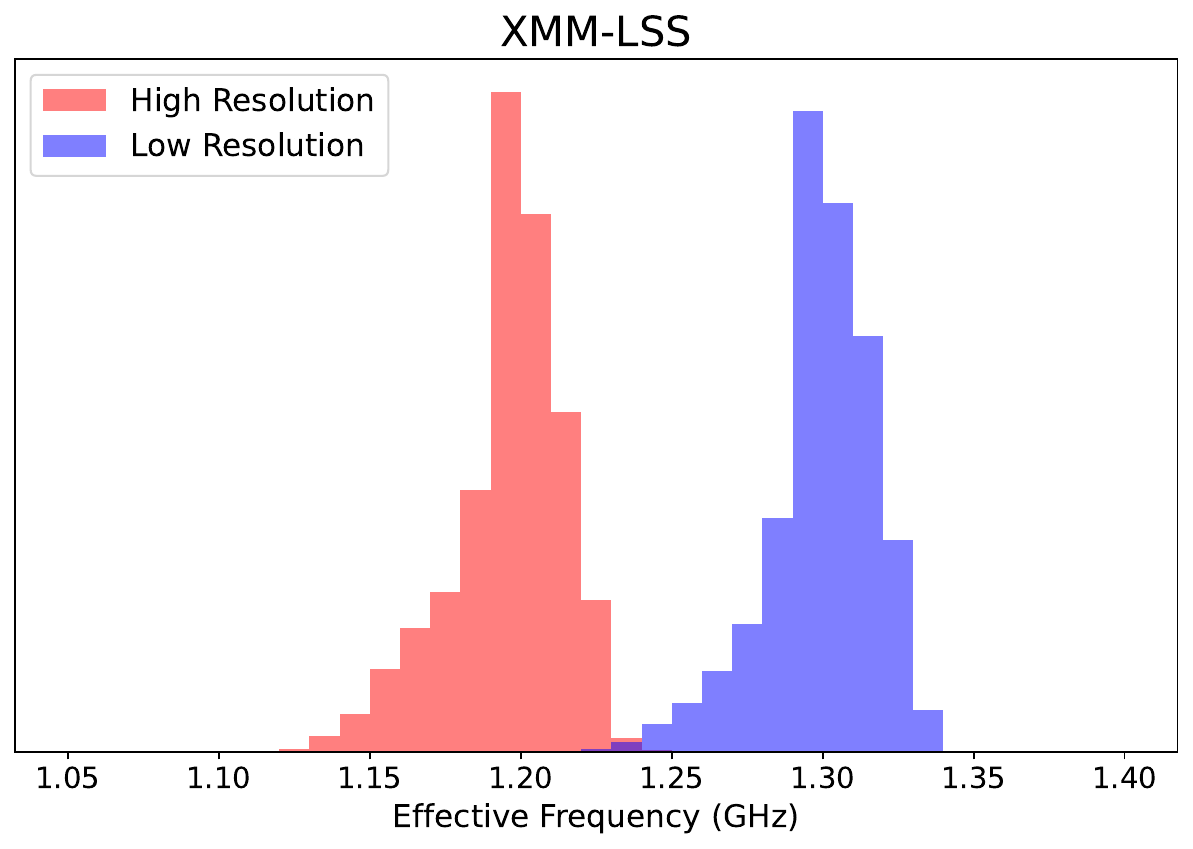}
    \end{minipage}%
     \caption{Effective frequency maps for COSMOS (left), CDFS-DEEP (centre) and XMM-LSS (right) fields for the high resolution images (top row) and low resolution images (middle row). Also shown {(bottom row)} are {the} histograms of the effective frequency pixel values within the images for the high (red) and low (blue) resolution images.}
     \label{fig:efffreq}
 \end{figure*}

 We present these effective frequency mosaics for each of the three fields in Figure~\ref{fig:efffreq}. The patterns caused by the overlap of the different pointings in the COSMOS and XMM-LSS fields {are} clearly visible and these maps are found to have typically lower effective frequencies in the outer regions of the image mosaics. The effective frequencies of the high resolution images have, in general, lower frequency values compared to the lower resolution images. This is due to the fact that the lower resolution images are produced with a higher weighting of the short spacings, which are more prone to RFI {\citep[see][]{Mauch2020}}. Since the majority of the RFI in MeerKAT's L-band occupies the lower half of the band, this serves to raise the effective frequency of the lower resolution maps. 

Whilst the effective frequency is reported within the catalogues of sources (see Section \ref{sec:radiocatalogues}), we do not apply any corrections to the flux density of the images or those recorded in the catalogue. Any such correction would need to assume a spectral shape for the source, given by the spectral index, $\alpha$\footnote{{Where a spectral index is assumed in this work we use the relation $S_{\nu}\propto \nu^{-\alpha}$ to relate the frequency, $\nu$, to the spectral flux density measurements, $S_{\nu}$ and will assume a value of $\alpha$=0.7, unless otherwise stated.}}. Different scientific studies and use cases may require the assumption of different spectral indices or may want to measure the spectral indices directly. We therefore leave it to the users of to appropriately apply any frequency corrections to the measured source flux densities, or images, as necessary for their work. {Not accounting for such an effect could lead up to an $\sim$8\% offset in the ratio of the flux densities from the minimum and maximum frequency values across the image, when scaled to a common frequency (and assuming $\alpha=0.7$).}

\begin{table*}
\caption{Tables describing the images and catalogues of the MIGHTEE Data Release 1 catalogues and images. Given are the area and resolution of the images, alongside the number of sources and Gaussian components detected with \textsc{PyBDSF}, as described in Section \ref{sec:radiocatalogues}. Additionally, the thermal rms noise within the image, the measured median rms sensitivity within the \textsc{PyBDSF} determined rms maps across the full image, and within a smaller $0.5^{\circ} \times 0.5^{\circ}$ central region of the image, are also provided. The latter estimate demonstrates the typical sensitivity in the most sensitive regions that are not affected by the sensitivity roll off in the outer, non-mosaicked, regions. {In the case of the mosaics, the thermal noise is estimated by mirroring the negative values in a histogram of pixel brightnesses and determining the standard deviation of a Gaussian fit to the resulting distribution. In the case of the CDFS-DEEP images the thermal noise is measured in a region far from the phase centre where the primary beam attenuation is sufficient to provide a source-free region.} {For the COSMOS and XMM-LSS fields where there is Early Science data from \protect \cite{Heywood2022} we also give a comparison of the area observed and number of Gaussian components detected.}}
\begin{tabular}{lcccccc}
& \multicolumn{2}{c}{CDFS-DEEP}  & \multicolumn{2}{c}{COSMOS} & \multicolumn{2}{c}{XMM-LSS} \\ \hline \hline
Resolution  [arcsec] & 5.5 & 7.3 & 5.2 & 8.9 & 5.0 & 8.9 \\ \hline 
Area [sq. deg] & 1.5 & 1.5 & 4.2 & 4.2 & 14.4 & 14.4 \\
{Thermal rms sensitivity [$\muup$Jy beam$^{-1}$]} & 0.9 & 0.6 & 2.2 & 1.6 & 3.4 & 1.5 \\
{Median measured rms sensitivity [$\muup$Jy beam$^{-1}$]} & 1.9 & 2.0 & 5.6 & 3.5 & 5.1 & 3.2 \\
{Median measured central rms sensitivity [$\muup$Jy beam$^{-1}$]} & {1.2} & {1.3} & 2.4 & 2.1 & 3.6 & 2.7 \\
Number of Sources & 21 152 & 17 866 & 20 886 & 28 267 & 72 187 & 97 684 \\
Number of Gaussian Components & 22 660 & 19 291 & 22 420 & 30 466 & 76 567 &  104 557 \\ 
Ratio of area c.f. \cite{Heywood2022} & - & - & 2.8 & 2.8 & 4.1 & 4.1  \\
Ratio of Gaussian components c.f. \cite{Heywood2022} & - & - & 4.9 & 3.0 &  7.1 & 5.1 \\ \hline 
\end{tabular}
\label{tab:catalogues}
\end{table*}

\section{Radio Catalogues}
\label{sec:radiocatalogues}
To detect radio sources and create a catalogue of their source properties we make use of the Python Blob Detection {and} Source Finder \citep[\textsc{PyBDSF};][]{PyBDSF}\footnote{\url{https://pybdsf.readthedocs.io/}}. The general practice of \textsc{PyBDSF} makes use of a sliding box to first derive {root mean square (rms)} and sky background maps for an image, after which islands of emission which exceed a user determined threshold are identified. For these islands, the emission is modelled using either a single Gaussian component or a combination of multiple Gaussian components, which are used to build up a source model. For the Early Science MIGHTEE data \citep{Heywood2022}, \textsc{PyBDSF} was used with a typical standard configuration where a 3$\sigma$ threshold {on flux within the image} to detect islands of emission {was used} to ensure high completeness for both point and extended sources, but only those sources which satisfied at least a 5$\sigma$ {peak emission} threshold criteria were included in the final source catalogue\footnote{Note this 5$\sigma$ threshold is based on image values, and so some catalogued sources may appear to have peak SNR slightly less than 5$\sigma$.}, alongside the associated Gaussian component catalogue. 

For this work, we use \textsc{PyBDSF} in a modified method {relative} {to the Early Science data} to produce the source and component catalogues. Such a change is employed in order to extract maximal information from the images, which are limited by confusion at the sensitivity and resolution of these data {products}. A similar approach is also being used in deep radio observations with LOFAR (Shimwell et al., in prep). As discussed in \cite{Heywood2022} and in Section \ref{sec:radiodata}, the MIGHTEE continuum imaging data are affected by confusion and as such, this will affect our ability to measure the noise in the image {reliably. When} determining the rms maps, \textsc{PyBDSF} uses a sliding box to generate a varying rms map across the full field. The size of such a box can either be determined by \textsc{PyBDSF} itself or specified by the user\footnote{see the guidance at \url{https://pybdsf.readthedocs.io/en/latest/process\_image.html\#term-rms\_box} for appropriate rms box scales to use.}. Such an rms map is generated prior to the identification (and hence removal) of any sources already within the image. Therefore, when there is a high source density, the measured rms will increase due to the sources responsible for elevating the confusion noise. 
An elevated rms due to this source confusion will adversely affect the detection of sources across the image, reducing the number of sources which reach the threshold needed to be considered `detected' by \textsc{PyBDSF}. Due to the high source density in the MIGHTEE images we therefore adopt a two-step process for the source finding and {catalogue generation}, {first} removing sources from the image to calculate a more accurate rms map and {using} this {rms map} for source detection. 

More formally, we first run \textsc{PyBDSF} on the image using the detection parameters:

\begin{verbatim}
bdsf.process_image(image, thresh_isl = 3., 
thresh_pix = 5., mean_map =`zero', rms_box=(120,30))
\end{verbatim}

\noindent where \texttt{image} is the file name of the image, \texttt{thresh\_isl} is the threshold for detecting an island of emission to  then be fit with Gaussian components, and \texttt{thresh\_pix} is the {criterion} which determines which sources are included within the final source catalogue. The \texttt{mean\_map} is the same shape as the input image, reflecting the background emission within the image. We set this to zero at all locations within the image. Finally, \texttt{rms\_box} is a parameter used to quantify the size of the box used to determine the rms {in terms of pixels (where our pixel scale within the images are 1.1\arcsec)}, as well as the step size used to move this sliding box across the image. Such values were chosen to ensure consistent box sizes were used across the three fields and to {avoid too large a box size which could smooth over noise variations}.

After running \textsc{PyBDSF}, we save the residual map which is generated from \textsc{PyBDSF}, i.e. the image with the source model of the Gaussian components of `detected sources' subtracted. This should therefore give a better representation of the background, with bright sources removed. We again use \textsc{PyBDSF}, with the parameters described above, on the residual image to generate the rms map. This rms image generated in the second run of \textsc{PyBDSF} should provide an improved estimate of the true rms of the image, with a reduced number of sources. Of course, due to the confused nature of the MIGHTEE data, there will be remaining sources of emission within the residual image and, as such, the rms will still be elevated, though this will be reduced compared to the rms measured using the initial \textsc{PyBDSF} run. This rms image is then used to improve the source detection for the final catalogues.

To obtain the final source and component catalogues, we perform a final run with \textsc{PyBDSF} in which we supply the original MIGHTEE image and use the parameters described above, but supply the rms map generated above as an input (\texttt{rms\_map}), instead of allowing \textsc{PyBDSF} to internally generate such a map. To do this we use:

\begin{verbatim}
rmsmean_map_filename = [mean_map, rms_map]
\end{verbatim}

\noindent where the \texttt{mean\_map} is again a map consisting solely of zeros over the image. From this final \textsc{PyBDSF} run, we obtain source and Gaussian component catalogues for each of the three fields for both the low and high resolution images. The total number of sources and components detected by \textsc{PyBDSF} are shown in Table \ref{tab:catalogues}.

As shown in Table~\ref{tab:catalogues}, we detect a total of 143~837 sources (154~314 Gaussian components) across the three fields (totalling 20.1 sq.~deg) in the lower resolution images and 114~225 sources (121~647 Gaussian components) in the corresponding higher-resolution images. 
We release both {the source} and Gaussian-component catalogues separately for the three fields and provide an example of the first {5} rows of the {source catalogue from the COSMOS field in Table~\ref{tab:sourcetab}}. The Gaussian catalogue contains the same columns as the source catalogue, with an additional Gaussian component ID. {Measured properties in the Gaussian catalogue such as flux densities, positions and shapes relate to the individual Gaussian components which are used to model a source, {unless otherwise stated, as described in \url{https://pybdsf.readthedocs.io}}. }

\section{Catalogue Validation}
\label{sec:validation}
In this section we present comparison of the astrometry and flux-density of our data to previous radio observations. As noted in \cite{Hale2023}, differences in the MeerKAT baseline distribution compared to other telescopes or survey  configurations could lead to emission being resolved {into multiple sources in one survey (where only one source is detected in another survey, or extended emission being resolved out entirely. This} may affect the comparison of flux densities of sources which are resolved. We therefore compare the COSMOS and XMM-LSS fields to the MIGHTEE Early Science data over these fields from \cite{Heywood2022}. These cover overlapping areas, but the mosaicking of many more pointings could affect the measured properties of sources even within the regions which were covered by the Early Science release, e.g. primary beam corrections between different observation tracks and smearing with distance from phase centres {is} exacerbated by the mosaicking. Moreover, the improved sensitivity within regions of {DR1 that were previously the outer edges of the Early Science data will now be less affected by the larger noise peaks/troughs in the Early science data, which would affect the accuracy of the flux density measurements.} This means that the measurement of flux densities for sources within the Early Science region of MIGHTEE (especially towards the outer edges) may differ from the measured properties of the same sources in the Data Release 1 region. {As we use different methods to detect and characterise sources (albeit both with \textsc{PyBDSF}), this may also introduce differences in the measured source properties. In their work, \cite{Heywood2022} make comparisons of the Early Science data to VLA observations of the COSMOS \citep{Schinnerer2010} and XMM-LSS fields \citep{Heywood2020} using catalogues extracted with the same method as in \cite{Heywood2022}. We will discuss their findings in the appropriate sections of astrometric (Section \ref{sec:ast}) and flux density (Section \ref{sec:flux}) accuracy.} Details of the location of each of the individual pointings that make the final mosaicked images are presented in Appendix~\ref{sec:appendix}. The outline of the Early Science regions compared to DR1 can be seen in Figure \ref{fig:images}.

\begin{figure}
    \centering
    \includegraphics[width=9cm]{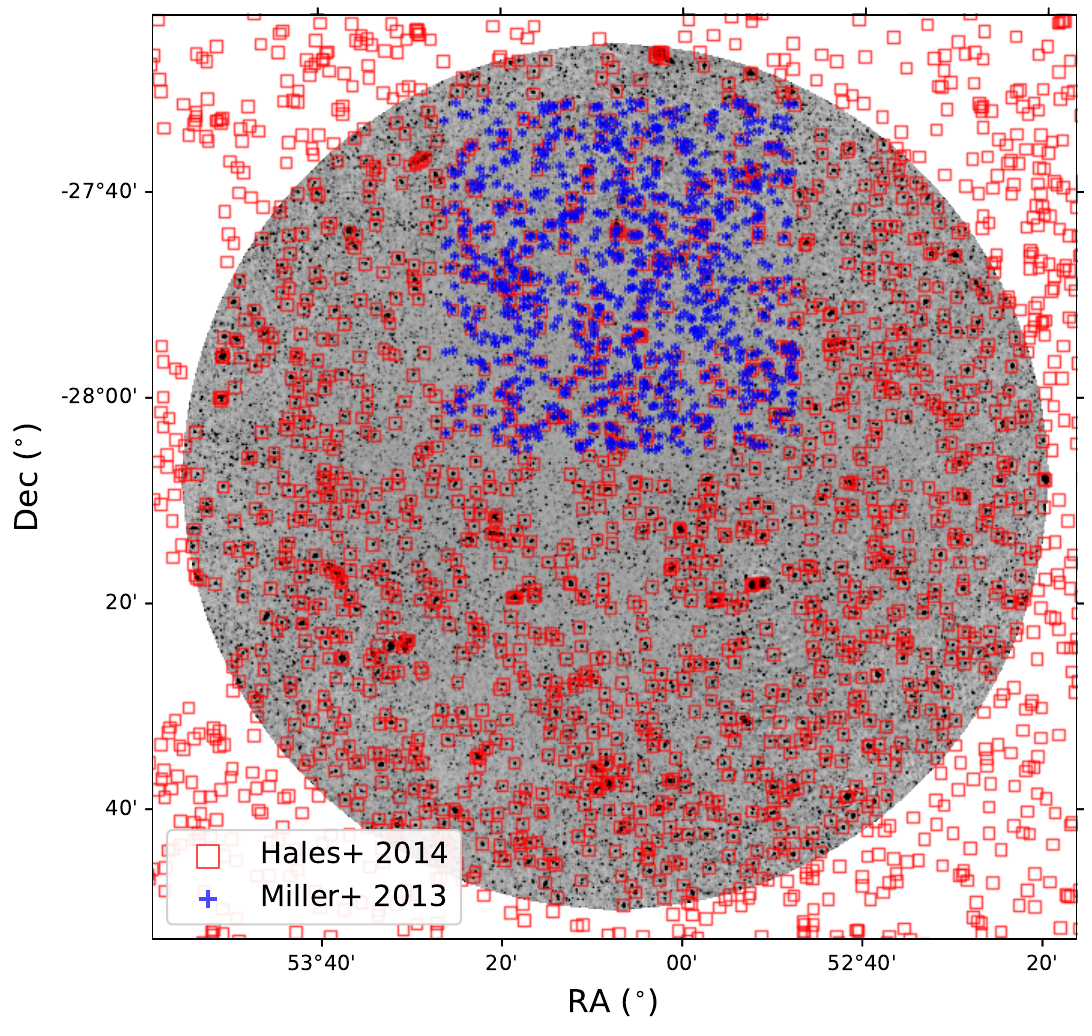}
    \caption{Comparison of the overlap region between sources in CDFS from MIGHTEE DR1 {(grey background image)} to previous observations with the VLA {\protect \citep[blue plus symbols; ][]{Miller2013}} and {from} ATCA with the ATLAS survey {\protect \citep[red squares; ][]{Franzen2015}.} }
    \label{fig:cdfs}
\end{figure}

For the deep data within the CDFS-DEEP field, there are no previous MIGHTEE data to compare with. We therefore compare to observations using the Very Large Array \citep[VLA; ][]{Miller2013} and the Australian Telescope Compact Array \citep[ATCA; through the Australia Telescope Large Area Survey, or ATLAS {Data Release 3};][]{Franzen2015}, both at 1.4 GHz\footnote{{We note that the ATLAS catalogue also has frequency variations across the field of view between $\sim$1.4$-$1.5 GHz and therefore we use the 1.4 GHz scaled flux densities from their catalogue.}}. The overlap between the coverage of the MIGHTEE CDFS-DEEP data and the data of \cite{Miller2013} and {\cite{Franzen2015}} is shown in Figure~\ref{fig:cdfs}. These data are significantly shallower than the MIGHTEE data, with the catalogues of {\cite{Franzen2015}} having significantly lower source density, but covering a larger area than that of \cite{Miller2013}.

 \begin{figure*}
     \begin{minipage}[b]{0.5\textwidth}
 \includegraphics[width=\textwidth]{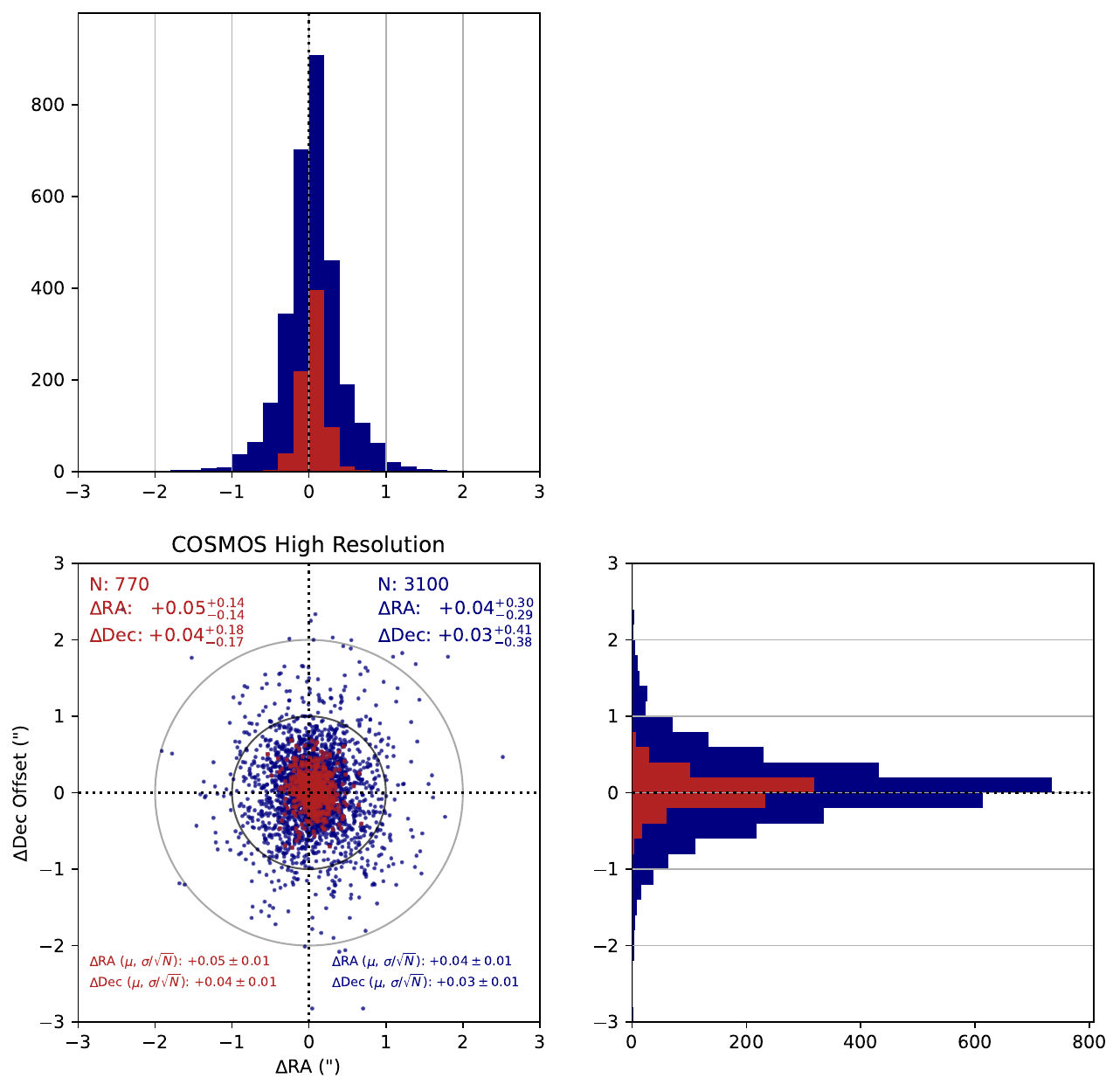}
    \subcaption{COSMOS High Resolution}
    \end{minipage}%
    \begin{minipage}[b]{0.5\textwidth}
\includegraphics[width=\textwidth]{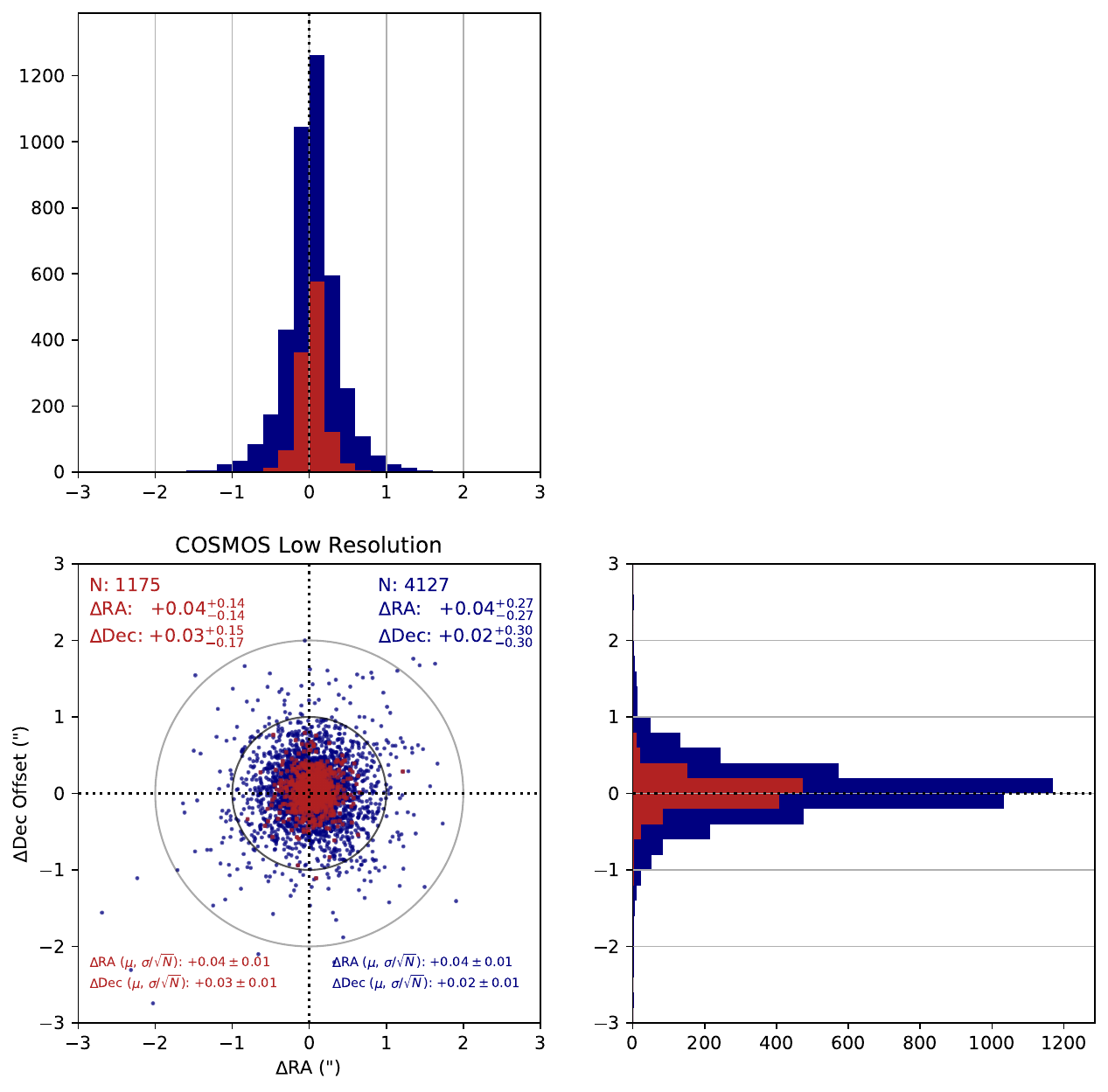}
    \subcaption{COSMOS Low Resolution}
    \end{minipage}%
    
     \begin{minipage}[b]{0.5\textwidth}
 \includegraphics[width=\textwidth]{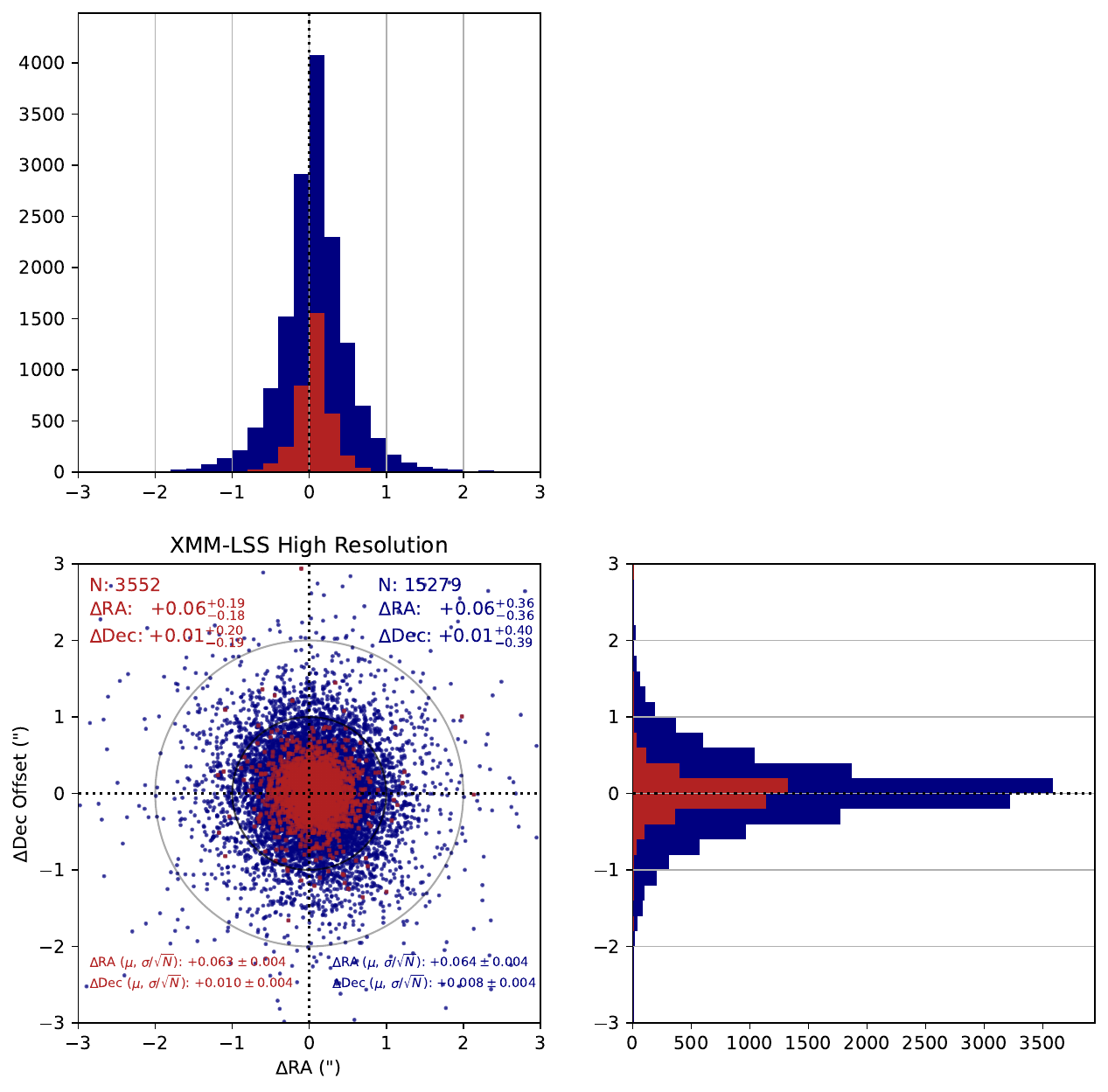}
    \subcaption{XMM-LSS High Resolution}
    \end{minipage}%
    \begin{minipage}[b]{0.5\textwidth}
\includegraphics[width=\textwidth]{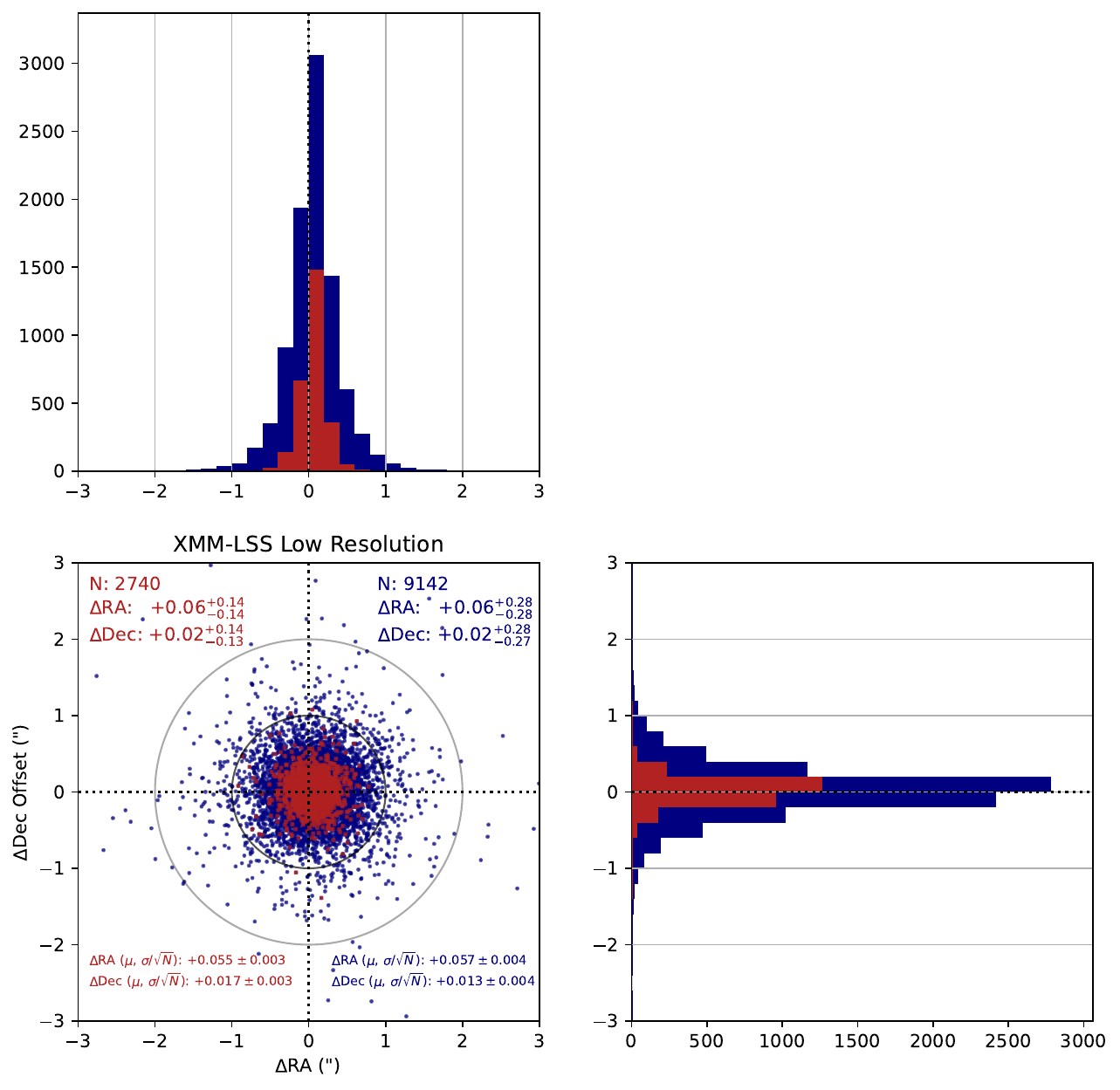}
    \subcaption{XMM-LSS Low Resolution}
    \end{minipage}%
     \caption{Comparisons of the astrometric offsets between DR1 and the Early Science data from \protect \cite{Heywood2022} for the COSMOS (upper) and XMM-LSS (lower) fields, for the high (left) and low resolution (right) catalogues. Shown are the offsets for both the single component sources, as discussed in Section \protect \ref{sec:validation} (navy) as well as with the additional SNR and unresolved criterion discussed in Section \protect \ref{sec:validation} (red). The numbers quoted in the upper part of the scatter panel presents the number of sources which contribute to the comparisons as well as the median offset in RA  and Dec (with $\Delta X = X_{\textrm{DR1}} - X_{\textrm{ES}}$, for both RA and Dec) and the associated errors from the 16th and 84th percentiles reported in arcsec. The colour of the text is chosen to match the data plotted. The circles of radii of 1\arcsec \ and 2\arcsec \ are included to guide the eye. The numbers provided in the lower part of the scatter plot quote the mean and standard error on the mean for the RA and Dec offsets. }
     \label{fig:astrometry}
 \end{figure*}

  \begin{figure*}
     \begin{minipage}[b]{0.5\textwidth}
 \includegraphics[width=\textwidth]{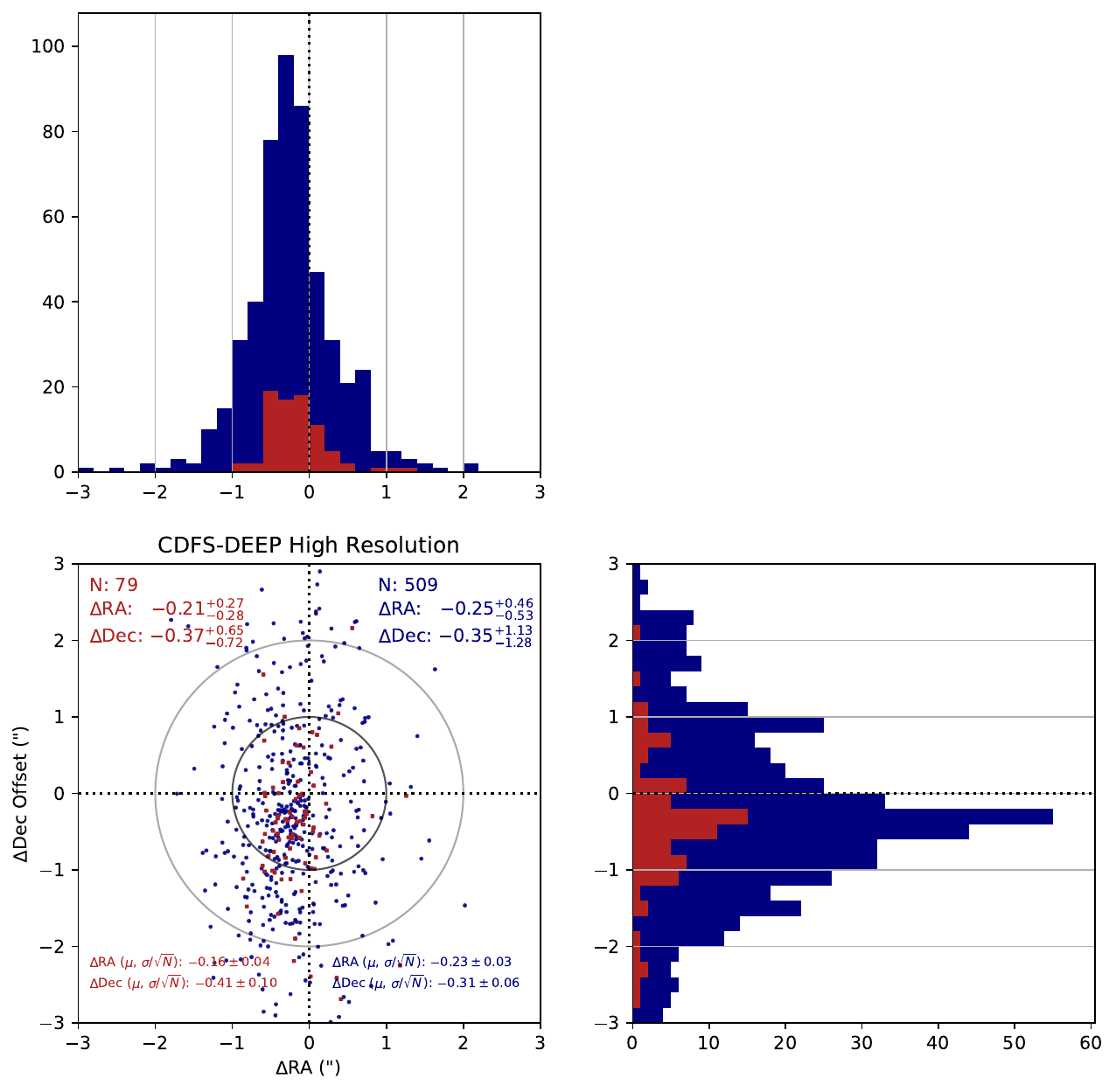}
    \subcaption{CDFS-DEEP High Resolution to  \protect {\cite{Franzen2015}}}
    \end{minipage}%
    \begin{minipage}[b]{0.5\textwidth}
\includegraphics[width=\textwidth]{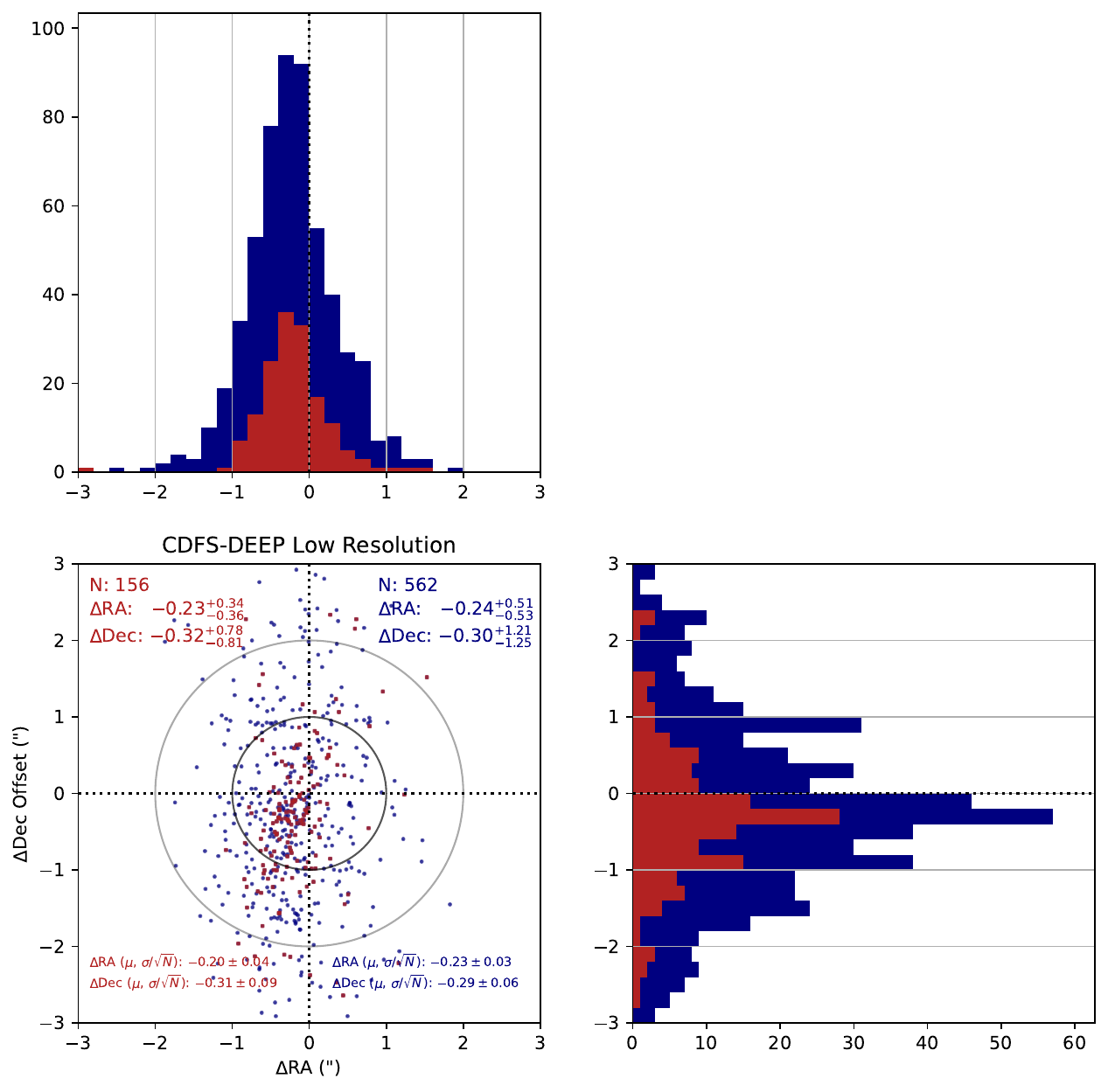}
    \subcaption{CDFS-DEEP Low Resolution to \protect {\cite{Franzen2015}}}
    \end{minipage}%
    
     \begin{minipage}[b]{0.5\textwidth}
 \includegraphics[width=\textwidth]{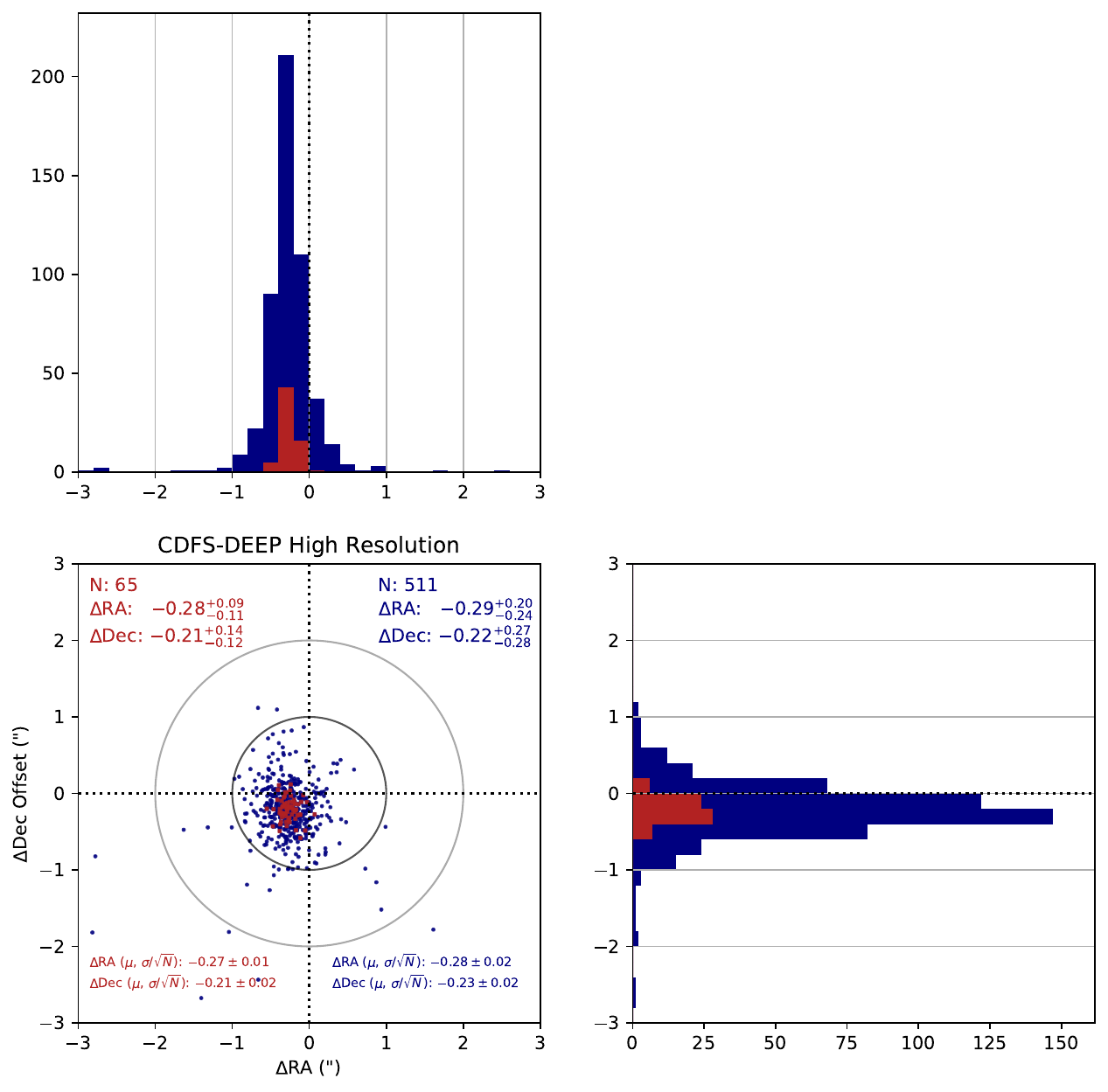}
    \subcaption{CDFS-DEEP High Resolution to  \protect \cite{Miller2013}}
    \end{minipage}%
    \begin{minipage}[b]{0.5\textwidth}
\includegraphics[width=\textwidth]{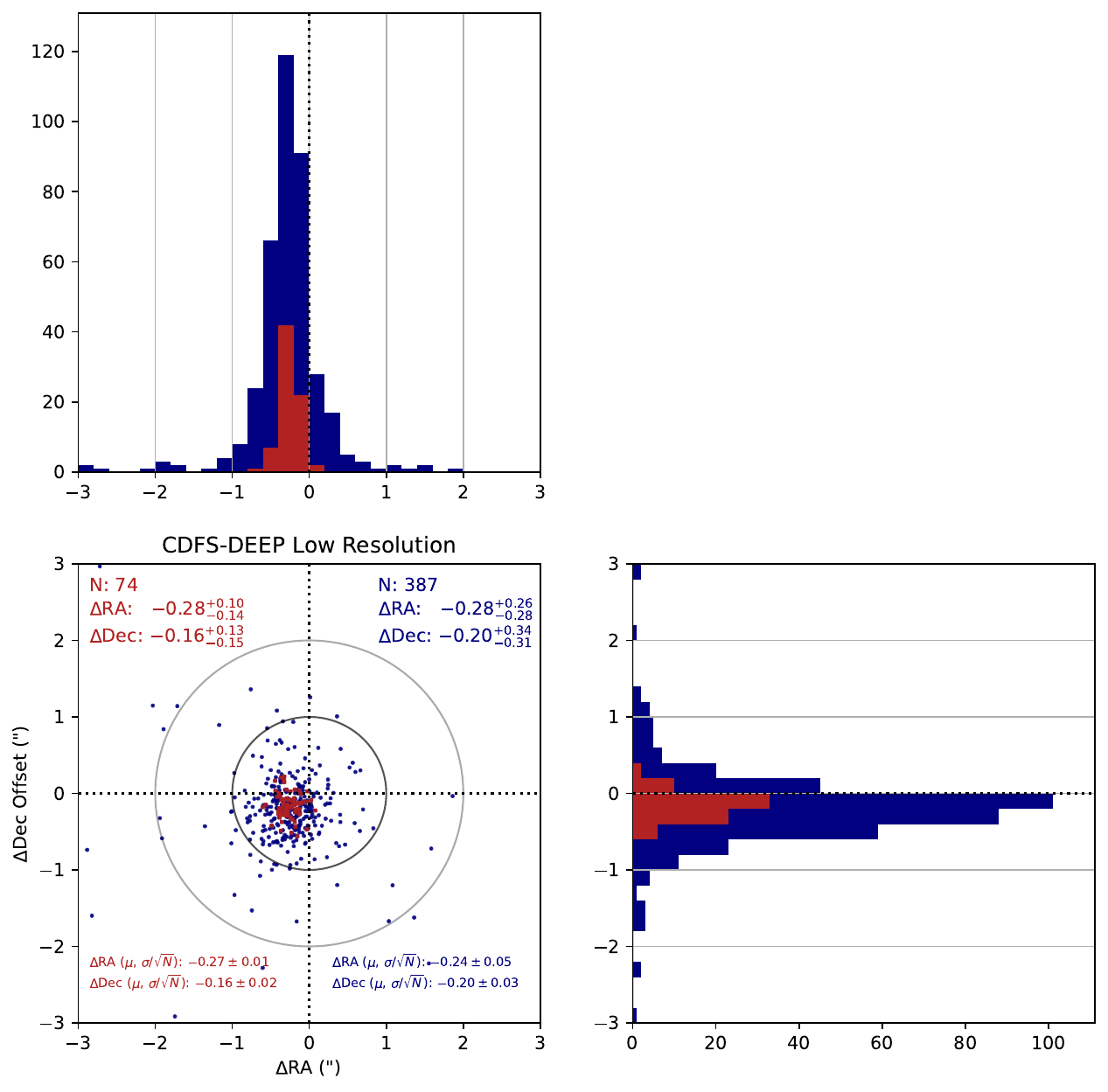}
    \subcaption{CDFS-DEEP Low Resolution to \protect \cite{Miller2013}}
    \end{minipage}%
     \caption{As for 
      \protect \ref{fig:astrometry} for the astrometric offsets, now for the CDFS-DEEP field for the high (left) and low resolution (right) catalogues compared to data {from ATCA \protect \citep[][upper]{Franzen2015} } and the VLA \protect \citep[][lower]{Miller2013}.}
     \label{fig:astrometrycdfs}
 \end{figure*}

\subsection{Matching of Catalogues}
\label{sec:catmatch}
To ensure a like-for-like comparison, we identify isolated sources within the DR1 catalogue (and the matching catalogue) which have no neighbours within the {2$\times$}FWHM of the restoring beam of the MIGHTEE data (which is typically the limiting resolution). This is to avoid sources which may be blended in one catalogue but not the other. The only exception to this is for comparison to the work of {\cite{Franzen2015}}, where the angular resolution is larger than that of MIGHTEE, {16\arcsec $\times$ 7\arcsec. For these comparisons the FWHM assumed is16\arcsec \ and the appropriate cut from that} is instead adopted. We then match the DR1 catalogues to the survey used for comparison, using a matching radius given by the angular resolution of the MIGHTEE data and only compare those sources which are defined as `single' sources in \textsc{PyBDSF} (\texttt{S\_Code=S}). 
We make a second catalogue with more stringent constraints, where only unresolved, high SNR sources are compared. To do this, we restrict to sources which have high SNR in both the DR1 data and comparison data (peak SNR $\geq$10) and only use sources which are defined as unresolved in MIGHTEE, using the method implemented in \cite{Hale2021} {and similar studies} \citep[e.g.][]{Bondi2008, Smolcic2017, Heywood2022}, where the SNR is used to define {an envelope between the integrated-to-peak flux density ratio ($S_I/S_P$) as a function of SNR,} within which sources in MIGHTEE are considered unresolved. {Sources that were identified as single sources in \textsc{PyBDSF} were used to define such an envelope, which would contain 95\% of unresolved sources, and the envelopes used to define unresolved MIGHTEE DR1 sources are given in Appendix \ref{sec:snrenv}, in Table \ref{tab:snr_env}.}

\subsection{Astrometry}
\label{sec:ast}
In Figure~\ref{fig:astrometry} we present a comparison of the astrometry between the DR1 and Early Science observations of \cite{Heywood2022}  for the COSMOS and XMM-LSS fields using the two comparison catalogues described above {for both the low and high resolution images}. {In the Early Science data, \cite{Heywood2022} find sub-pixel offsets in RA and DEC, with mean offsets (RA, DEC) of (-0.27\arcsec, -0.19\arcsec) in the COSMOS field, and (-0.20\arcsec, -0.43\arcsec) in the XMM-LSS field. The astrometric comparison to multi-wavelength data is also presented in \cite{Whittam2023}, with a mean offset of (-0.24\arcsec, -0.40\arcsec) in a sub-region of the COSMOS field.}  As shown in Figure \ref{fig:astrometry} the median astrometric offsets for our more permissive comparison catalogue are typically limited to $\lesssim$0.06\arcsec, {with the uncertainties derived from the 16th and 84th percentiles $\sim0.3-0.4$\arcsec. The density of sources which can be seen for the astrometric offsets demonstrate that most offsets are well within} 1.1\arcsec \ (the image pixel size {in both the high and low resolution images}), suggesting that any effects of smearing due to the mosaicking of pointings are having a negligible impact on the measured locations of sources. The comparison of the more restrictive matched catalogues (with the additional SNR and unresolved criterion) show a similar offset magnitude compared to the more permissive catalogue, {with a reduction in the percentile derived uncertainties to $\sim0.15-0.2$\arcsec.} 
These comparisons demonstrate that the astrometry in DR1 is consistent with that of the Early Science data. 

For CDFS-DEEP we compare the MIGHTEE data to VLA \citep{Miller2013} and ATCA \citep{Franzen2015} imaging in Figure~\ref{fig:astrometrycdfs}. We note that there are far fewer sources to compare with in this field, due to the lower sensitivity and smaller area of the ATCA and VLA data, respectively. {Such surveys from the VLA and ATCA have been compared to each other in \citep{Franzen2015} and find mean offsets in RA and DEC of $\sim$0.1\arcsec for sources with SNR$\geq$20 in the ATCA data.}
The median offset to the ATCA data is {$\sim0.25$\arcsec \ in RA and $\sim0.35$\arcsec \ in Dec, much smaller than the pixel resolution of both the high and low resolution images. Comparing with the VLA data, we find a similar offset in RA, but a smaller offset in Dec of $\sim0.20$\arcsec.} Both are well within the 1.1\arcsec \ pixel size of the MIGHTEE data. 
Whilst the majority of sources have consistent astrometry, within a pixel, when compared to the work of \cite{Miller2013}, there are a significant number of sources which appear to have larger astrometric uncertainties when compared to the work of {\cite{Franzen2015}}, which appear to be typically more dispersed in Declination. As this is significantly less notable in the comparison with \cite{Miller2013}, {this} may relate to the ATLAS observations, which have an elongated beam in the north-south direction (due to the array configuration) {and so any offsets may be more likely in this direction}. {\citep{Franzen2015} do not see such large declination offsets compared to the data of \cite{Miller2013}, which suggests this relates to differences in the selection criteria used to match sources to MIGHTEE.  } 

{As in the work of \cite{Whittam2023}, the astrometric accuracy of these catalogues will be understood better through the cross-matching of the MIGHTEE detected sources to multi-wavelength host galaxies. Such analysis will be presented in future work. }

 \begin{figure*}
     \begin{minipage}[b]{0.5\textwidth}
 \includegraphics[width=\textwidth]{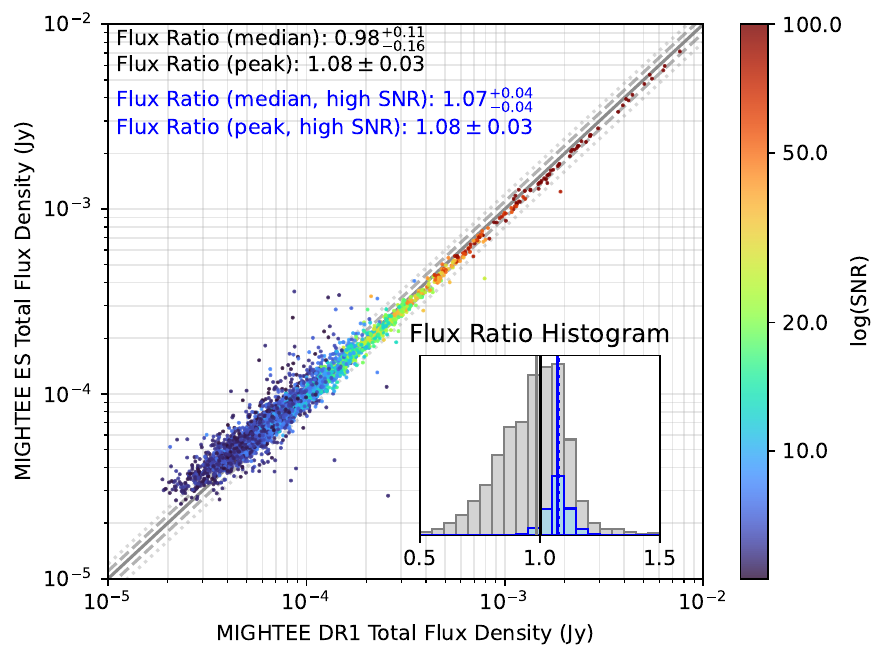}
    \subcaption{COSMOS High Resolution}
    \end{minipage}%
    \begin{minipage}[b]{0.5\textwidth}
\includegraphics[width=\textwidth]{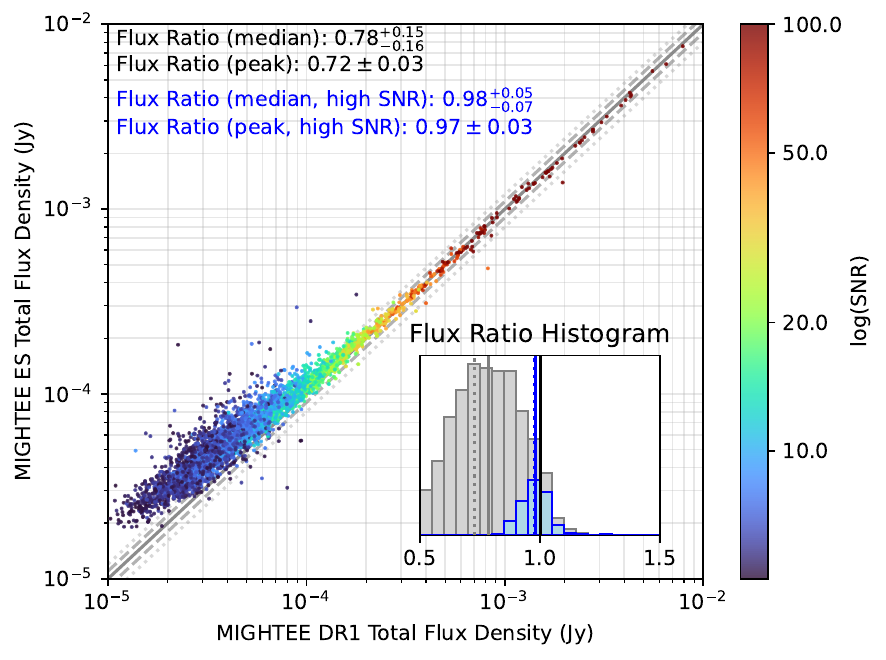}
    \subcaption{COSMOS Low Resolution}
    \end{minipage}%
    
    \begin{minipage}[b]{0.5\textwidth}
\includegraphics[width=\textwidth]{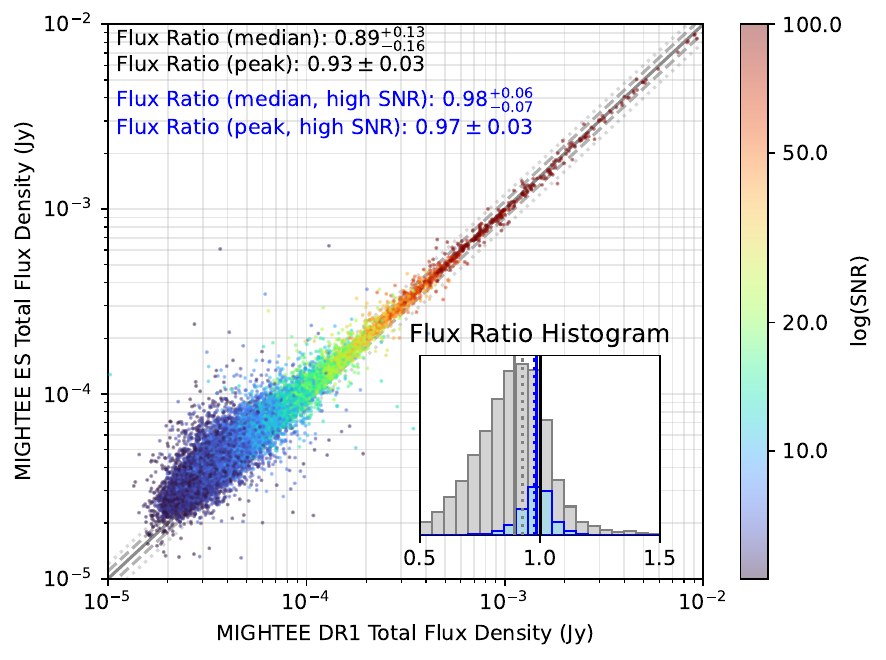}
    \subcaption{XMM-LSS High Resolution}
    \end{minipage}%
        \begin{minipage}[b]{0.5\textwidth}
\includegraphics[width=\textwidth]{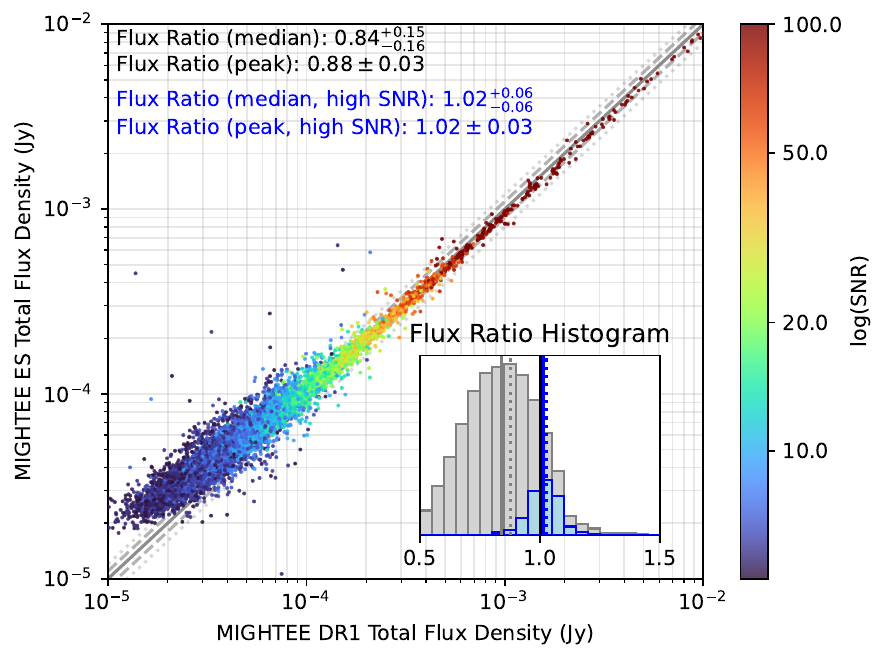}
    \subcaption{XMM-LSS Low Resolution}
    \end{minipage}%
     \caption{{Comparisons of the integrated flux densities {(scaled to 1.4 GHz)} between DR1 and the Early Science (ES) data from \protect \cite{Heywood2022} for the COSMOS (upper) and XMM-LSS (lower) fields, for the high (left) and low resolution (right) catalogues for the single component sources matched with {the matching criteria} discussed in Section \protect \ref{sec:validation}. Sources are coloured by their peak SNR in the ES data and the inset shows the histogram of the ratio of the flux density measurement in the ES data compared to DR1 for all sources plotted (grey) and for sources with SNR in ES$\geq20$ (blue). For each histogram the vertical lines are the ratio of 1 (black), and the coloured lines are the peak (dotted) and median (solid) values of the ratio, using the same colour scheme as the histogram. The black solid line indicates a 1-to-1 ratio, grey dashed lines indicate ratios of the flux densities of 0.9 and 1.1 and the grey dotted lines indicate flux density ratios of 0.8 and 1.2.}}
     \label{fig:fluxscale1}
 \end{figure*}

 \begin{figure*}
     \begin{minipage}[b]{0.5\textwidth}
 \includegraphics[width=\textwidth]{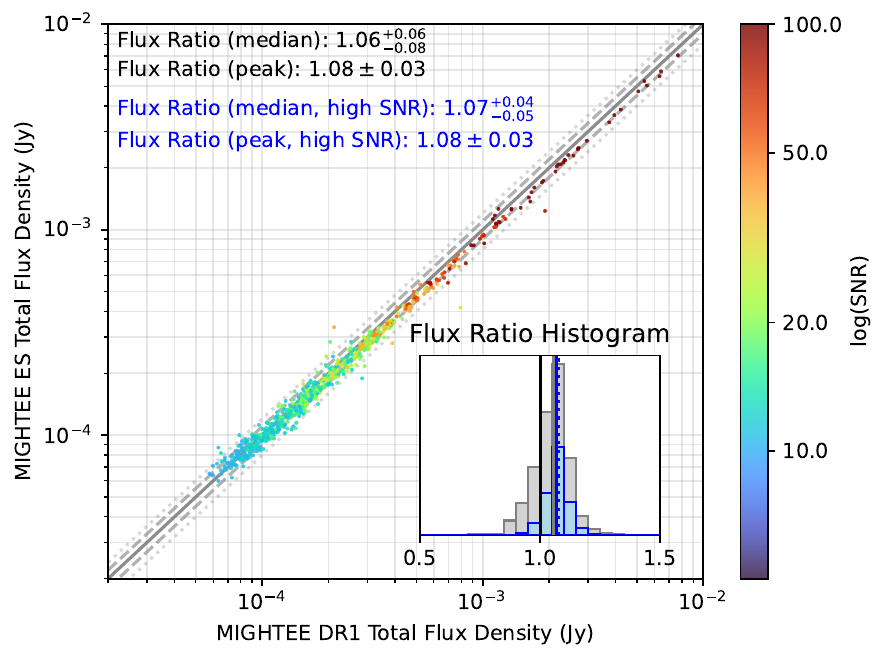}
    \subcaption{COSMOS High Resolution}
    \end{minipage}%
    \begin{minipage}[b]{0.5\textwidth}
\includegraphics[width=\textwidth]{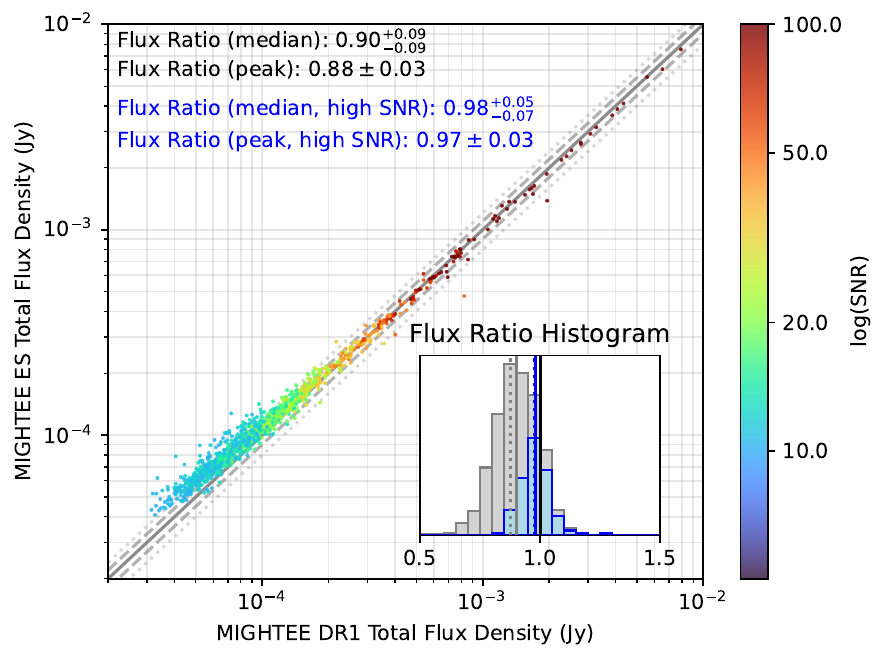}
    \subcaption{COSMOS Low Resolution}
    \end{minipage}%
    
    \begin{minipage}[b]{0.5\textwidth}
\includegraphics[width=\textwidth]{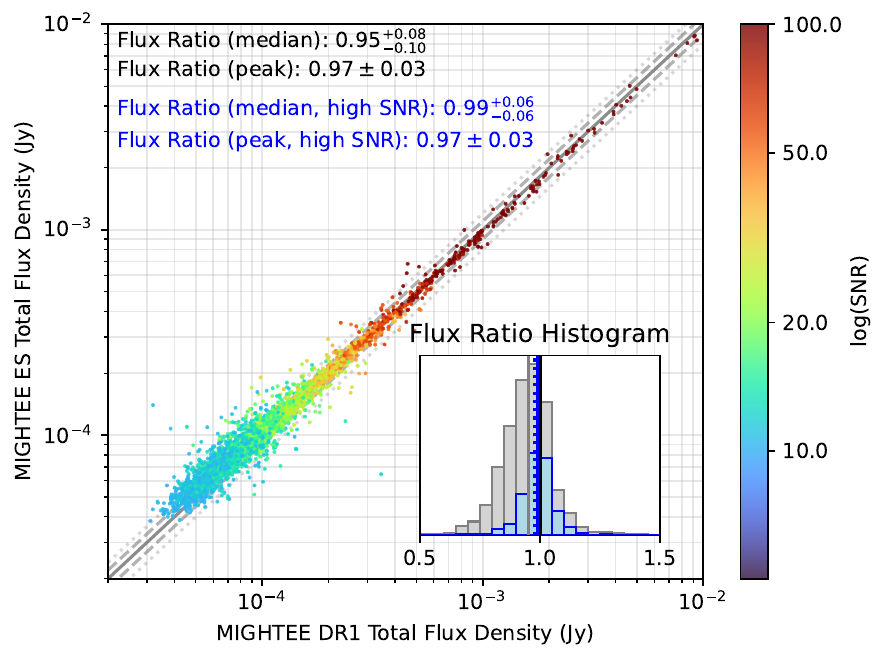}
    \subcaption{XMM-LSS High Resolution}
    \end{minipage}%
        \begin{minipage}[b]{0.5\textwidth}
\includegraphics[width=\textwidth]{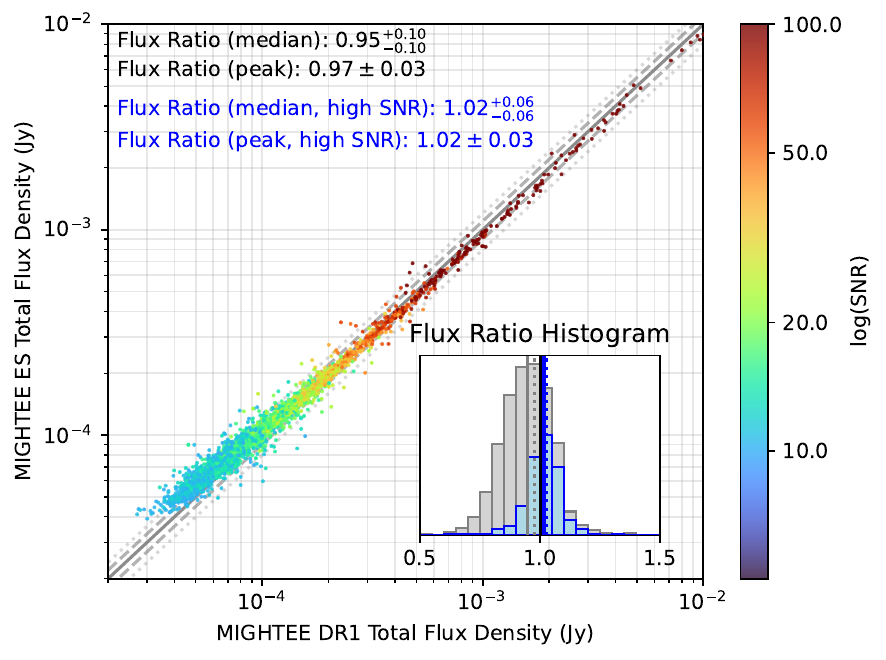}
    \subcaption{XMM-LSS Low Resolution}
    \end{minipage}%
     \caption{As for \protect Figure \ref{fig:fluxscale1} the additional SNR and unresolved criteria applied to the COSMOS and XMM-LSS sources, as discussed in Section \protect \ref{sec:validation}.}
     \label{fig:fluxscale2}
 \end{figure*}

 \begin{figure*}
     \begin{minipage}[b]{0.5\textwidth}
 \includegraphics[width=\textwidth]{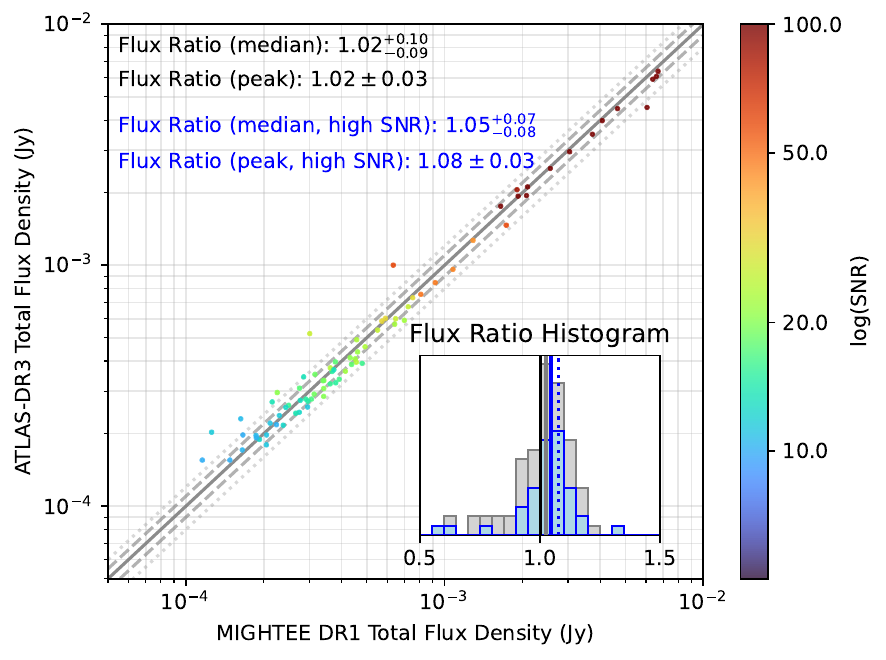}
    \subcaption{CDFS-DEEP High Resolution to \protect {\cite{Franzen2015}}}
    \end{minipage}%
    \begin{minipage}[b]{0.5\textwidth}
\includegraphics[width=\textwidth]{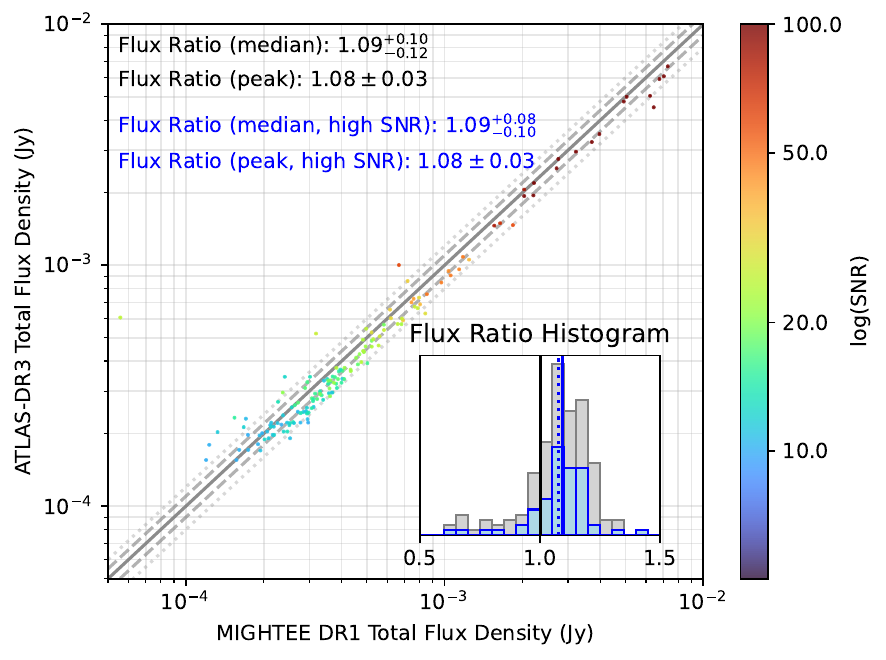}
    \subcaption{CDFS-DEEP Low Resolution to \protect {\cite{Franzen2015}}}
    \end{minipage}%
    
     \begin{minipage}[b]{0.5\textwidth}
 \includegraphics[width=\textwidth]{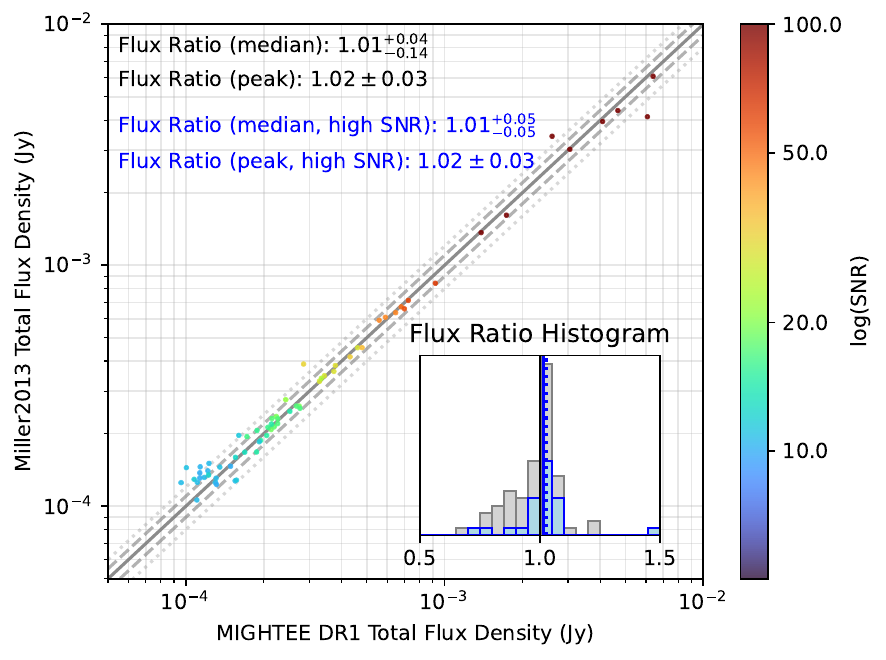}
    \subcaption{CDFS-DEEP High Resolution to \protect \cite{Miller2013}}
    \end{minipage}%
    \begin{minipage}[b]{0.5\textwidth}
\includegraphics[width=\textwidth]{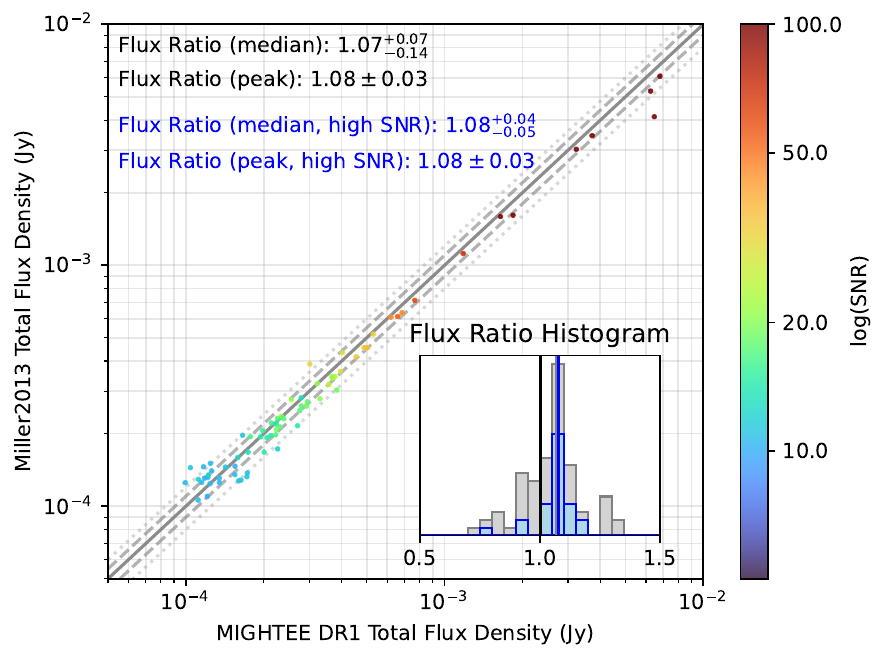}
    \subcaption{CDFS-DEEP Low Resolution to \protect \cite{Miller2013}}
    \end{minipage}%
    
     \caption{Comparisons of the integrated flux densities {(scaled to 1.4 GHz)} between MIGHTEE CDFS-DEEP data to {\protect \cite{Franzen2015}} (upper panels) and \protect \cite{Miller2013} (lower panels), for the high (left) and low resolution (right) catalogues for the catalogues made with the more stringent criteria discussed in Section \protect \ref{sec:validation}. Sources are coloured by their peak SNR in the non-MIGHTEE data and the inset shows the histogram of the ratio of the flux density measurement in the non-MIGHTEE data compared to DR1.  }
     \label{fig:fluxscalecdfs}
 \end{figure*}

\subsection{Flux-Density Comparisons}
\label{sec:flux}
In this Section we compare the flux-density scale of these Data Release 1 images to that of the Early Science data in the COSMOS and XMM-LSS fields in Figure~\ref{fig:fluxscale1} for the more permissive cross-matched catalogue, and in Figure~\ref{fig:fluxscale2} when the additional SNR and unresolved criteria are applied. {As discussed, the Early Science data have been compared to previous surveys from the Very Large Array (VLA) in \cite{Heywood2022}, where they run the same source finder, \textsc{PyBDSF}, over the VLA image data to make source comparisons using a consistent source finding process. \cite{Heywood2022} identify median flux density ratios between the MIGHTEE and VLA data of 0.95 (1.0) for the COSMOS (XMM-LSS) field.} These more restrictive cuts are likely to have a much larger effect on the flux-density comparisons. This is because low SNR sources are much more likely to be affected by confusion, Eddington bias \citep{Eddington1913} and measurement biases \citep[see discussions for Early Science data in][]{Hale2023} which will all affect the measured flux density of sources. This will be especially important in the outer regions of the Early Science data where the rms is larger and so the measurement of faint sources is more challenging. In DR1 the effect of mosaicing together multiple pointings reduces the noise in the outer regions of the Early Science area, which are now fully within the DR1 mosaic, as can be seen in Figure \ref{fig:images_es_comp}. {As these DR1 images have different effective frequency maps compared to the Early Science data, we scale the data to a common frequency of 1.4 GHz. We similarly use a frequency of 1.4 GHz when making comparison to the data of \cite{Miller2013} and {\cite{Franzen2015}} in the CDFS field.}

As can be seen in Figure \ref{fig:fluxscale1}, at the highest flux densities the comparison between ES and DR1 follows an approximate 1-to-1 relationship, with larger deviations from this found at lower flux densities. {This scatter reduces when the additional SNR and unresolved criterion are applied (Figure \ref{fig:fluxscale2}), with median ratios of the flux densities $\lesssim$ 3\%. In the COSMOS high resolution images, though, this median offset is more similar to 7\%, with the reason for the larger offset unclear.} In Figures ~\ref{fig:fluxscale1} and \ref{fig:fluxscale2}, sources are coloured based on their peak SNR in the Early Science data, which highlights that the sources which significantly deviate from the 1-to-1 relation are at the lowest SNR ($\lesssim10$). These sources appear to have an excess flux density in the Early Science data compared to DR1. This can be understood due to the higher noise of the Early Science data, which may cause an apparent boost in the flux densities due to Eddington bias compared {with DR1. This is because fainter sources in the Early science data may be preferentially boosted to be detected above the detection threshold, given the larger noise in the Early science data.} This {causes} the median ratio of the flux densities (shown in the histogram inset) {to} appear to deviate from the expected ratio of 1 by up to $\sim$20\%. In comparison, with the more stringent cuts applied to the data to ensure sources are unresolved and at high SNR, the median flux density ratio is closer to a ratio of 1 with median offsets typically constrained to within $\sim$10\%. The boosting of flux in the ES data compared to DR1 is more apparent in the low resolution images where, as discussed, the combination of confusion and Eddington bias will be more prevalent.

For CDFS-DEEP, we again compare to the work of \cite{Miller2013} and {\cite{Franzen2015}}. This introduces challenges in comparing flux densities due to the different baseline configurations between MeerKAT compared to the VLA and ATCA, which could lead to extended emission being resolved out, particularly in the VLA data. The observations of \cite{Miller2013} are much higher resolution (2.8\arcsec $\times$ 1.6\arcsec) whilst {\cite{Franzen2015}} has a more elongated beam, but at more comparable resolution (12\arcsec $\times$ 6\arcsec). For comparisons with \cite{Miller2013}, it may therefore be the case that sources which are unresolved in MIGHTEE still have emission resolved out in the VLA data of \cite{Miller2013}. Furthermore, there are also differences in the source finding algorithms that were used to generate the source catalogues which may also affect detections and characterisations of sources \citep[{see comparisons in e.g.}][]{Hopkins2015,Bonaldi2021, Boyce2023b}.

The comparison of the flux density of MIGHTEE CDFS-DEEP sources, when the more stringent cuts are applied, are shown in Figure~\ref{fig:fluxscalecdfs}. This results in a small number of sources available for comparison. We find {a good} agreement between our observations and the catalogue of \cite{Miller2013}, especially when making comparisons to the high resolution MIGHTEE data, with a median offset of $\lesssim5$\%. However, we find a larger offset in the source flux density compared to {\cite{Franzen2015}}, with the MIGHTEE sources appearing to have an excess in flux density compared to those in the ATCA catalogue, with a median offset similar to {$\sim$10\%}. The reasons for any remaining differences {are} unclear, although we note the difference in resolution, baseline distributions and source extraction software could all have an impact. {We note, though, that in their work, \cite{Franzen2015} find a typical 2$-$3\% offset to the surveys of \cite{Miller2013}, again suggesting that our work has a different selection of sources for comparison with MIGHTEE than used in their work. }

\section{Combined 1.4 GH\MakeLowercase{z} Source Counts}
\label{sec:sourcecounts}
In this section we use our source catalogues to measure the Euclidean normalised source counts within the three deep fields\footnote{Unless otherwise stated, the completeness and sources counts presented have been scaled to 1.4 GHz.}. We take a similar {approach to a number of works which consider the completeness in radio images \citep[see e.g.][]{Smolcic2017, Hale2021, Shimwell2022} and in the work of \cite{Hale2023} by} injecting sources into an image and considering the recovery of sources by the source finder, \textsc{PyBDSF}. However, such confused images present challenges to accurately measure and account for incompleteness. For example, any sources injected in the image would add to the confusion {noise, whereas} injecting sources into the residual image will mean that the effects of source confusion {and source blending} are {likely to be} underestimated. Furthermore, injecting a realistic number of sources into the residual image that is consistent with the observed source counts, will add to the background of $<1\sigma$ sources which have not been removed from the image when determining the rms map, {significantly increasing the sky density of these faint sources}. Finally, one could also create {images which do not make use of the MIGHTEE images (or residual images) and instead theoretically account for the rms variations and primary beam attenuation based on theoretical predictions from the telescopes known properties. However, such images} also present challenges in accurately including observational limitations of e.g. artefacts, deconvolution issues as well as accurately including the true morphology of sources in the image. 

In this work, we take the approach of directly injecting sources into the image and recovering sources. This allows us to better account for effects of confusion within this deep imaging however, as the image is already confused, it already contains a significant number of sources both above and below the detection threshold. {This approach is likely to be the most accurate in estimating and accounting for the image and source detection systematics, provided that the number of sources injected into the image is restricted to minimise additional contribution to the confusion noise.} For such simulations, we make use of the Tiered Radio Extragalactic Continuum Simulations \citep[T-RECS][]{Bonaldi2019, Bonaldi2023} and inject a number of sources directly into the original image ({2000} in CDFS-DEEP, {2500} in COSMOS and {7500} in XMM-LSS, {or $\sim$10\% of the detected sources per field}) using a limiting 1.4 GHz flux density of 4 $\muup$Jy in COSMOS and XMM-LSS and 2 $\muup$Jy in CDFS-DEEP {(as CDFS-DEEP is deeper, see Table \ref{tab:catalogues})}. As the effective frequency changes across the field of view, we scale the flux density of simulated sources using the effective frequency (and a spectral index of $\alpha = 0.7$) at the random locations. Each source is then injected in the image {using} to different source models depending on the source type, discussed in \cite{Bonaldi2019,Bonaldi2023}. This makes use of models from the \texttt{galsim} package \citep{galsim} and injects using the prescriptions used in the \texttt{simuclass} package\footnote{\url{https://github.com/itrharrison/simuclass-public/}} {\citep{Harrison2020}}.

\begin{table}
\begin{tabular}{c | c c | c c | c c} \hline
& \multicolumn{2}{c}{CDFS-DEEP} & \multicolumn{2}{c}{COSMOS} & \multicolumn{2}{c}{XMM-LSS} \\
 Label & $S_{\textrm{min}}$ & N& $S_{\textrm{min}}$ & N & $S_{\textrm{min}}$ & N  \\
   & ($\muup$Jy) & & ($\muup$Jy) & & ($\muup$Jy) &  \\ \hline
 Default & 2 & 2000 & 4 & 2500 &4 & 7500 \\
  High Flux 0 & 10 & 1000 & 10 & 2500 & 10 & 7500 \\
    High Flux 1 & 25 & 1000 & 25 & 2500 & 25 & 7500 \\
    High Flux 2 & 100 & 1000 & 100 & 2500 & 100 & 7500 \\ \hline
\end{tabular}
\caption{Details of the number (N) and minimum 1.4 GHz flux density limit ($S_{\textrm{min}}$) of sources injected into the images of the three fields (CDFS-DEEP, COSMOS and XMM-LSS) fields respectively for the completeness simulations used in Section \ref{sec:sourcecounts}.}
\label{tab:sc_sims}
\end{table}

After injecting these sources into the image, the final rms map discussed in Section~\ref{sec:radiocatalogues} is used to re-extract the sources, using the {final stage of the} source finding prescription discussed in Section \ref{sec:radiocatalogues}.
{In} this work, our procedure of {supplying the rms map used in the DR1 source catalogue generation} ensures that the number of simulated sources injected in the image are not artificially inflating the measured noise in the image which would be measured by \textsc{PyBDSF}. {We repeat this process to generate a total of {4000} simulations for each field.}

{As we supply the \textsc{PyBDSF} {rms map which was used for source detection to produce the final continuum catalogue}, the detected sources will be a combination of the original sources in the image and the simulated image. Therefore, when measuring the completeness we use a different approach to that of \cite{Hale2023}. We do not impose any matching radius when comparing our simulated sources that we detect to those injected in the image. Instead we calculate the completeness by comparing the flux distributions of the total detected sources with the total  number of injected sources as well as those within the image already. This allows us to better measure the true completeness, accounting for the blending and grouping of sources.}  {More formally}, we consider the total completeness ($C$) within a given flux-density bin ($S$ to $S + dS$) {to be}:

\begin{equation}
C(S, S+dS) = \frac{N_{\textrm{Output sims}}(S, S+dS) - N_{\textrm{sims}}  N_{\textrm{Image}}(S, S+dS)}{N_{\textrm{Input sims}}(S, S+dS)},
\label{eq:comp}
\end{equation}

\noindent where $N_{\textrm{Output sims}}(S, S+dS)$ is the number of sources detected by \textsc{PyBDSF} across all simulations with measured flux densities {in the given flux density bin ($S \rightarrow S+dS$)}.  $N_{\textrm{Image}}(S, S+dS)$ is the number of sources within the Data Release 1 catalogues (Section \ref{sec:radiocatalogues}) in the given flux-density bin and
$N_{\textrm{sims}}$ is the number of simulations {(4000 here for the default simulations)}. $N_{\textrm{Input sims}}(S, S+dS)$ is the total number of simulated sources {injected} across all of these simulations {with simulated input flux densities} within the given flux-density bin. The numerator accounts for the fact that the output catalogue for each simulation is comprised of both the simulated sources and the real sources. We note that the flux bins considered in this calculation are at 1.4 GHz and so the simulated input flux densities and measured output flux densities are first scaled to 1.4 GHz using the effective frequency maps, prior to the calculation of the completeness. 

Calculating the total completeness using such a method accounts for the combined effects of (i) incompleteness due to sensitivity variations, (ii) Eddington bias \citep{Eddington1913}, (iii) incompleteness due to the source finder,  and (iv) measurement errors {introduced from} the source finder. It also accounts for the effects of confusion and the blending of sources which are in close proximity, as well as the effects of source splitting where for e.g. the lobes of {AGN are} identified as two separate sources, instead of being attributed to a single source. This would not be properly probed if an exclusion radius around sources already in the original image was imposed, which is {important to  measure the true incompleteness} of the catalogues. 

We note that, due to the expected source counts distribution, the majority of sources injected into the image will be faint. Therefore to better understand the completeness at brighter flux densities, we generate additional simulations where we adopt a higher minimum flux density for the simulated sources. The additional flux density limits used and number of sources injected into the image are presented in Table \ref{tab:sc_sims}. To combine the simulations together, we use the `Default' simulations (Table~\ref{tab:sc_sims}) up to 10 times the minimum of the `High Flux 0' simulations. For each of the subsequent `High Flux' simulations we use them to determine the completeness above 10$\times$ the minimum flux density limit of the simulation, {having run 1000 iterations for each simulation}. 

{We estimate the associated uncertainties on the completeness measurements through combining subsets of simulations ($N_{\textrm{sims, subset}}$) into independent samples. To determine how many simulations should be combined (denoted $N_{\textrm{sims, subset}}$), we calculate the number of sources we expect to observe over the image area for each field, using the counts distribution of T-RECS \citep[][]{Bonaldi2019, Bonaldi2023}. We then compare how many simulations need to be combined (in each flux density bin used for completeness, {up to 10 mJy}) such that the total number of sources matches the expected number of sources in T-RECS. {Using the median of the integer number of simulations which should be combined, we generate a number of different `combined' samples.} }
We calculate the completeness in each of these {combined} samples, using Equation \ref{eq:comp} and use the standard deviation between these sub-samples for the uncertainty, as a function of flux density. Typically this results in $\sim$30$-$40 independent samples per field to calculate the completeness when using the default simulations and increases when the `High Flux' simulations are used. {For comparison, we also consider the completeness we would measure from the Default simulations up to the highest flux densities ($>0.1$ mJy). Above 0.1 mJy all the simulations described in Table~\ref{tab:sc_sims} are combined together to estimate the completeness and their associated errors.}

\begin{figure*}
\includegraphics[width=17cm]{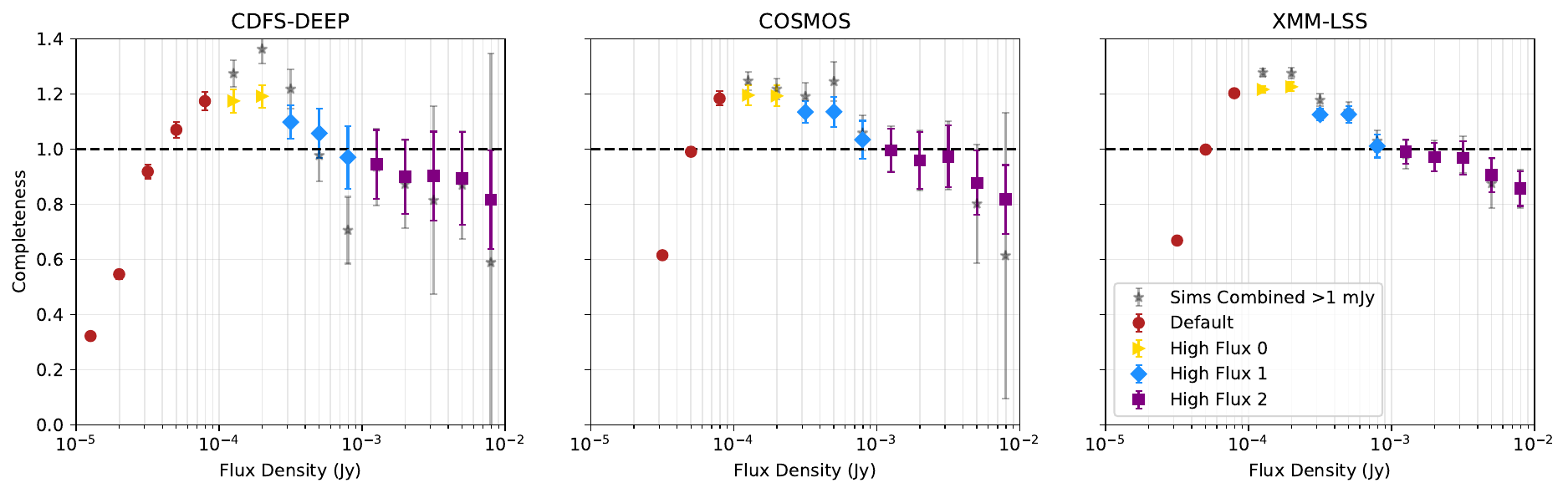}
\caption{Measured completeness as a function of flux density as described in Section \ref{sec:sourcecounts} for the CDFS-DEEP (left) COSMOS (centre) and XMM-LSS (right) fields. The completeness values associated with the different simulations in the analysis, as presented in Table \ref{tab:sc_sims} are indicated by different markers: Default (red circles), High Flux 0 (gold triangles), High Flux 1 (blue diamonds) and High Flux 2 (purple squares). {Finally the completeness when the default simulations is combined with the completeness from all the simulations in Table \ref{tab:sc_sims} together are also shown (grey stars).}}
\label{fig:completeness}
\end{figure*}

The completeness measurements are shown in Figure \ref{fig:completeness}, colour coded by the simulations used. For each of the fields, these can be seen to rise steeply before peaking around a completeness of $\sim$1.2,  and then declining to the expected value around a completeness of 1. {This shows some differences between the completeness from combining the simulations together at 10$\times$ the minimum flux density limit of the higher flux density simulations, compared to only at the highest flux densities ($>0.1$ mJy). We discuss the effect on the source counts below.} {At the highest flux densities, the completeness can appear to drop below a value of 1. This may be in part due to bright, jetted source being split into multiple components, but could also relate to the source models used by the simulation which if, for example, led to too large sources, may affect their completeness \citep[see e.g. the discussion of source sizes in TRECS from][]{Asorey2021}}. 

\begin{figure*}
 \includegraphics[width=15cm]{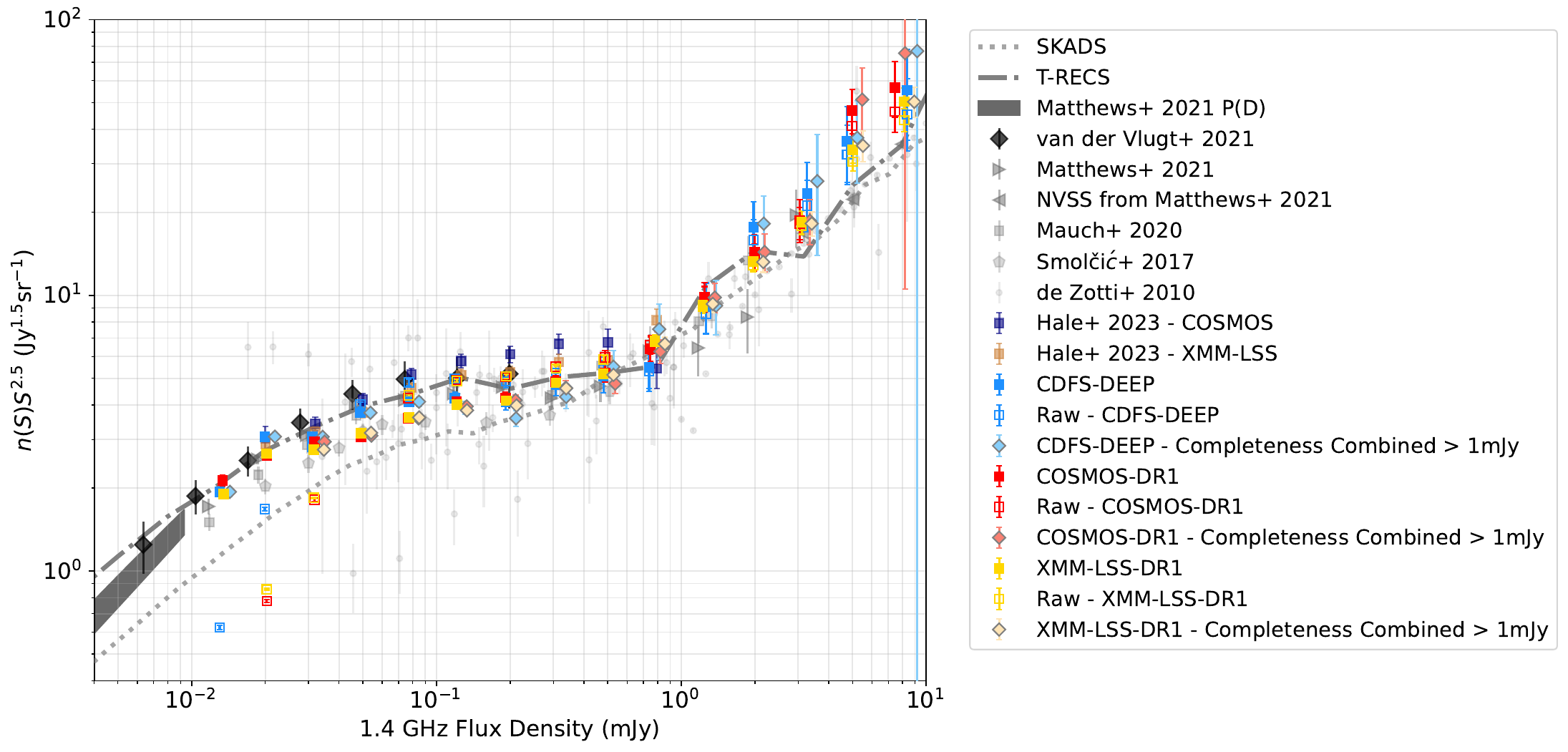}
 \caption{Euclidean normalised 1.4 GHz source counts for the COSMOS {(red)}, CDFS-DEEP {(blue)} and XMM-LSS (yellow) fields for the MIGHTEE Data Release 1 for both the corrected (filled markers) and raw (open markers) counts.  {The corrected source counts where the completeness is instead calculated from combining simulations above 1 mJy are also given as coloured diamonds with grey outlines and are artificially offset in flux density for visibility.} Also shown are source counts from the T-RECS simulations \protect \cite{Bonaldi2019, Bonaldi2023} (grey dot-dashed line), the SKADS simulated catalogues \protect \cite{Wilman2008} (grey dotted line) as well as counts from previous works of: \protect \cite{deZotti2010} (light grey circles), \protect \cite{Smolcic2017} (light grey pentagons), \protect \cite{Mauch2020} (grey squares), \protect \cite{Matthews2021} (grey triangles) and \protect \cite{vandervlugt2021} (black diamonds) as well as the sub-threshold counts from the P(D) analysis of \protect \cite{Matthews2021}. Source counts are presented at the median flux density of observed sources within the bin. }
 \label{fig:sc}
 \end{figure*}

{For the final corrected source counts for each field, we divide the source counts {from sources within the catalogue of each field (with the flux densities scaled to 1.4 GHz)} by the completeness measured in the given flux density bin. We only consider flux densities within the range from 5$\times$ the minimum flux density of the `Default' simulations (20 $\muup$Jy for COSMOS and XMM-LSS, 10 $\muup$Jy for CDFS-DEEP) to 10mJy. }
To determine the uncertainty on the source counts, we combine in quadrature the errors from both the Poissonian uncertainties from the number counts using the method of \cite{Gehrels} with the uncertainties from the completeness simulations (we discuss the effects of cosmic variance in Section \ref{sec:sc_cv}). The raw and corrected source counts for each field are presented in Figure \ref{fig:sc}, alongside the model source counts from T-RECS \citep{Bonaldi2023} and previous source counts from  various surveys, converted to 1.4 GHz, from the works of: \cite{deZotti2010, Smolcic2017, Mauch2020, Matthews2021, vandervlugt2021} and from the MIGHTEE Early Science data \citep{Hale2023}. An estimate of the sub-threshold source counts using a P(D) analysis in the MeerKAT DEEP2 field from \cite{Matthews2021} is also shown. {The results from the two methods used to combine the completeness simulations typically show small differeneces in the sources counts.} 
A table of the raw and corrected source counts is provided in Table \ref{tab:sc_tab}.

 Figure \ref{fig:sc} shows that the corrected source counts for the deep pointing within CDFS are in good agreement with previous measurements of the source counts from \cite{vandervlugt2021, Matthews2021, Hale2023} (which are some of the deepest source counts currently available) and the simulated catalogues of \cite{Bonaldi2019}. The source counts from COSMOS and XMM-LSS are lower than for CDFS-DEEP, across the flux density range of $\sim$50$-$200 $\muup$Jy, but in  agreement with \cite{Smolcic2017}. Notably, in this range the raw source counts are generally larger than the corrected source counts measurements, due to a completeness value$>$1. 
 {As discussed in \cite{Hale2023}, a completeness greater than 1 reflects the movement of sources within flux density bins, leading to an apparent excess of sources compared to the number of simulated sources that were injected. This will be in part due to Eddington bias, which preferentially boosts the more abundant fainter flux-density objects to high fluxes, whereas fewer sources are shifted to fainter fluxes resulting in a net positive gain in certain flux-density bins. This is also exacerbated by} the merging/splitting of sources when they are in close proximity to each other {(as may be the case in these images)} {and will also be affected by the fitting of source by \textsc{PyBDSF}}. These effects are  dependent on the underlying true source counts, as discussed in \cite{Hale2023}, thus a reasonable model of the underlying sources counts is important to determine the correct completeness.

{The results from DR1} are also generally lower in the range $S_{1.4} \sim 0.1-1.0$ mJy compared to the Early Science results in the {COSMOS} {field \citep{Hale2023}}. We note that the source counts in that regime for the COSMOS Early Science field were, in general, larger than both the models from T-RECS and some deep source count observations from previous surveys. This may be as a result of {differences in accounting} for the effects of confusion and the blending of sources in {this work and} the work of \cite{Hale2023} or may relate to differences in the flux scale when comparing the flux densities in the Early Science and DR1 data, as can be seen in Figure \ref{fig:fluxscale1} and \ref{fig:fluxscale2}. {As demonstrated, these figure} appear to suggest an excess in the flux density of the Early Science data compared to {DR1 for those sources which are at low SNR, which will lead to differences in the raw source counts}. However, whilst in \cite{Hale2023} evidence was presented that a small number of sources with extended host galaxies were detected in MIGHTEE and not in the VLA 3 GHz COSMOS survey and so {\cite{Hale2023} attributed the larger source counts observed in their work compared to \cite{Smolcic2017} to be in part due to resolution bias. However,} it may be the case that a combination of effects are in play. The VLA 3 GHz COSMOS survey may indeed be less sensitive to extended emission within the field, however it is also significantly less affected by the effects of confusion. In this work we have allowed confusion to be more readily included in our completeness calculations through not forcing any radial constraints on simulated sources, prior to the calculation of completeness. Therefore, confusion may be better accounted for in this work {compared to the Early Science source counts of \cite{Hale2023}} and so this may explain the lower source counts in the COSMOS and XMM-LSS fields. However, the work of \cite{vandervlugt2021} {also} use much higher resolution observations than in this work {but agree better with the results from the CDFS-DEEP field in this work.} The source counts in CDFS-DEEP are also in better agreement with that of \cite{Matthews2021} and \cite{Bonaldi2023}. {In reality, a combination of effects will contribute to the differences observed in the source counts presented in this work compared to previous studies both with the MIGHTEE Early Science Data and other work. These include, differences in the source finding strategies, cosmic variance, flux-scale offsets, and differences in accounting for incompleteness and frequency variations across the image. Most notably, in this work accounting for the effects of confusion and source blending is challenging and, whilst we have presented a method to account for this, there may be remaining systematics. }

\subsection{Sample Variance}
\label{sec:sc_cv}
As can be seen from Figure \ref{fig:sc} and discussed in Section \ref{sec:sourcecounts}, there are differences between the three fields which are a result of sample variance in the number of sources in different regions of the sky, although we note that there may be residual observational effects, e.g. flux offsets between fields, as well as the scatter due to the Poisson statistics alone. 
One of the key advantages of MIGHTEE is its ability to probe large regions of sky that {will probe independent sight lines and therefore different parts of the large-scale structure}. Larger sky surveys such as FIRST \citep{FIRST, Helfand2015} and LoTSS \citep{Shimwell2019, Shimwell2022} allow larger areas to be probed and sample variance to be overcome on much larger scales. {However, these surveys are typically shallower than MIGHTEE, so can only constrain sample variance for brighter populations.} 
In this section we compare the {variance in the source counts as a function of areal extent. We make comparisons to the work of} \cite{Heywood2013}, who used the SKA Simulated Skies \citep[SKADS][]{Wilman2008} catalogue to probe {the expected sample variance}.

\begin{figure*}
    \centering
    \includegraphics[width=16cm]{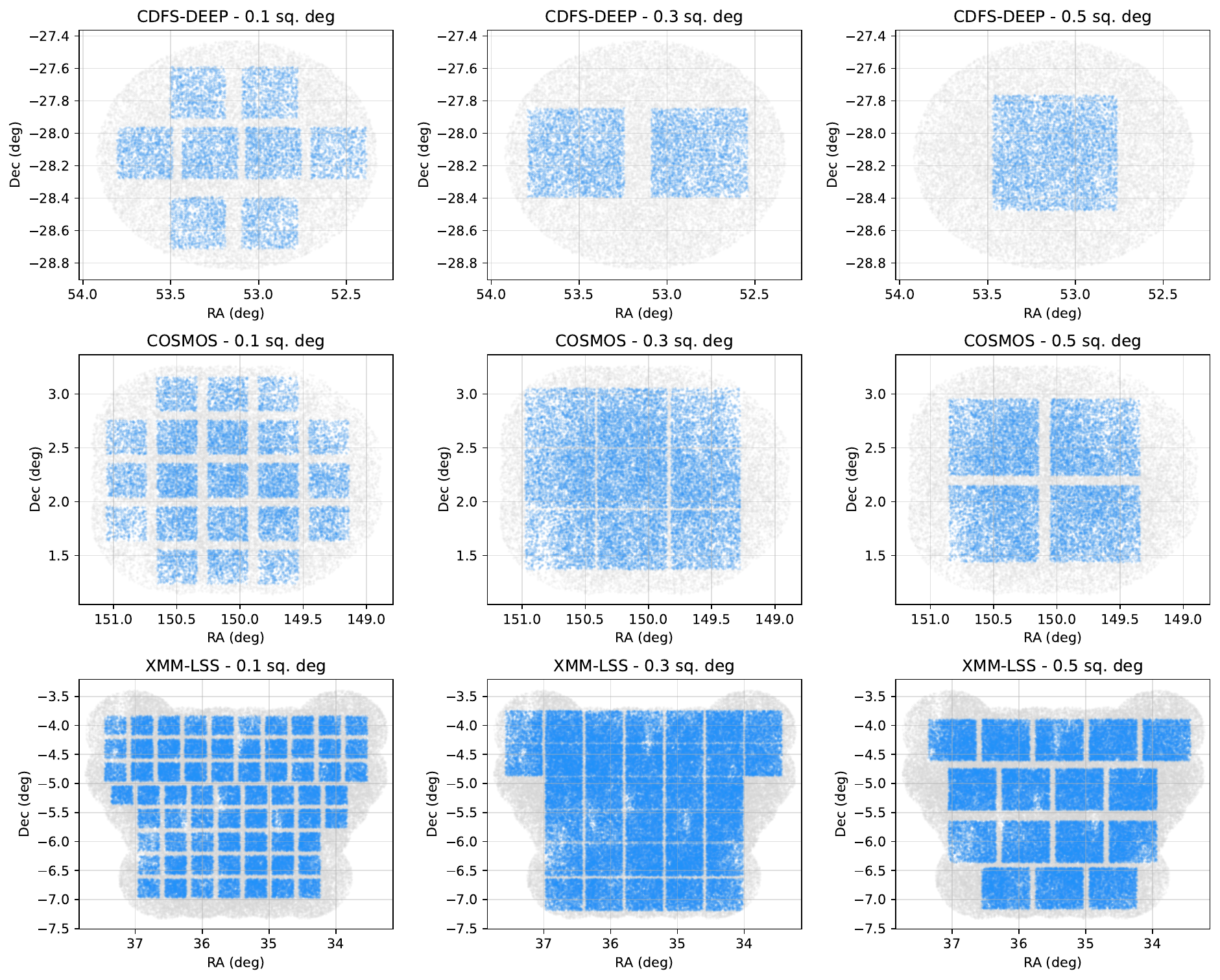}
    \caption{{Regions used to define the cosmic variance within the fields. Grey points indicate the MIGHTEE radio sources across the full fields and the blue points indicate those in the smaller regions used to probe the sample variance. Shown are such regions for CDFS-DEEP (top row), COSMOS (middle row) and XMM-LSS (bottom row) for the 0.1, 0.3 and 0.5 sq. deg regions from left to right.}}
    \label{fig:cv_regions}
\end{figure*}

To do this, we consider the angular areas used in \cite{Heywood2013} which can be sampled within the MIGHTEE area, across each of the three fields (i.e. area$\lesssim$ 1.5 sq. deg) and where we can also define $\geq$10 sub-regions. This restricts us to areas of 0.1, 0.3, and 0.5 sq. deg. {For each area considered, we} then estimate uncertainty due to sample variance using our three fields. {In doing so we average the effects of clustering and large-scale structure across a range of redshifts.} Figure~\ref{fig:cv_regions} presents the sub-regions we use for this analysis. Following the same method as described in Section~\ref{sec:sourcecounts}, we calculate the raw source counts along with the completeness over each sub-region\footnote{{Where we combine the different simulations using the 10$\times$ minimum flux density limit of the high flux density simulations of Table \ref{tab:sc_sims}.}}. {Using this we estimate the fractional sample variance from the mean and standard deviation of the completeness corrected source counts over these sub-regions.}

\begin{figure}
    \centering
    \includegraphics[width=8cm]{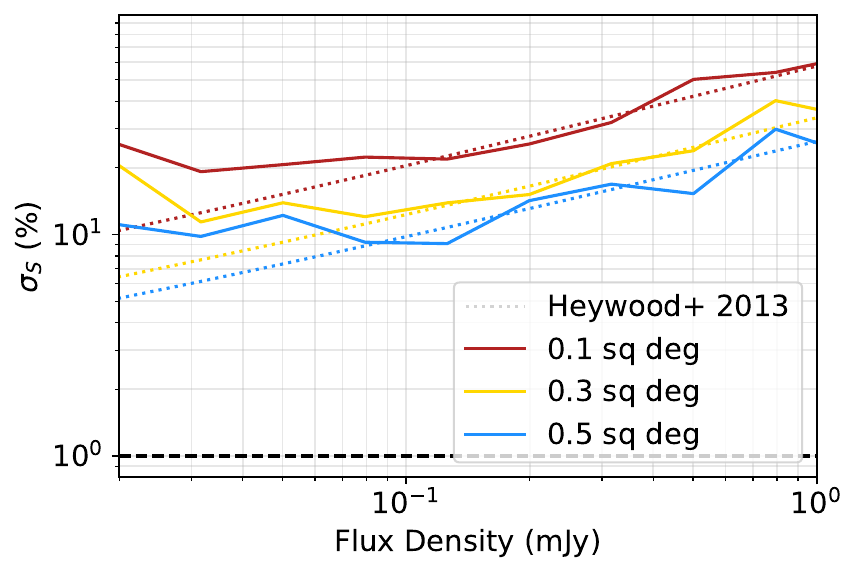}
    \includegraphics[width=8cm]{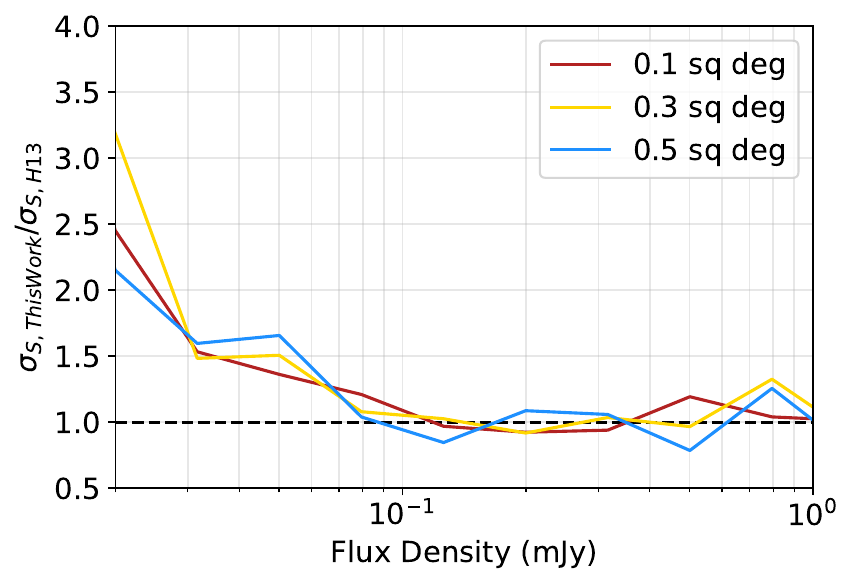}
    \caption{{Comparison of the observed sample variance relative error {($\sigma/\mu$)} from field to field variations compared to the predictions from \protect \cite{Heywood2013} for patches of areas: 0.1 sq. deg (red), 0.3 sq. deg (gold) and 0.5 sq. deg (blue). The upper panel show the sample variance relative error of this work (solid lines) given as a percentage compared to the models of \protect \cite{Heywood2013} (dotted) and the lower panel presents the ratio of our work compared to \protect \cite{Heywood2022} for each of the areas studied.}}
    \label{fig:cv_std}
\end{figure}

A comparison of the fractional sample variance ($\sigma$/$\muup$) measured across the sub-regions is shown in Figure \ref{fig:cv_std} (top panel) as a function of flux density for the three areal regions considered (0.1, 0.3 and 0.5 sq. deg). We find a typical sample variance of 10-20 per cent for the {areal sub-regions at $S_{1.4} \leq 100\muup$Jy}.
The sample variance increases in all cases to brighter flux densities as expected, given the lower source density and the expected higher galaxy bias for these {sources \citep[e.g.][]{Lindsay2014, Magliocchetti2017, Hale2018, Mazumder2022}, which will be dominated by AGN populations as opposed to the SFGs which dominate at fainter flux densities \citep[see e.g.][]{Smolcic2017b, Whittam2022, Best2023}.}

We also show this as a ratio {to} the {work} of \cite{Heywood2013} {in Figure} \ref{fig:cv_std} (bottom panel) over the same flux density and areas. We find that the fractional {sample variance} is {comparable when measured to that} {by \cite{Heywood2013} from SKADS, across a a wide range of flux densities ($\gtrsim$ 0.1 mJy), increasint to a factor of 2$-$3 at fainter flux densities}. \cite{Heywood2013} defined the sample variance ($\sigma_{S, \%}$) as the combination of the relative percentage errors due to clustering (or cosmic variance, $\sigma_{CV, \%}$) and the error due to the Poisson counts ($\sigma_{P, \%}$) of sources within the field:

\begin{equation}
    \sigma_{S, \%}^2 = \sigma_{CV, \%}^2 + \sigma_{P, \%}^2.
\end{equation}

However, we note that this is an idealised situation in which there are no incompleteness effects across the full area investigated. {Our observations also include uncertainty from the completeness simulations on our measurements of the source counts. Therefore, to compare with the clustering uncertainty presented in \cite{Heywood2013} we should combine the variance from the Poissonian counts, {cosmic variance} \emph{and} completeness uncertainties.} Thus, a direct comparison to the relative uncertainties due to cosmic variance (or clustering of sources) is challenging to measure. This is especially true here, as our completeness corrections {vary} for different sub-regions, principally due to the position within the field {and their proximity to bright sources and regions of higher noise}. {Therefore we are unable to conclude that the excess in sample variance at faint flux densities can be wholly attributed to an underestimate in the work of \cite{Heywood2013}. }

\section{Discussion and Conclusions}
\label{sec:conclusions}
In this work we present catalogues and images of the COSMOS, XMM-LSS and CDFS fields from the MIGHTEE survey {Data Release 1}, using the MeerKAT telescope. These observations total $\sim$20 sq. deg across the three fields with 1.5 sq. deg in CDFS-DEEP, 4.2 sq. deg in COSMOS and 14.4 sq. deg in XMM-LSS. Images are released at two angular resolutions, one prioritising resolution ($\sim$5\arcsec) and the second prioritising sensitivity at lower resolution\footnote{{Though for the CDFS-DEEP imaging, both are of comparable sensitivity.}} ($\sim7-9$\arcsec). {This results in images with central rms sensitivities of $\sim$1.3$-$2.7 $\muup$Jy beam$^{-1}$ in the lower resolution images and $\sim$1.2$-$3.6 $\muup$Jy beam$^{-1}$ in the higher resolution images. This increases to $\sim$2.0$-$3.5 $\muup$Jy beam$^{-1}$ (lower resolution) and $\sim$1.9$-$5.6 $\muup$Jy beam$^{-1}$ (higher resolution) across the full area.} Source finding and characterisation using \textsc{PyBDSF} identified a total of {143~817} sources in the lower resolution images and 114~225 sources in the higher resolution images, using a multi-stage source finding process that more accurately characterises the rms {across the areas observed}. {These observations present the largest areal observations {at GHz frequencies, which reach} an rms sensitivity {$\sim$ a few $\muup$Jy/beam depth}. This allows a {high density} of sources to be obtained, with $\gtrsim6000$ ($\gtrsim12000$) sources per sq. deg in the maximal sensitivity for the COSMOS and XMM-LSS (CDFS-DEEP) fields respectively, whilst also allowing the effects of sample variance to be overcome.}

We compare the astrometry and flux-density measurements of these catalogues to the Early Science MIGHTEE data of \cite{Heywood2022} for the COSMOS and XMM-LSS fields, and previous observations of \cite{Miller2013} and {\cite{Franzen2015}} for the CDFS-DEEP fields. These comparisons show that astrometric offsets are typically constrained to within a pixel of the MIGHTEE data. Comparison of source flux densities show agreement with Early Science observations at high SNR, with an excess seen in the flux density measurements of the Early Science data compared to the observations presented in this work, {for sources that were low SNR in the ES data}. We attribute this to effects of the higher noise {in the Early science data, which exacerbate Eddington bias in the ES catalogue at faint flux densities in these confused images, together with an improvement in the PyBDSF methodology used in this work}. \todo{For the CDFS field, differences in the flux densities are seen between previous CDFS field observations {of \cite{Franzen2015} using the ATCA telescope compared to the sources presented in this work. Such offsets are reduced when compared to \cite{Miller2013} at high SNR.} The reasons for this are unclear, though we note that there are significant differences in both the telescopes used, their configurations and the source finding algorithms used, which will all contribute {to observed variations}.}

{Next,} we consider the source counts distribution within the fields, compared to the results of the Early Science data of \cite{Hale2023}, simulations from T-RECS \citep{Bonaldi2019, Bonaldi2023} and previous deep radio observations. We calculate the raw and completeness corrected source counts  using a method that better accounts for the merging of sources within the highly confused fields discussed in this work. Our corrected counts for the CDFS-DEEP field show good agreement with previous observations and simulations \citep[e.g.][]{Matthews2021, Bonaldi2019, vandervlugt2021} in the faintest flux density bins ($\lesssim$100 $\muup$Jy), with the COSMOS and XMM-LSS fields exhibiting lower source counts to these works and more consistent with \cite{Smolcic2017}. These investigations demonstrate the challenges in accurately determining the source counts within such deep fields where the source density is high {as well as variations which can be observed between fields}. {High resolution imaging such as from the Square Kilometre Array Observatory (SKAO) and the International LOFAR Telescope \citep{Morabito2022} will be crucial in the future to overcome the effects of confusion in increasingly deeper observations of deep fields.}

{Finally, we} also use the large area covered by {MIGHTEE DR1} to measure the sample variance between sub-regions in the fields and compare these to the expectations from  \cite{Heywood2013}. We find a typical sample variance of 10-20 per cent for {the areal sub-regions considered for $S_{1.4} \leq 100\muup$Jy. The sample variance increases to brighter flux densities as expected, given the lower source density and the expected higher galaxy bias for these sources. Our results suggest a larger sample variance (up to a factor of $\sim2-3$) at the faintest flux densities $\lesssim$ 0.1 mJy compared to the predictions of \cite{Heywood2013}, with comparable variance found at brighter flux densities. However,} our measurements combine both uncertainties considered in their work (from Poissonian counts and source clustering) as well as the {effects arising from the spread of calculated completeness correction factors found. }

Alongside this paper we release images of the three fields (at both the resolutions discussed) and rms maps for the images alongside the catalogues presented in this paper. These are available from \url{https://doi.org/10.48479/7msw-r692}. Such catalogues will be elevated with future work, currently in progress, which will use the Gaussian and source catalogues to produce a host galaxy associated catalogue using a combination of visual identification \citep[as used in other radio surveys e.g.][]{Banfield2015, Prescott2018, Williams2019, Whittam2023} as well as statistical matching through the likelihood ratio method \citep[see e.g.][]{McAlpine2012, Kondapally2022, Whittam2023}. Furthermore, a future data release from the MIGHTEE field will extend the observations across the CDFS field and extend to the ELAIS-S1 field and will be the final L-band continuum observations from the MIGHTEE survey.

\section*{Acknowledgements}

The MeerKAT telescope is operated by the South African Radio Astronomy Observatory, which is a facility of the National Research Foundation, an agency of the Department of Science and Innovation. {We acknowledge the use of the ilifu cloud computing facility – www.ilifu.ac.za, a partnership between the University of Cape Town, the University of the Western Cape, Stellenbosch University, Sol Plaatje University and the Cape Peninsula University of Technology. The ilifu facility is supported by contributions from the Inter-University Institute for Data Intensive Astronomy (IDIA – a partnership between the University of Cape Town, the University of Pretoria and the University of the Western Cape), the Computational Biology division at UCT and the Data Intensive Research Initiative of South Africa (DIRISA).} 

This research made use of Astropy,\footnote{\url{https://www.astropy.org/}} a community-developed core Python package for Astronomy \citep{astropy1, astropy2}; \texttt{TOPCAT} \citep{topcat1, topcat2}; ds9 \citep{ds9, ds92}; \texttt{matplotlib} \citep{matplotlib}; \texttt{NumPy} \citep{numpy1,numpy2}; \texttt{SciPy} \citep{scipy}; and {\texttt{tqdm} \citep{tqdm}.} This work has made use of the Cube Analysis and Rendering Tool for Astronomy \citep[CARTA;][]{comrie2021}. This research has made use of NASA's Astrophysics Data System. This research made use of Montage, which is funded by the National Science Foundation under Grant Number ACI-1440620, and was previously funded by the National Aeronautics and Space Administration's Earth Science Technology Office, Computation Technologies Project, under Cooperative Agreement Number NCC5-626 between NASA and the California Institute of Technology. This research has made use of NASA's Astrophysics Data System.

{We thank SVW, FT and the anonymous referee for the comments which have been helpful in improving the manuscript.} CLH acknowledges support from the Leverhulme Trust through an Early Career Research Fellowship. CLH, MJJ and IHW also acknowledge support from the Oxford Hintze Centre for Astrophysical Surveys which is funded through generous support from the Hintze Family Charitable Foundation. IH and MJJ acknowledge the support of the STFC consolidated grant [ST/S000488/1] and [ST/W000903/1] and from a UKRI Frontiers Research Grant [EP/X026639/1]. IH acknowledges support from the South African Radio Astronomy Observatory which is a facility of the National Research Foundation (NRF), an agency of the Department of Science and Innovation. IH thanks the Rhodes Centre for Radio Astronomy Techniques and Technologies (RATT) for the provision of computing facilities that were used for processing some of the MIGHTEE data. PNB is grateful for support from the UK STFC via grant ST/V000594/1. FXA acknowledges the support from the National Natural Science Foundation of China (12303016) and the Natural Science Foundation of Jiangsu Province (BK20242115) RB acknowledges support from an STFC Ernest Rutherford Fellowship [grant number ST/T003596/1]. IHa acknowledges support from the European Research Council (ERC) under the European Union's Horizon 2020 research and innovation programme (Grant agreement No. 849169). DJBS acknowledges support from the UK Science and Technology Facilities Council (STFC) under grants ST/V000624/1 and ST/Y001028/1. MV acknowledges financial support from the Inter-University Institute for Data Intensive Astronomy (IDIA), a partnership of the University of Cape Town, the University of Pretoria and the University of the Western Cape, and from the South African Department of Science and Innovation's National Research Foundation under the ISARP RADIOMAP Joint Research Scheme (DSI-NRF Grant Number 150551) and the CPRR HIPPO Project (DSI-NRF Grant Number SRUG22031677). 

\section*{Data Availability}
Upon publication, the radio images and catalogues will be accessible through: \url{https://doi.org/10.48479/7msw-r692}



\bibliographystyle{mnras}
\bibliography{mightee_dr1} 



\appendix

\section{List of MeerKAT observations}
\label{sec:appendix}
Tables \ref{tab:cosmos_obs}, \ref{tab:xmm_obs} and \ref{tab:cdfs_obs} provide details of the 709.2~h of MeerKAT observations that were used to produce the data products presented here, for each of the three target fields. 

\begin{table*}
\begin{minipage}{176mm}
\centering
\caption{MeerKAT observations that were used to produce the COSMOS mosaics. {Columns included are the date of observation, observation ID, field, central RA and Dec of the pointing, time in hours of the full observation track and on-source time, number of spectral channels, number of antennas used in the observation and finally the primary and secondary calibrator sources. }}
\begin{tabular}{lllllllrlll} \hline
Date       & ID         & Field  & RA          & Dec        & Track   & On-source & N$_{\mathrm{chan}}$ & N$_{\mathrm{ant}}$ & Primary & Secondary \\ 
           &            &        & (J2000)     & (J2000)    & (h) & (h)   &                     &                    & \\ \hline
2018-04-19     & 1524147354     & COSMOS          & 10\hhh00\mmm29\sss        & 02\ddd12\dmm21\farcs0     & 8.65           & 6.10            & 4096           & 64             & J0408-6545     & 3C237          \\
2018-05-06     & 1525613583     & COSMOS          & 10\hhh00\mmm29\sss        & 02\ddd12\dmm21\farcs0     & 8.39           & 5.10            & 4096           & 62             & J0408-6545     & 3C237          \\
2019-07-16     & 1563267356     & COSMOS\_1       & 09\hhh59\mmm46\sss        & 02\ddd01\dmm44\farcs6     & 7.00           & 6.33           & 4096           & 59             & J0408-6545     & J1008+0740     \\
2019-07-27     & 1564215117     & COSMOS\_2       & 09\hhh59\mmm46\sss        & 02\ddd22\dmm57\farcs4     & 7.95           & 6.98           & 4096           & 61             & J0408-6545     & J1008+0740     \\
2019-07-28     & 1564301832     & COSMOS\_3       & 10\hhh01\mmm11\sss        & 02\ddd01\dmm44\farcs6     & 7.96           & 6.97           & 4096           & 60             & J0408-6545     & J1008+0740     \\
2019-08-16     & 1565939836     & COSMOS\_4       & 10\hhh01\mmm11\sss        & 02\ddd22\dmm57\farcs4     & 7.99           & 6.97           & 4096           & 58             & J0408-6545     & J1008+0740     \\
2019-08-23     & 1566542621     & COSMOS\_1       & 09\hhh59\mmm46\sss        & 02\ddd01\dmm44\farcs6     & 7.97           & 6.98           & 4096           & 61             & J0408-6545     & J1008+0740     \\
2020-03-28     & 1585413022     & COSMOS\_5       & 09\hhh59\mmm04\sss        & 02\ddd12\dmm21\farcs0     & 8.00           & 6.25           & 32768          & 59             & J0408-6545     & J1008+0740     \\
2020-03-29     & 1585498873     & COSMOS\_6       & 10\hhh01\mmm54\sss        & 02\ddd12\dmm21\farcs0     & 8.00            & 6.25           & 32768          & 59             & J0408-6545     & J1008+0740     \\
2020-03-31     & 1585671638     & COSMOS\_7       & 10\hhh00\mmm29\sss        & 01\ddd51\dmm08\farcs2     & 8.00            & 6.25           & 32768          & 60             & J0408-6545     & J1008+0740     \\
2020-04-02     & 1585844155     & COSMOS\_8       & 10\hhh00\mmm29\sss        & 02\ddd33\dmm33\farcs8     & 8.00            & 6.25           & 32768          & 60             & J0408-6545     & J1008+0740     \\
2020-04-30     & 1585928757     & COSMOS\_9       & 10\hhh01\mmm54\sss        & 02\ddd33\dmm33\farcs8     & 8.00           & 6.25           & 32768          & 60             & J0408-6545     & J1008+0740     \\
2020-04-04     & 1586016787     & COSMOS\_10      & 09\hhh59\mmm04\sss        & 02\ddd33\dmm33\farcs8     & 8.03           & 6.25           & 32768          & 60             & J0408-6545     & J1008+0740     \\
2020-04-06     & 1586188138     & COSMOS\_11      & 09\hhh58\mmm21\sss        & 02\ddd22\dmm57\farcs4     & 8.00           & 6.25           & 32768          & 59             & J0408-6545     & J1008+0740     \\
2020-04-07     & 1586274966     & COSMOS\_12      & 09\hhh58\mmm21\sss        & 02\ddd01\dmm44\farcs6     & 8.00           & 6.25           & 32768          & 60             & J0408-6545     & J1008+0740     \\
2020-04-12     & 1586705155     & COSMOS\_13      & 09\hhh59\mmm04\sss        & 01\ddd51\dmm08\farcs2     & 8.00           & 6.25           & 32768          & 59             & J0408-6545     & J1008+0740     \\
2020-04-13     & 1586791316     & COSMOS\_14      & 10\hhh01\mmm53\sss        & 01\ddd51\dmm08\farcs2     & 8.00           & 6.25           & 32768          & 60             & J0408-6545     & J1008+0740     \\
2020-04-26     & 1587911796     & COSMOS          & 10\hhh00\mmm29\sss        & 02\ddd12\dmm21\farcs0     & 7.98           & 6.25           & 32768          & 59             & J0408-6545     & J1008+0740     \\
2021-04-07     & 1617809470     & COSMOS\_1       & 09\hhh59\mmm46\sss        & 02\ddd01\dmm44\farcs6     & 8.00           & 6.25           & 32768          & 60             & J0408-6545     & J1008+0740     \\
2021-05-02     & 1619963656     & COSMOS\_2       & 09\hhh59\mmm46\sss        & 02\ddd22\dmm57\farcs4     & 7.96           & 6.25           & 32768          & 62             & J0408-6545     & J1008+0740     \\
2021-05-15     & 1621083675     & COSMOS\_4       & 10\hhh01\mmm11\sss        & 02\ddd22\dmm57\farcs4     & 7.97           & 6.25           & 32768          & 61             & J0408-6545     & J1008+0740     \\
2021-05-30     & 1622376680     & COSMOS\_3       & 10\hhh01\mmm11\sss        & 02\ddd01\dmm44\farcs6     & 8.00            & 6.70           & 32768          & 61             & J0408-6545     & J1008+0740     \\
\hline
\end{tabular}
\label{tab:cosmos_obs}
\end{minipage}
\end{table*}

\begin{table}
\centering
\caption{MeerKAT observations that were used to produce the CDFS-DEEP image. For all CDFS observations the primary calibrator was PKS B0408$-$65, and the secondary was J0240$-$2309. The pointing centre for all of these observations was J2000 03\hhh32\mmm30.4\sss $-$28\ddd07\mmm57\farcs0. {These observations all have 32768 channels and the column descriptions are given in Table \ref{tab:cosmos_obs}.}}
\begin{tabular}{lllll} \hline 
Date       & ID         &   Track   & On-source &  N$_{\mathrm{ant}}$  \\ 
           &            &  (h)      & (h)       &                      \\ \hline
 2019-12-12    & 1576162858     &  9.91          &  7.26          &  59            \\
 2020-01-03    & 1578058860     &  9.92          &  7.26          &  61            \\
 2020-01-24    & 1579878660     &  7.47          &  5.60          &  61            \\
 2020-01-26    & 1580039158     &  9.90          &  7.25          &  61            \\
 2020-02-08    & 1581162358     &  9.85          &  7.25          &  62            \\
 2020-02-09    & 1581248760     &  10.18         &  7.59          &  62            \\
 2020-02-22    & 1582371217     &  9.81          &  7.26          &  58            \\
 2020-04-25    & 1587795059     &  9.94          &  7.24          &  59            \\
 2020-09-28    & 1601325069     &  9.08          &  6.61          &  60            \\
 2020-09-29    & 1601409818     &  9.09          &  6.29          &  58            \\
 2020-10-03    & 1601756163     &  9.08          &  6.28          &  61            \\
 2020-10-05    & 1601928962     &  9.09          &  6.28          &  59            \\
 2020-10-07    & 1602098167     &  9.06          &  6.27          &  58            \\
 2020-10-08    & 1602183844     &  9.06          &  6.27          &  59            \\
 2020-10-09    & 1602270065     &  9.06          &  6.27          &  60            \\
 2020-10-14    & 1602704665     &  9.06          &  6.60          &  59            \\
 2020-10-15    & 1602792067     &  9.08          &  6.27          &  60            \\
 2020-10-16    & 1602873187     &  9.06          &  6.60          &  60            \\
 2020-10-17    & 1602964865     &  9.10          &  6.26          &  59            \\ \hline
\end{tabular}
\label{tab:cdfs_obs}
\end{table}

\begin{table*}
\begin{minipage}{176mm}
\centering
\caption{MeerKAT observations that were used to produce the XMM-LSS mosaics. {Column descriptions are given in Table \ref{tab:cosmos_obs}.}}
\begin{tabular}{lllllllrlll} \hline
Date       & ID         & Field  & RA          & Dec        & Track   & On-source & N$_{\mathrm{chan}}$ & N$_{\mathrm{ant}}$ & Primary & Secondary \\ 
           &            &        & (J2000)     & (J2000)    & (h) & (h)   &                     &                    & \\ \hline
2018-10-06     & 1538856059     & XMMLSS\_12       & 02\hhh17\mmm51\sss        & $-$04\ddd49\dmm59\farcs0  & 8.02           & 6.20            & 4096           & 59             & J1939-6342     & J0201-1132     \\
2018-10-07     & 1538942495     & XMMLSS\_13       & 02\hhh20\mmm42\sss        & $-$04\ddd49\dmm59\farcs0  & 8.07           & 6.22           & 4096           & 59             & J1939-6342     & J0201-1132     \\
2018-10-08     & 1539028868     & XMMLSS\_14       & 02\hhh23\mmm22\sss        & $-$04\ddd49\dmm59\farcs0  & 8.03           & 6.19           & 4096           & 60             & J1939-6342     & J0201-1132     \\
2018-10-11     & 1539286252     & XMMLSS\_12       & 02\hhh17\mmm51\sss        & $-$04\ddd49\dmm59\farcs0  & 8.05           & 6.23           & 4096           & 63             & J1939-6342     & J0201-1132     \\
2018-10-12     & 1539372679     & XMMLSS\_13       & 02\hhh20\mmm42\sss        & $-$04\ddd49\dmm59\farcs0  & 8.03           & 5.92           & 4096           & 62             & J1939-6342     & J0201-1132     \\
2018-10-13     & 1539460932     & XMMLSS\_14       & 02\hhh23\mmm22\sss        & $-$04\ddd49\dmm59\farcs0  & 8.00            & 6.24           & 4096           & 62             & J1939-6342     & J0201-1132     \\
2019-07-27     & 1564271932     & XMMLSS\_15       & 02\hhh26\mmm22\sss        & $-$04\ddd37\dmm58\farcs8  & 8.00            & 7.00            & 4096           & 59             & J1939-6342     & J0201-1132     \\
2019-08-02     & 1564788958     & XMMLSS\_12p5     & 02\hhh19\mmm16\sss        & $-$04\ddd49\dmm58\farcs8  & 7.99           & 6.97           & 4096           & 62             & 0408-6545      & J0201-1132     \\
2019-08-03     & 1564874467     & XMMLSS\_13p5     & 02\hhh22\mmm06\sss        & $-$04\ddd49\dmm58\farcs8  & 8.07           & 7.00            & 4096           & 62             & J1939-6342     & J0201-1132     \\
2020-01-06     & 1578317762     & XMMLSS\_12       & 02\hhh17\mmm51\sss        & $-$04\ddd49\dmm59\farcs0  & 8.00          & 6.50            & 32768          & 62             & J1939-6342     & J1008+0740     \\
2020-08-10     & 1597099565     & XMMLSS\_6\_8     & 02\hhh15\mmm47\sss        & $-$04\ddd49\dmm58\farcs8  & 8.00           & 6.85           & 32768          & 60             & J1939-6342     & J0201-1132     \\
2020-08-13     & 1597359662     & XMMLSS\_8\_8     & 02\hhh19\mmm54\sss        & $-$04\ddd49\dmm58\farcs8  & 8.00            & 6.83           & 32768          & 60             & J1939-6342     & J0201-1132     \\
2020-08-14     & 1597445461     & XMMLSS\_9\_8     & 02\hhh21\mmm58\sss        & $-$04\ddd49\dmm58\farcs8  & 8.00           & 6.83           & 32768          & 61             & J1939-6342     & J0201-1132     \\
2020-08-15     & 1597534262     & XMMLSS\_10\_8    & 02\hhh24\mmm02\sss        & $-$04\ddd49\dmm58\farcs8  & 8.00           & 6.83           & 32768          & 62             & J1939-6342     & J0201-1132     \\
2020-08-16     & 1597617063     & XMMLSS\_11\_8    & 02\hhh26\mmm06\sss        & $-$04\ddd49\dmm58\farcs8  & 8.00            & 6.83           & 32768          & 62             & J1939-6342     & J0201-1132     \\
2020-08-17     & 1597703462     & XMMLSS\_12\_8    & 02\hhh28\mmm09\sss        & $-$04\ddd49\dmm58\farcs8  & 8.00           & 6.83           & 32768          & 61             & J1939-6342     & J0201-1132     \\
2020-08-24     & 1598306465     & XMMLSS\_6\_9     & 02\hhh16\mmm48\sss        & $-$04\ddd23\dmm16\farcs8  & 8.00            & 6.85           & 32768          & 58             & J1939-6342     & J0201-1132     \\
2020-08-27     & 1598564761     & XMMLSS\_7\_9     & 02\hhh18\mmm52\sss        & $-$04\ddd23\dmm16\farcs8  & 8.00            & 6.83           & 32768          & 59             & J1939-6342     & J0201-1132     \\
2020-08-30     & 1598823546     & XMMLSS\_8\_9     & 02\hhh20\mmm56\sss        & $-$04\ddd23\dmm16\farcs8  & 7.99           & 6.83           & 32768          & 59             & J1939-6342     & J0201-1132     \\
2020-09-03     & 1599168664     & XMMLSS\_9\_9     & 02\hhh22\mmm59\sss        & $-$04\ddd23\dmm16\farcs8  & 7.99           & 6.83           & 32768          & 58             & J1939-6342     & J0201-1132     \\
2020-09-10     & 1599770763     & XMMLSS\_10\_9    & 02\hhh25\mmm03\sss        & $-$04\ddd23\dmm16\farcs8  & 7.99           & 6.83           & 32768          & 58             & J1939-6342     & J0201-1132     \\
2020-09-11     & 1599858231     & XMMLSS\_11\_9    & 02\hhh27\mmm07\sss        & $-$04\ddd23\dmm16\farcs8  & 7.99           & 6.83           & 32768          & 58             & J1939-6342     & J0201-1132     \\
2020-09-21     & 1600722156     & XMMLSS\_6\_7     & 02\hhh16\mmm49\sss        & $-$05\ddd16\dmm40\farcs8  & 8.05           & 6.85           & 32768          & 60             & J1939-6342     & J0201-1132     \\
2020-09-23     & 1600893068     & XMMLSS\_7\_7     & 02\hhh18\mmm53\sss        & $-$05\ddd16\dmm40\farcs8  & 7.97           & 6.78           & 32768          & 59             & J1939-6342     & J0201-1132     \\
2020-09-25     & 1601066490     & XMMLSS\_8\_7     & 02\hhh20\mmm57\sss        & $-$05\ddd16\dmm40\farcs8  & 7.97           & 6.78           & 32768          & 60             & J1939-6342     & J0201-1132     \\
2020-10-01     & 1601583156     & XMMLSS\_9\_7     & 02\hhh23\mmm01\sss        & $-$05\ddd16\dmm40\farcs8  & 7.98           & 6.78           & 32768          & 59             & J1939-6342     & J0201-1132     \\
2020-10-02     & 1601667467     & XMMLSS\_10\_7    & 02\hhh25\mmm04\sss        & $-$05\ddd16\dmm40\farcs8  & 7.97           & 6.78           & 32768          & 61             & J1939-6342     & J0201-1132     \\
2020-10-18     & 1603049464     & XMMLSS\_11\_7    & 02\hhh27\mmm08\sss        & $-$05\ddd16\dmm40\farcs8  & 8.00            & 6.78           & 32768          & 59             & J1939-6342     & J0201-1132     \\
2021-03-20     & 1616233334     & XMMLSS\_7\_6     & 02\hhh17\mmm52\sss        & $-$05\ddd43\dmm22\farcs8  & 3.99           & 3.12           & 32768          & 63             & J0408-6545     & J0201-1132     \\
2021-04-17     & 1618640777     & XMMLSS\_7\_6     & 02\hhh17\mmm52\sss        & $-$05\ddd43\dmm22\farcs8  & 8.00            & 6.74           & 32768          & 63             & J0408-6545     & J0201-1132     \\
2021-04-18     & 1618726873     & XMMLSS\_8\_6     & 02\hhh19\mmm56\sss        & $-$05\ddd43\dmm22\farcs8  & 8.00            & 6.74           & 32768          & 64             & J0408-6545     & J0201-1132     \\
2021-04-23     & 1619161272     & XMMLSS\_9\_6     & 02\hhh21\mmm59\sss        & $-$05\ddd43\dmm22\farcs8  & 8.00            & 6.74           & 32768          & 60             & J0408-6545     & J0201-1132     \\
2021-04-26     & 1619245873     & XMMLSS\_10\_6    & 02\hhh24\mmm03\sss        & $-$05\ddd43\dmm22\farcs8  & 8.00           & 6.74           & 32768          & 63             & J0408-6545     & J0201-1132     \\
2021-04-25     & 1619330180     & XMMLSS\_11\_6    & 02\hhh26\mmm07\sss        & $-$05\ddd43\dmm22\farcs8  & 8.00            & 6.74           & 32768          & 63             & J0408-6545     & J0201-1132     \\
2021-04-26     & 1619416874     & XMMLSS\_7\_5     & 02\hhh18\mmm54\sss        & $-$06\ddd10\dmm04\farcs8  & 8.35           & 6.74           & 32768          & 61             & J0408-6545     & J0201-1132     \\
2021-05-02     & 1619933473     & XMMLSS\_10\_5    & 02\hhh25\mmm06\sss        & $-$06\ddd10\dmm04\farcs8  & 8.03           & 6.74           & 32768          & 62             & J0408-6545     & J0201-1132     \\
2021-05-04     & 1620109872     & XMMLSS\_9\_5     & 02\hhh23\mmm02\sss        & $-$06\ddd10\dmm04\farcs8  & 8.03           & 6.74           & 32768          & 59             & J0408-6545     & J0201-1132     \\
2021-05-09     & 1620536533     & XMMLSS\_8\_5     & 02\hhh20\mmm58\sss        & $-$06\ddd10\dmm04\farcs8  & 8.00            & 6.70            & 32768          & 62             & J0408-6545     & J0201-1132     \\
2021-05-14     & 1620967872     & XMMLSS\_8\_4     & 02\hhh19\mmm57\sss        & $-$06\ddd36\dmm46\farcs8  & 8.00           & 6.70            & 32768          & 59             & J0408-6545     & J0201-1132     \\
2021-05-22     & 1621656680     & XMMLSS\_7\_4     & 02\hhh17\mmm53\sss        & $-$06\ddd36\dmm46\farcs8  & 8.00            & 6.70            & 32768          & 63             & J0408-6545     & J0201-1132     \\
2021-05-23     & 1621742776     & XMMLSS\_10\_4    & 02\hhh24\mmm05\sss        & $-$06\ddd36\dmm46\farcs8  & 8.00            & 6.70            & 32768          & 62             & J0408-6545     & J0201-1132     \\
2021-06-05     & 1622863620     & XMMLSS\_11\_4    & 02\hhh26\mmm09\sss        & $-$06\ddd36\dmm46\farcs8  & 8.00            & 6.70            & 32768          & 61             & J0408-6545     & J0201-1132     \\
2021-06-06     & 1622949019     & XMMLSS\_6\_9p5   & 02\hhh15\mmm47\sss        & $-$04\ddd05\dmm60\farcs0  & 8.00           & 6.70            & 32768          & 61             & J0408-6545     & J0201-1132     \\
2021-06-12     & 1623469482     & XMMLSS\_9\_4     & 02\hhh22\mmm01\sss        & $-$06\ddd36\dmm46\farcs8  & 8.00            & 6.70            & 32768          & 63             & J0408-6545     & J0201-1132     \\
2021-06-27     & 1624760792     & XMMLSS\_12\_9p5  & 02\hhh28\mmm09\sss        & $-$04\ddd05\dmm60\farcs0  & 8.00            & 6.70            & 32768          & 58             & J0408-6545     & J0201-1132     \\
\hline
\end{tabular}
\label{tab:xmm_obs}
\end{minipage}
\end{table*}

\section{Extract from Catalogue}

{Table \ref{tab:sourcetab} shows the} first five entries from the source catalogue of the {low resolution imaging of the COSMOS field}.

\begin{table*}
\centering
\caption{{An example source catalogue, showing the first five lines for the low resolution (8.9\arcsec) COSMOS catalogue. The columns for each source included from \textsc{PyBDSF} are: The ID of the source (Source\_id); the ID of the island of emission associated with the source (Isl\_id); the RA and Dec of the source  (in J2000 co-ordinates), the integrated flux density (Total\_flux) and peak flux density (Peak\_flux); source size information of the major and minor axes and position angle (Maj, Min and PA) and deconvolved source sizes (indicated by DC\_); rms across the island (Isl\_rms) and a code to describe the type of source (S\_Code where S = Single, M = Multiple, C = Complex). All errors are indicated by columns which begin with E\_. Further details of the \textsc{PyBDSF} columns can be found at \protect \url{https://pybdsf.readthedocs.io}. We additionally include a column with the effective frequency of the source, in MHz, at the source position (Eff\_freq) and a column which lists the number of Gaussian components that the source is comprised of (NGaus).}}
\begin{tabular}{ccccccccc} \hline
Source\_id & Isl\_id & RA & E\_RA & DEC & E\_DEC & Total\_flux & E\_Total\_flux & Peak\_flux \\ 
 &  & (\degree) & (\degree)  & (\degree)  & (\degree)  & (Jy) & (Jy) & (Jy beam$^{-1}$) \\ \hline \hline
0 & 0 & 151.173585 & 0.000063 & 2.216888 & 0.000053 & 0.0001580 & 0.0000135 & 0.0001373 \\
1 & 1 & 151.172523 & 0.000092 & 1.854424 & 0.000112 & 0.0000810 & 0.0000121 & 0.0000706 \\
2 & 3 & 151.173070 & 0.000222 & 2.541834 & 0.000289 & 0.0000756 & 0.0000183 & 0.0000452 \\
3 & 4 & 151.171141 & 0.000083 & 1.845267 & 0.000160 & 0.0001294 & 0.000016 & 0.0000747 \\
4 & 5 & 151.169030 & 0.000294 & 1.774813 & 0.000245 & 0.0000428 & 0.0000138 & 0.0000330 \\ \hline \\ \\ 
\end{tabular}
\begin{tabular}{ccccccccc} \hline
E\_Peak\_flux & Maj & E\_Maj & Min & E\_Min & PA & E\_PA & DC\_Maj & E\_DC\_Maj \\
(Jy beam$^{-1}$)  & (\degree) & (\degree) & (\degree)  & (\degree)  & (\degree)  & (\degree) & (\degree) & (\degree) \\ \hline \hline
0.0000071 & 0.002774 & 0.000149 & 0.002534 & 0.000125 & 87.855644 & 24.131809 & 0.001260 & 0.000149 \\
0.0000064 & 0.002795 & 0.000265 & 0.002507 & 0.000215 & 173.143515 & 35.577114 & 0.001306 & 0.000265 \\
0.0000073 & 0.004090 & 0.000790 & 0.002500 & 0.000333 & 34.249398 & 17.301097 & 0.003258 & 0.000790 \\
0.0000062 & 0.003945 & 0.000379 & 0.002682 & 0.000194 & 174.329007 & 10.942307 & 0.003075 & 0.000379 \\
0.0000065 & 0.003461 & 0.000820 & 0.002285 & 0.000377 & 126.972961 & 23.713182 & 0.002423 & 0.000820 \\ \hline \\ \\ 
\end{tabular}
\begin{tabular}{ccccccccc} \hline
DC\_Min & E\_DC\_Min & DC\_PA & E\_DC\_PA & Isl\_rms & S\_Code & Eff\_freq & NGaus &  \\
(\degree) & (\degree) & (\degree)  & (\degree)  & (Jy beam$^{-1}$)  &  & (MHz) &  \\ \hline \hline
0.000558 & 0.000125 & 87.855644 & 24.131809 & 0.0000069 & S & 1225.10 & 1 \\
0.000420 & 0.000215 & 173.143515 & 35.577114 & 0.0000062 & S & 1222.32 & 1 \\
0.000380 & 0.000333 & 34.249398 & 17.301097 & 0.0000070 & S & 1223.50 & 1 \\
0.001043 & 0.000194 & 174.329007 & 10.942307 & 0.0000059 & S & 1222.16 & 1 \\
0.000000 & 0.000377 & 126.972961 & 23.713182 & 0.0000064 & S & 1236.36 & 1 \\  \hline
\end{tabular}
\label{tab:sourcetab}
\end{table*}

\section{{Signal-to-Noise Envelopes Applied}}
\label{sec:snrenv}
{To identify unresolved sources we fit an upper envelope of the form:}

\begin{equation}
{\frac{S_{I}}{S_{P}} = A + B \times \textrm{SNR}^{-C}},
\label{eq:snr_env}
\end{equation}

\noindent {where $S_{I}$ is the integrated flux density, $S_{P}$ is the peak flux density, SNR is the peak signal-to-noise ratio and $A$, $B$ and $C$ are constants which are fit for as in the method of \cite{Hale2021}. The parameters used for the catalogue validation is given in Table \ref{tab:snr_env}.}

\begin{table*}
\caption{{Parameters $A$, $B$ and $C$ used to define the upper signal-to-noise envelopes as in Equation \ref{eq:snr_env}, to identify unresolved sources, as used in Section \ref{sec:validation}. }}
\begin{tabular}{ccccc}
\hline
Field & Image Resolution & $A$ & $B$ & $C$ \\
 & (\arcsec) & & &  \\ \hline
{CDFS-DEEP} &  5.5 & 1.02 & 0.80 & 0.75 \\
CDFS-DEEP & 7.3 & 1.04 & 0.60 & 0.55 \\ \hline

COSMOS & 5.2 & 1.06 & 1.10 & 0.75\\
COSMOS & 8.9 & 1.01 & 0.70 & 0.60 \\ \hline

XMM-LSS &  5.0 & 1.06 & 1.10 & 0.70 \\
XMM-LSS & 8.9 & 1.00 & 0.60 & 0.55\\ \hline

\label{tab:snr_env}
\end{tabular}
\end{table*}

\section{Table of source counts}

{Table \ref{tab:sc_tab} presents the 1.4 GHz Euclidean normalised source counts in each of the three fields observed.}

\bsp	

\begin{landscape}
 \begin{table}
\caption{{Details of the 1.4 GHz Euclidean normalised source counts presented in Section \protect \ref{sec:sourcecounts} and Figure \ref{fig:sc}. Given are the flux density ranges for the source counts, the median flux density of sources within the bin considered, the raw number counts and source counts as well as the corrected source counts for each of the three fields studied: CDFS-DEEP, COSMOS and XMM-LSS {both through (i) combining all completeness simulations of the high flux density simulations at $\geq10\times$ the minimum input flux density and (ii) combining completeness simulations together for $>$1 mJy. This is released as supplementary material alongside the paper, in which the terms `nerr' an `perr' are used to indicate the negative and positive errors respectively.} }}
\addtolength{\tabcolsep}{-0.4em}
\begin{tabular}{c| ccc | ccc | ccc | ccc | ccc}
-------------------- & \multicolumn{3}{c}{----------- CDFS-DEEP -----------}  & \multicolumn{3}{c}{------------ COSMOS ------------} & \multicolumn{3}{c}{----------- XMM-LSS -----------} & \multicolumn{3}{c}{---- {Completeness (High Flux $\geq$10$\times S_{\textrm{min}}$)} ----} & \multicolumn{3}{c}{------- { Completeness (combined >1 mJy)} -------} \\
Flux  & Median & Raw & Raw  & Median & Raw & Raw  & Median & Raw & Raw & Corrected & Corrected & Corrected  & Corrected & Corrected & Corrected \\
Density & Flux & Number & Source  & Flux & Number & Source & Flux & Number & Source& Source & Source & Source  & Source  & Source  & Source \\
Range  & Density& Counts &  Counts & Density& Counts &  Counts &  Density& Counts &  Counts & Counts & Counts & Counts & Counts & Counts & Counts \\ \
  & &  &   &  &  &  &   & &  & CDFS-DEEP & COSMOS & XMM-LSS  & CDFS-DEEP & COSMOS & XMM-LSS \\ \
($\muup$Jy) & ($\muup$Jy) & &  (Jy$^{1.5}$sr$^{-1}$) & ($\muup$Jy) & &  (Jy$^{1.5}$sr$^{-1}$) &  ($\muup$Jy) & &  (Jy$^{1.5}$sr$^{-1}$) & (Jy$^{1.5}$sr$^{-1}$)&  (Jy$^{1.5}$sr$^{-1}$) & (Jy$^{1.5}$sr$^{-1}$) & (Jy$^{1.5}$sr$^{-1}$) & (Jy$^{1.5}$sr$^{-1}$) & (Jy$^{1.5}$sr$^{-1}$) \\ \hline
10-15 & 13 & 3058 & $0.62^{-0.01}_{+0.01}$ &- & - & - &- & - & - & $1.94^{-0.08}_{+0.08}$ & - & - & $1.94^{-0.08}_{+0.08}$ & - & - \\ 
15-25 & 19 & 4122 & $1.68^{-0.03}_{+0.03}$ &- & - & - &- & - & - & $3.07^{-0.11}_{+0.11}$ & - & - & $3.07^{-0.11}_{+0.11}$ & - & - \\ 
25-39 & 31 & 3468 & $2.82^{-0.05}_{+0.05}$ &31 & 6068 & $1.81^{-0.02}_{+0.02}$ &31 & 21074 & $1.84^{-0.01}_{+0.01}$ & $3.07^{-0.10}_{+0.10}$ & $2.94^{-0.07}_{+0.07}$ & $2.76^{-0.03}_{+0.03}$ & $3.07^{-0.10}_{+0.10}$ & $2.94^{-0.07}_{+0.07}$ & $2.76^{-0.03}_{+0.03}$ \\ 
39-63 & 48 & 2476 & $4.01^{-0.08}_{+0.08}$ &49 & 5152 & $3.07^{-0.04}_{+0.04}$ &49 & 18141 & $3.16^{-0.02}_{+0.02}$ & $3.75^{-0.13}_{+0.13}$ & $3.10^{-0.07}_{+0.07}$ & $3.17^{-0.04}_{+0.04}$ & $3.75^{-0.13}_{+0.13}$ & $3.10^{-0.07}_{+0.07}$ & $3.17^{-0.04}_{+0.04}$ \\ 
63-100 & 77 & 1493 & $4.83^{-0.12}_{+0.13}$ &76 & 3564 & $4.24^{-0.07}_{+0.07}$ &77 & 12458 & $4.34^{-0.04}_{+0.04}$ & $4.11^{-0.16}_{+0.16}$ & $3.58^{-0.10}_{+0.10}$ & $3.60^{-0.05}_{+0.05}$ & $4.11^{-0.16}_{+0.16}$ & $3.58^{-0.10}_{+0.10}$ & $3.60^{-0.05}_{+0.05}$ \\ 
100-158 & 118 & 772 & $4.98^{-0.18}_{+0.19}$ &120 & 2080 & $4.93^{-0.11}_{+0.11}$ &121 & 7041 & $4.89^{-0.06}_{+0.06}$ & $4.24^{-0.21}_{+0.22}$ & $4.12^{-0.15}_{+0.15}$ & $4.02^{-0.06}_{+0.06}$ & $3.91^{-0.20}_{+0.21}$ & $3.95^{-0.14}_{+0.14}$ & $3.83^{-0.06}_{+0.06}$ \\ 
158-251 & 192 & 379 & $4.88^{-0.25}_{+0.26}$ &191 & 1072 & $5.07^{-0.15}_{+0.16}$ &193 & 3667 & $5.08^{-0.08}_{+0.09}$ & $4.09^{-0.25}_{+0.26}$ & $4.25^{-0.19}_{+0.19}$ & $4.14^{-0.09}_{+0.09}$ & $3.58^{-0.23}_{+0.24}$ & $4.16^{-0.18}_{+0.18}$ & $3.98^{-0.09}_{+0.09}$ \\ 
251-398 & 307 & 203 & $5.21^{-0.37}_{+0.39}$ &306 & 583 & $5.50^{-0.23}_{+0.24}$ &307 & 1961 & $5.42^{-0.12}_{+0.13}$ & $4.75^{-0.43}_{+0.44}$ & $4.85^{-0.26}_{+0.27}$ & $4.82^{-0.14}_{+0.14}$ & $4.28^{-0.39}_{+0.41}$ & $4.62^{-0.27}_{+0.28}$ & $4.60^{-0.14}_{+0.14}$ \\ 
398-630 & 480 & 105 & $5.38^{-0.52}_{+0.58}$ &488 & 316 & $5.95^{-0.33}_{+0.35}$ &480 & 1061 & $5.85^{-0.18}_{+0.19}$ & $5.09^{-0.66}_{+0.70}$ & $5.24^{-0.39}_{+0.40}$ & $5.20^{-0.21}_{+0.21}$ & $5.51^{-0.75}_{+0.79}$ & $4.78^{-0.38}_{+0.39}$ & $5.15^{-0.22}_{+0.22}$ \\ 
630-1000 & 738 & 52 & $5.32^{-0.73}_{+0.84}$ &746 & 176 & $6.61^{-0.50}_{+0.54}$ &779 & 623 & $6.86^{-0.27}_{+0.29}$ & $5.48^{-0.99}_{+1.08}$ & $6.39^{-0.64}_{+0.67}$ & $6.78^{-0.39}_{+0.40}$ & $7.53^{-1.66}_{+1.76}$ & $6.25^{-0.60}_{+0.63}$ & $6.66^{-0.37}_{+0.38}$ \\ 
1000-1584 & 1260 & 42 & $8.57^{-1.32}_{+1.54}$ &1245 & 131 & $9.82^{-0.86}_{+0.94}$ &1221 & 415 & $9.11^{-0.45}_{+0.47}$ & $9.06^{-1.83}_{+2.02}$ & $9.87^{-1.17}_{+1.23}$ & $9.20^{-0.61}_{+0.63}$ & $9.16^{-1.96}_{+2.13}$ & $9.81^{-1.18}_{+1.24}$ & $9.30^{-0.67}_{+0.68}$ \\ 
1584-2511 & 1973 & 39 & $15.88^{-2.53}_{+2.97}$ &1989 & 92 & $13.76^{-1.43}_{+1.59}$ &1965 & 294 & $12.88^{-0.75}_{+0.80}$ & $17.65^{-3.86}_{+4.23}$ & $14.35^{-2.14}_{+2.26}$ & $13.26^{-1.06}_{+1.09}$ & $18.19^{-4.43}_{+4.77}$ & $14.35^{-2.20}_{+2.32}$ & $13.20^{-1.08}_{+1.11}$ \\ 
2511-3981 & 3264 & 26 & $21.12^{-4.11}_{+5.00}$ &3044 & 61 & $18.21^{-2.32}_{+2.64}$ &3100 & 204 & $17.83^{-1.25}_{+1.34}$ & $23.40^{-6.19}_{+6.95}$ & $18.71^{-3.21}_{+3.46}$ & $18.42^{-1.74}_{+1.81}$ & $25.93^{-11.96}_{+12.46}$ & $18.63^{-3.35}_{+3.59}$ & $18.20^{-1.77}_{+1.84}$ \\ 
3981-6309 & 4756 & 20 & $32.41^{-7.18}_{+8.98}$ &4983 & 69 & $41.10^{-4.93}_{+5.57}$ &5018 & 175 & $30.52^{-2.30}_{+2.49}$ & $36.23^{-10.54}_{+12.15}$ & $46.83^{-8.38}_{+8.88}$ & $33.68^{-3.45}_{+3.60}$ & $37.26^{-11.79}_{+13.33}$ & $51.25^{-15.06}_{+15.40}$ & $34.87^{-4.42}_{+4.55}$ \\ 
6309-10000 & 8358 & 14 & $45.27^{-11.94}_{+15.60}$ &7455 & 39 & $46.35^{-7.39}_{+8.67}$ &8126 & 124 & $43.15^{-3.87}_{+4.23}$ & $55.45^{-19.04}_{+22.67}$ & $56.68^{-12.51}_{+13.68}$ & $50.37^{-5.83}_{+6.17}$ & $76.86^{-100.85}_{+102.28}$ & $75.51^{-64.98}_{+65.40}$ & $50.34^{-6.10}_{+6.42}$ \\ 

 \\ 
 \hline
\label{tab:sc_tab}
  \end{tabular}
 \end{table}
\end{landscape}

\label{lastpage}
\end{document}